# Protein residue networks from a local search perspective

Susan Khor

slc.khor@gmail.com

(Date: Nov 28, 2015)

**Abstract**

We examined protein residue networks (PRNs) from a local search perspective to understand why PRNs are highly clustered when having short paths is important for protein functionality. We found that by adopting a local search perspective, this conflict between form and function is resolved as increased clustering actually helps to reduce path length in PRNs. Further, the paths found via our EDS local search algorithm are more congruent with the characteristics of intra-protein communication. EDS identifies a subset of PRN edges called short-cuts that are distinct, have high usage, impacts EDS path length, diversity and stretch, and are dominated by short-range contacts. The short-cuts form a network (SCN) that increases in size and transitivity as a protein folds. The structure of a SCN supports its function and formation, and the function of a SCN influences its formation. Several significant differences in terms of SCN structure, function and formation is found between "successful" and "unsuccessful" MD trajectories, with SCN transitivity playing a central role. By connecting the static and the dynamic aspects of PRNs, the protein folding process becomes a problem of network/graph formation with the purpose of forming suitable pathways within proteins.

*Outline of manuscript:*

Note: Only materials and results related to sections 3.1 to 3.6 have been published in the JCN journal article which has the same title as this manuscript. Materials and results from section 3.7 and section 4 are each expanded in other manuscripts.

## 1. Introduction

The genetic system is a heritable mechanism for producing proteins which are the building-blocks of life. In general, proteins attain their structure necessary for functioning via the different possible physical and chemical interactions amongst their amino acid molecules within the constraints of their environment. One way of understanding protein structure is through their *contact maps*, which abstracts away the physical and chemical details and puts the spotlight on contacts between amino acid molecules of a protein. More formally, the contact map of a protein is the adjacency matrix **A** of a graph $G$ representing the protein as a set of nodes $V$ and a set of edges $E$. Typically, each node in $V$ represents an amino acid, and an edge is placed between a node-pair if they satisfy certain conditions, e.g. if they are within an acceptable Euclidean distance from each other. We call such a graph (constructed in accordance with section 2.1) a *Protein Residue Network* (PRN). Other criteria have been used to construct the network of interacting amino acids within proteins. To avoid confusion, we will refer to the general class of networks induced by protein contact maps as *residue interaction networks* (RINs), of which our PRNs are a specific form.

Pursuant to the introduction of a model of the small-world phenomenon in social networks [1], networks induced by protein contact maps were classified as small-world networks [2]. This study is performed on globular proteins, although both fibrous and membrane proteins are qualitatively small-world networks also [3, 4]. A *small-world network* (SWN) combines the order inherent in regular graphs, with the arbitrary connectivity of pure random graphs (where any two vertices have a non-zero probability to link with each other) to supply the short-cut edges or long-range connections. The second half of this definition has undergone refinement. Briefly, for a SWN to be navigable by a local search algorithm, the short-cut edges need not be long-ranged or random, but multi-scaled [5, 6, 7]. More formally, a graph with $N$ nodes and average degree $K$ is identified as having SWN structure if: (i) its clustering coefficient $C$ is significantly larger than the clustering coefficient of a comparable Erdos-Renyi (ER) random graph with $C_{ER} \sim K / N$; and (ii) its characteristic path-length $L$ approaches that of a comparable ER random graph with $L_{ER} \sim \ln(N) / \ln(K)$. For constant $K$, the latter property implies that $L$ increases logarithmically with $N$. Most real-world networks (food webs being an exception) can be considered sparse, i.e. their mean vertex degree $K$ remains constant with increase in network size $N$ [8 p.19, 9 p. 134]. The increase in $L$ can be slower than $\ln(N)$ for small-world networks with power-law degree distribution [10].

The clustering coefficient $C$ reflects the probability that two unique nodes $u$ and $v$ which are directly connected to a third other node $w$, are themselves connected in the network forming a triangle. Typically, the clustering coefficient for a network with $N$ nodes is $C = \dfrac{1}{N} \sum_i^N C_i$ where $C_i = \dfrac{2e_i}{k_i(k_i - 1)}$ is the clustering coefficient of a node $i$ with degree $k_i$ and $e_i$ is the number of links amongst $i$'s $k_i$ direct neighbor nodes [1]. Links in an ER graph are independent of each other, so $C_{ER} = p = \frac{2M}{N(N-1)} = \frac{K}{(N-1)}$ where $p$ is the probability of connecting two nodes in the ER graph, $M$ is the number of links and $K = \frac{2M}{N}$. Typically, $L$ is the average length of paths between all unique node-pairs in a network: $L = \frac{2}{N(N-1)} \sum_{i<j}^{N} \lambda(i,j)$ where $\lambda(i,j)$ is the length of a path (number of edges in a path) from node $i$ to node $j$ found through a global search such as breadth-first search (BFS). All networks considered in this paper are simple graphs.

The description of proteins as small-world networks is intuitive since the combination of order and randomness in a SWN parallels the coexistence of the highly ordered alpha helical and beta sheet secondary structures with random coils in protein molecules. Further, the short characteristic path length of a SWN appeals directly to the need for rapid communication between distantly located sites in a protein. Such efficient long-range intra-protein communication underpins allosteric interactions between cooperative binding sites which are crucial for proteins to be functional [11, 12]. Proteins may have more than one binding site, which may be (un)-occupied in concert. More commonly, activation of a site regulates the binding receptivity of other sites on the same protein. The ability for such long-range interactions has also been observed in non-allosteric proteins, and is believed to be a fundamental phenomenon of all globular proteins [11, 13]. Indeed, the short characteristic path length of a SWN is a widely mentioned and frequently applied topological feature of RINs [2, 14, 15, 16, 17].

With a few exceptions [3, 18, 19], there is less discussion about the role clustering plays in proteins. This may be because clustering in RINs is seen as an inevitable consequence of protein packing in 3D space. However, the presence of clustering in RINs needs to be explained *and related to the characteristic path length* of RINs if we are to fully understand why proteins have SWN structure and to fully exploit a network model of proteins. We propose that the role of clustering in RINs can be better understood in the light of a *greedy local search*. In a greedy local search, information available to the algorithm to decide the next step in a path is confined to the neighbourhood within a small radius of the current node, and the algorithm moves to the neighbouring node that is closest to the target node. As such a greedy local search is *directed* and in our case, as in Kleinberg's [5, 6], by proximity to the target. There exist decentralized search algorithms that do not require target location information [e.g. 20]. Where necessary, $L_G$ refers to the characteristic path length of a RIN found with a global search strategy, and $L_W$ to one obtained with a local search strategy.

The majority of previous research on RINs has implicitly defined characteristic path length and other network statistics derived from path length such as betweenness centrality and closeness centrality, in terms of shortest paths found via a global search strategy such as BFS on unweighted RINs or Dijkstra's algorithm on weighted RINs [17, 19]. An exception is [21] which uses a Markov random walk on RINs. There are three issues with using global search on RINs (since PRNs are un-weighted, we will discuss global search in terms of BFS).

First, "[vibrational] energy flow in globular proteins resembles transport on a percolation cluster with channels through which energy flows easily and dead-end regions where energy flow stalls" [13]. Vibrational energy is due to stretching and twisting of molecular bonds, and is also generated to compensate for loss in translational and rotational degrees of freedom when two molecules associate [12]. The transport of such energy through specific pathways is believed to be a mechanism underlying allosteric interactions in proteins. Anisotropic energy flow has been experimentally observed to occur efficiently between two allosterically linked binding sites: FA1 in subdomain IB and Sudlow site I in subdomain IIA, in the multi-domain BSA protein [22].

In a BFS on PRNs, there is very little variation in the lengths of paths, i.e. all nodes are easily reached from every other node (section 3.1). Thus BFS models a communication strategy where there is little specificity in inter-nodal communication. But protein sites are not created equal: certain sites are more actively involved in protein activity and are more evolutionarily conserved than others [23]. While there is a fair amount of redundant pathways, specific communication pathways within a protein molecule have been traced, and some of these pathways exist to slow down communication as a way to absorb or localize the after effects of undesirable perturbations to maintain protein stability [15].

Second, a BFS progresses by radiating outward from a source node in all possible directions. Thus, BFS is directionless, and unlike energy flow in proteins which is anisotropic and sub-diffusive [13]. The lack of direction increases the volume of space explored by BFS, i.e. the number of sites or nodes visited during a search. While the average length of shortest paths found with BFS increases only logarithmically with network size (section 3.1), the cost of BFS in terms of the average number of unique nodes visited during the search increases linearly with network size (section 3.2). The linear search cost of BFS is known [8 p.44]. In contrast, a greedy local search is more focused and less diffusive. Hence it is less costly and it turns out that for PRNs, it is possible for a Euclidean distance directed greedy local search with backtrack (our EDS algorithm is described in section 2.5) to produce paths with average length that increases logarithmically with network size, and at a cost that also increases logarithmically with network size. The small-world property of PRNs is preserved with local search, and as such PRNs are *navigable* small-world networks.

Third, the clustering coefficient of folded proteins is significantly larger than $C_{ER}$, and a high level of clustering forms a barrier to short inter-nodal path-lengths $L_G$. The direct relationship between $C$ and $L_G$ follows directly from $L_{ER} \sim \ln(N) / \ln(K)$. ER graphs with typical low $p$ have little to no clustering and are therefore locally tree-like. The average degree $K$ then approximates the branching factor of a search tree. By introducing cycles into the search tree, i.e. increasing clustering, some of the branches now loop back to a previous tree level. In other words, clustering reduces the effective branching factor, or effective $K$. Consequently, $L_G$ becomes larger than $L_{ER}$.

However, if short inter-nodal distances are important for protein functionality, why should proteins take on a highly clustered conformation that prevents them from having shorter pathways? Molecular dynamics (MD) simulation on the 2EZN protein confirms the presence of a significant gap in $C$ values between the set of PRNs for configurations at equilibrium, and the set of PRNs for non-equilibrium configurations (Fig. 1). This contradiction does not arise with a local view of search where suitable clustering is an enabler, not an impediment, to shorter path lengths. The inverse relationship between $C$ and $L_W$ is more attuned to what happens when a protein folds or unfolds. MD simulations of the 2EZN protein reveal that $C$ generally increases (decreases) and $L$ generally decreases (increases) as the protein folds (unfolds) (Fig. 2).

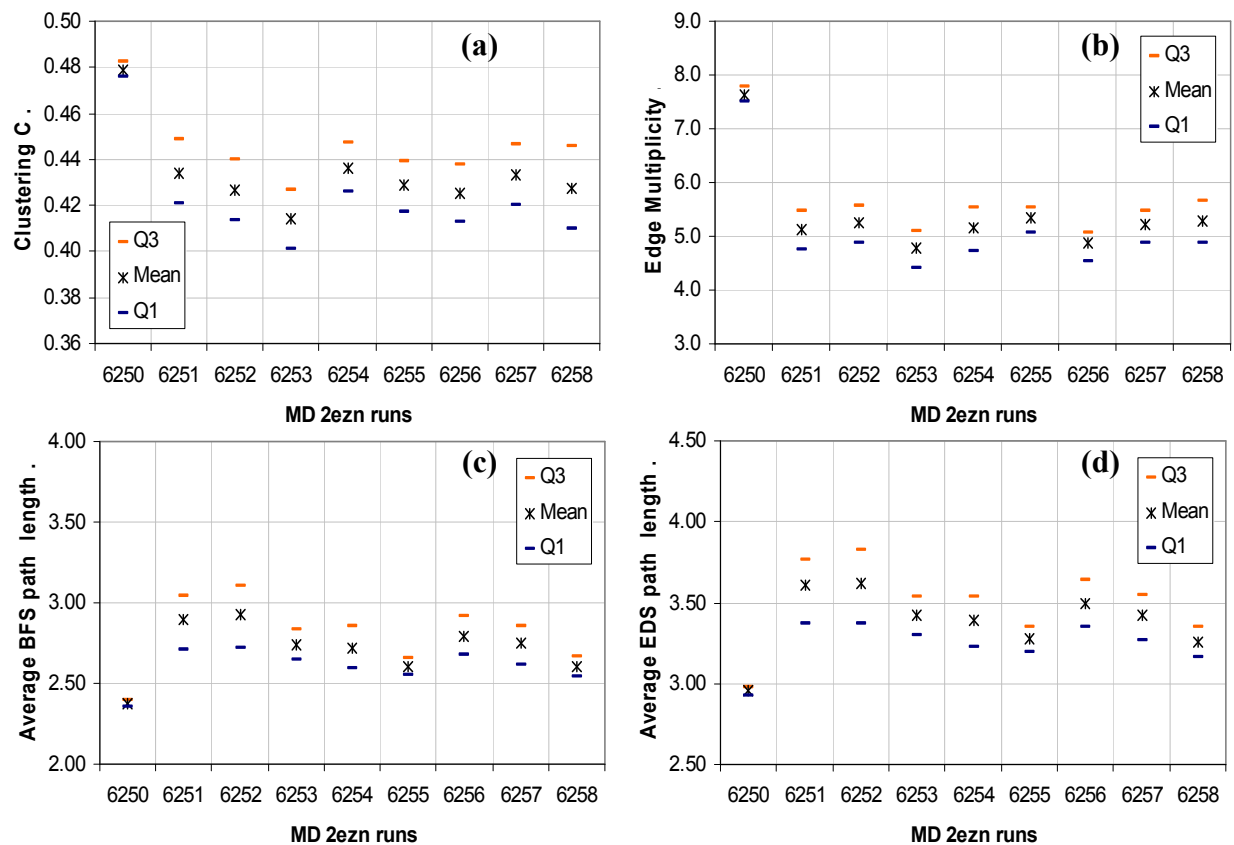

**Fig. 1 A topological gap exists between native and non-native configurations. Native configurations are much more clustered and have much shorter paths on average. (a & b)** Native state PRNs have significantly higher clustering and stronger transitivity (section 2.3) than non-native state PRNs. **(c & d)** Regardless of the type of search algorithm (BFS or EDS), paths in native state PRNs are significantly shorter on average than paths in non-native state PRNs. Averages are computed over all snapshots (PRNs) per MD run (section 2.7). Run 6250 is a simulation of the native dynamics of the 2EZN protein, while the remaining runs are simulations of non-equilibrium dynamics of 2EZN. Q1 and Q3 denote values for the first and third quartiles respectively.

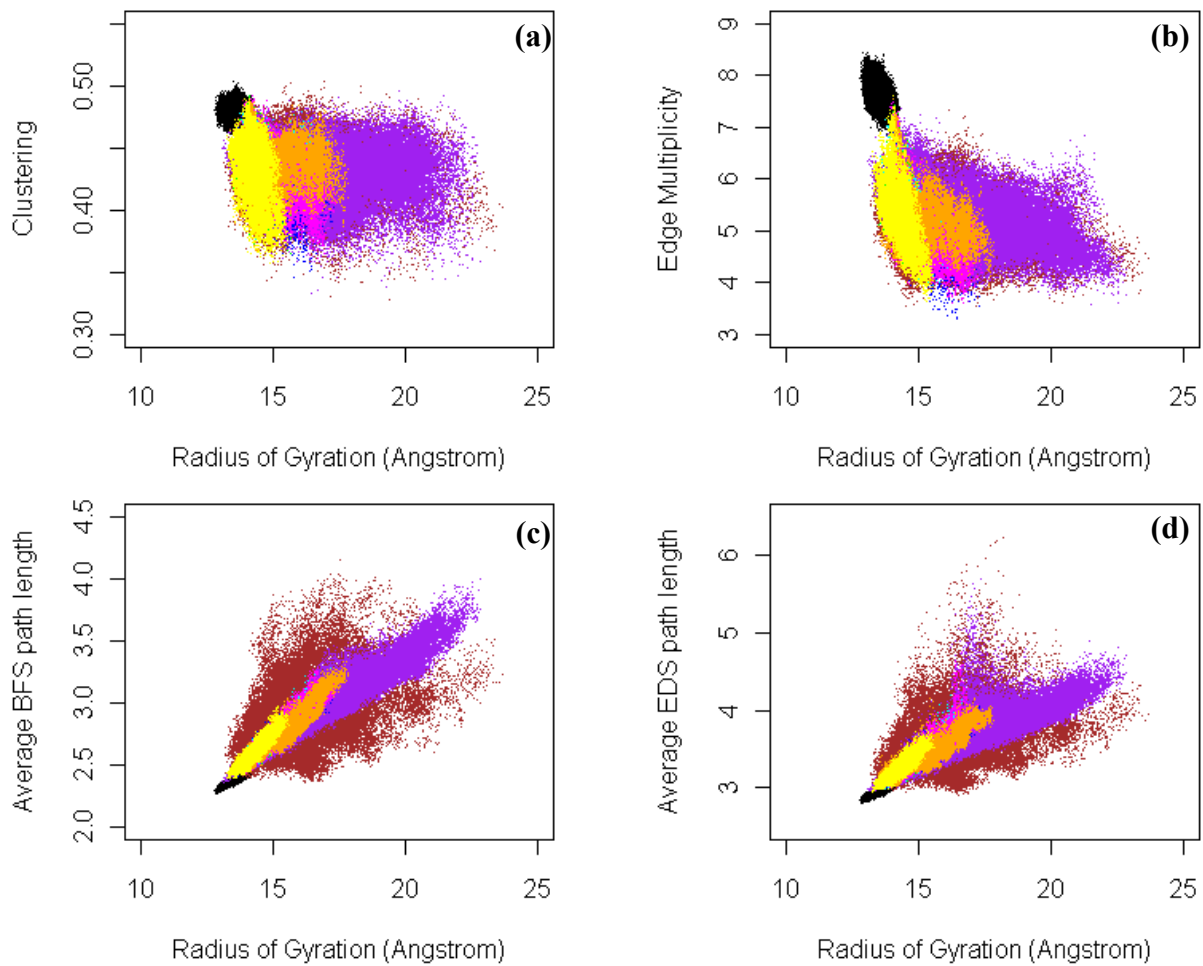

**Fig. 2 A local search perspective of PRNs is more reflective of what happens when a protein folds or unfolds due to its inverse relationship between clustering and average path length.** Higher clustering and stronger transitivity has an inflationary effect on BFS path lengths, but a deflationary effect on EDS path lengths (section 3.1). Since clustering **(a)** and transitivity **(b)** increases while average path length decreases **(c & d)** as a protein becomes more compact (radius of gyration decreases), we propose that a local search perspective of PRNs, e.g. with EDS, makes more sense than a global search perspective, e.g. with BFS. The scatter plots show the network statistic for each snapshot in a run. Black points denote the native state statistics (section 2.7).

## 2. Materials and Method

Since we are proposing the use of local search to investigate PRNs, this study focuses quite heavily on the differences between global search implemented as BFS (Breath-First Search) and local search implemented as EDS (section 2.5). The differences are measured primarily in terms of path length (section 3.1) and search cost (section 3.2), but other differences stemming from path length differences, as well as some unexpected similarities are reported in section 3. To observe the impact of clustering on local search, a null model in the form of the MGEO networks, which are identical to PRNs in several respects including the 3D coordinates of nodes, is designed (section 2.2).

In all figures, the average (mean), standard deviation, median, first quartile and third quartile values for a variable are indicated by "avg", "sd", "mid", "Q1" and "Q3" respectively. A "BFS" prefix denotes that the statistic is measured with a set of paths found through Breadth-First Search. An "EDS" prefix denotes that the statistic is measured with a set of paths found with the EDS algorithm. Unless stated otherwise, significance tests are made with either R's t.test or Wilcox.test, in paired form where applicable, and a p-value $< 0.05$ is required for significance. Eigenvalues and eigenvectors were calculated with GNU Octave 3.8.1 on Linux. In figures where the x-axis is labeled "Nodes", read as $N$ (number of nodes).

## 2.1 PRN construction

Define a Protein Residue Networks (PRN) as a simple undirected graph $G = (V, E)$ with $|V| = N$ and $|E| = M$. The PRNs are constructed from the PDB coordinates files [24] with side-chain considerations following the method in [25 & 26]. Each element in the set of nodes $V$ represents an amino acid molecule (residue) in a protein sequence. Nodes are labeled by the residue id (*rid*) given in the coordinates file. Two nodes $u$ and $v$ are linked *iff* $| u - v | \geq 2$ and their interaction strength $I_{uv}$ is $\geq 5.0\%$. $I_{uv} = \dfrac{n_{uv} \times 100}{\sqrt{R_u \times R_v}}$ where $n_{uv}$ is the number of distinct atom-pairs $(i, j)$ such that $i$ is an atom of residue $u$, $j$ is an atom of residue $v$ and the Euclidean distance between atoms $i$ and $j$, $ed(i, j)$ is $\leq 7.5$ Å. All the atoms of a residue, including those of the backbone, are considered when calculating $n_{uv}$ (this departs from [25] where only the side-chain atoms are used to calculate $I_{uv}$). $R_u$ and $R_v$ are normalization values by residue type. They are obtained from Table 1 in [26] (and expanded in Appendix A of this paper to accommodate alternatives that appear in the MD datasets). Peptide bonds are excluded, i.e. links are prohibited between nodes $i$ and $i+1$. Besides being faithful to the model in [25 & 26], this constraint is also consistent with the model in [49] which we refer to in section 3.7. Other structural consequences to excluding peptide bonds are expanded upon in Appendix H, most notably is the change in clustering and in the number of short-cut edges.

The threshold value-pair of 5.0% and 7.5 Å was chosen after some initial experiments to permit most intra-protein domain-based interactions identified in 3did [27] to be edges in most of the 2000 initial PRNs sampled at random from the list of proteins appearing in the 3did catalog (see Appendix C for overview of protein selection for network construction). A pair of PFAM domains is deemed able to interact with each other if they have at least five estimated contacts (hydrogen bonds, electrostatic or van der Waals interaction) between them [27]. The 3did links are the intra- or inter-chain residue-residue interactions between contacting PFAM domains of a protein as listed in the 3did catalog.

Each PRN satisfies the following criteria: (i) there are no missing atoms in the PDB coordinate file, (ii) all domain-based interactions cataloged in 3did are represented as links, (iii) the PRN is a single connected component, and (iv) there is a power-law relation between the number of nodes $N$ and link density. The last condition reflects the sparseness of PRNs. The link density of PRNs is $\frac{2M}{(N-1)(N-2)} \sim KN^{-1}$ and average node degree $K$ is constant for PRNs (Fig. 5a).

The PRNs (and MGEO networks from section 2.2) are reduced to the same set of 166 proteins after screening out outliers to permit pairwise comparison. Compared to the BFS paths, the EDS paths show much greater variation in length which section 3.1 argues is a more realistic model for proteins. However, the extremely large values skew path-length statistics and spills over to other path-length related statistics such as link usage and betweenness centrality. Therefore the networks were screened to limit the effect of outliers by excluding networks whose average BFS or EDS path exceeds $ln(max(N)) = 7.482119$ where $max(N)$ is the number of nodes in the largest PRN in the set of 204 PRNs.

The edges or links of a PRN are partitioned into two sets according to their sequence distance. A link connecting nodes $u$ and $v$ is long-range if $u$ and $v$ are more than 10 residues apart on the protein sequence, i.e. $|u - v| > 10$ [28]. *Long-range* edges (*LE*) represent interactions between residues which are far apart on the protein sequence but close to each other in the tertiary structure. *Short-range* edges (*SE*) represent interactions between residues that are close to each other in the primary and the tertiary structures.

Since $K$ is constant with $N$ (Fig. 5a), the number of edges $M$ is linearly related with the number of nodes $N$. The number of short-range edges (*SE*) is also linearly related with $N$, which means that the number of long-range edges (*LE*) also relates linearly with $N$. The effect of $N$ on both *SE* and *LE* cancels out each other to a large extent, and the ratio $|LE|/|SE|$ increases only slightly with $N$ (Fig. 3b). The number of 3did edges also increases with $N$ in general (Fig. 3c), but the linear correlation is weaker, 0.6091. The 3did links are mostly long-range (Fig. 3c).

## 2.2 MGEO network construction

The MGEO networks are constructed in a similar manner as random geometric networks. Random geometric networks with random coordinates were proposed in [30] as null models for RINs [1]. However, the MGEO networks use the PDB coordinates of the Cα atoms. To increase the number of single component MGEO networks, the nodes are first connected to form a "backbone" as follows: link $u$, the most recent node to join the backbone, to a node $v$ which is not already in the backbone and is closest in Euclidean distance to $v$ subject to $|u - v| \geq 2$. The backbone starts with a single node chosen uniformly at random. Due to the way the backbone is constructed, there is no guarantee that the backbone will connect all the nodes in a network. For example, in a network comprising five nodes, if the backbone is built with edge (1, 5) then (5, 3), it is impossible for nodes 2 or 4 to join the backbone. The backbone edges tend to be short (Fig. 3d). Next, the 3did edges are added. Finally, $m$ edges are added so that a MGEO network will have identical link density as its PRN (Fig. 3a). $m = M - B - D$ where $M$ and $D$ are respectively the number of links and 3did edges of a PRN, and $B$ is the number of links used to construct the backbone. There may be overlap between backbone links and 3did edges.

---

[1] Although [29] reported that random geometric graphs (RGG) have many network statistics that are quantitatively similar to RINs, we note significant differences between RGGs and PRNs in terms of correlation between Euclidean distance and sequence distance of links, graph connectivity, the $|LE|/|SE|$ ratio, and the number of short-cut edges generated by EDS. In particular, that there is a non-zero positive correlation between link Euclidean distance and link sequence distance is an important assumption for our results. Details in Appendix B.

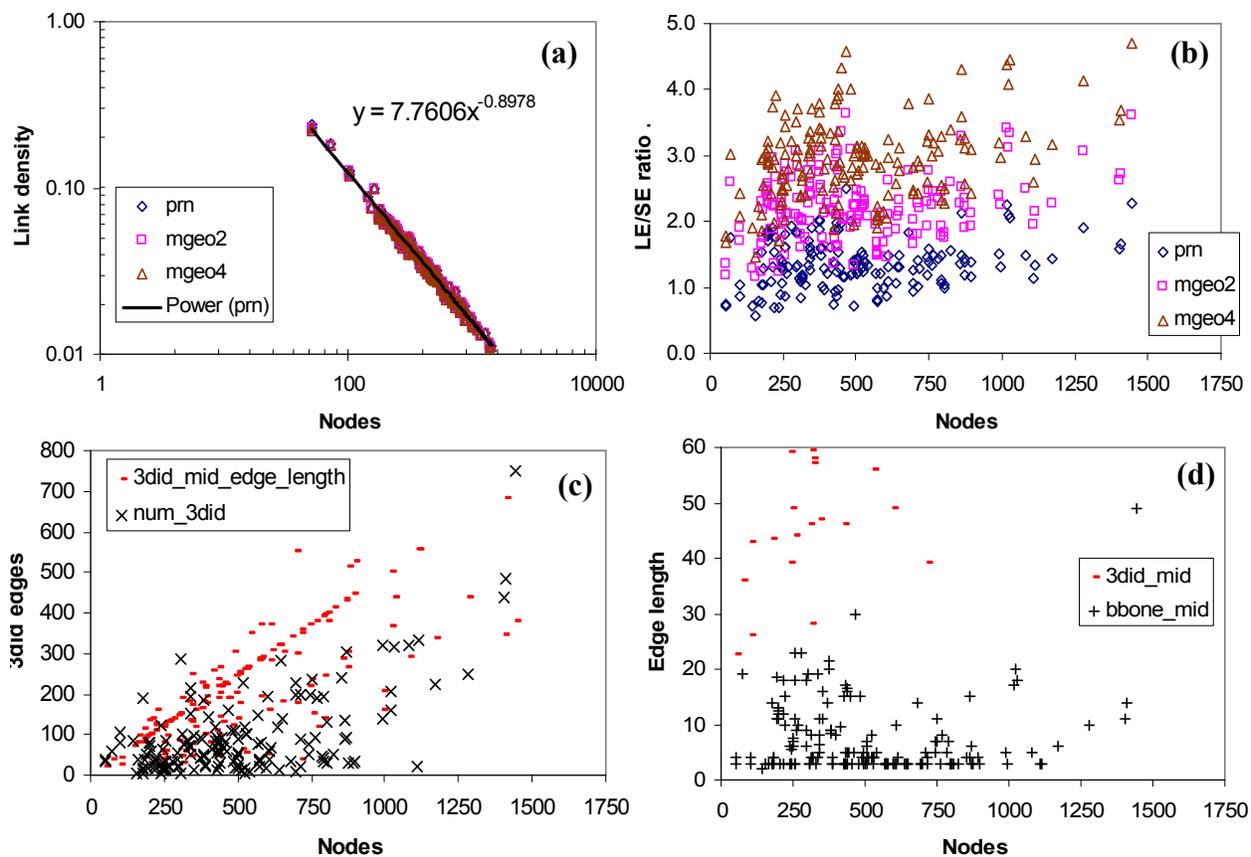

**Fig. 3 Edge statistics. (a)** Link density of PRNs and MGEO networks. **(b)** PRNs have significantly smaller LE/SE ratio than MGEO networks. *LE/SE* is the ratio of long-range links to short-range links. **(c)** Both the number of 3did edges (*num_3did*), and the median sequence distance of 3did links (*3did_mid*) tend to increase with *N*. The smallest 3did_mid point is above 20, which implies that 3did links tend to be long-range. **(d)** The median sequence distance of bbone edges (*bbone_mid*) in MGEO networks are mostly less than 20.

The $m$ edges comprise the shortest (Euclidean distance between the Cα atoms) links such that $|u - v| \geq 2$ and that satisfy the *skip* condition, which defines how many shortest links to ignore between two link inclusion events. The skip condition is introduced to control clustering. Skipping a larger number of shortest links results in a MGEO network with lower *C*. Thus, the MGEO4 networks, where only every fifth shortest link is included, are significantly less clustered than the MGEO2 networks which include only every third shortest link (Fig. 5b). The contact maps of two PRNs and their MGEO4 networks are shown in Fig. 4.

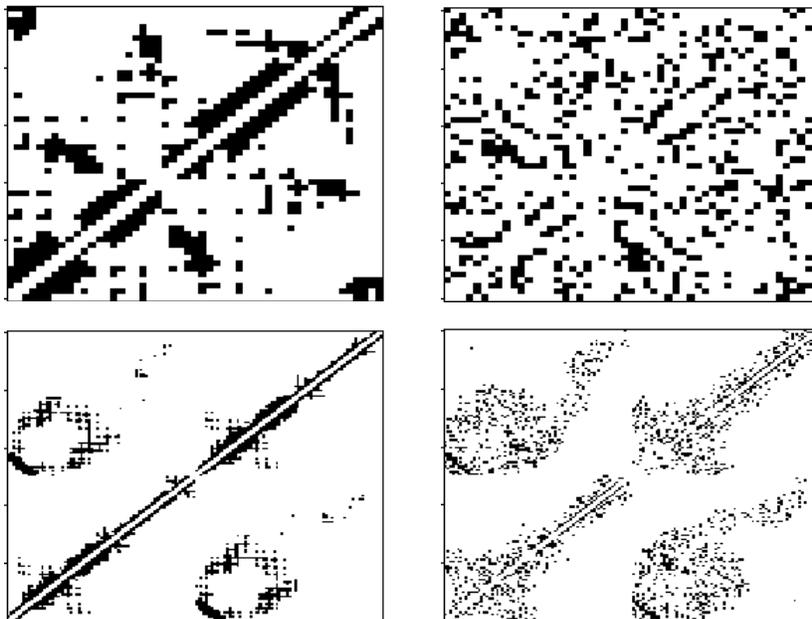

**Fig. 4 Contact maps of two PRNs (left) and their respective MGEO4 networks (right).** The pair at the top is for protein 1B19 which has 51 nodes and 282 edges. At the bottom, is 2ADL which has 144 nodes and 904 edges. Dark cell denotes an edge. A white cell denotes a non-edge.

## 2.3 Structural properties of PRNs and MGEO networks

The average degree for PRNs is constant with $N$ (Fig. 5a), and PRNs have Gaussian degree distribution which can be explained by the excluded volume argument [30]. As expected, the MGEO networks have identical average node degree as their respective PRNs, and their degree distributions are also Gaussian. Both the MGEO networks have significantly smaller clustering coefficients than their respective PRNs (Fig. 5b). The MGEO4 networks are also significantly less clustered than the MGEO2 networks. These differences in one-vertex clustering extend to dyadic or two-vertex clustering as demonstrated by the significant differences in *edge multiplicity* (Figs. 5c & 5d).

Edge multiplicity (*EM*) was introduced in [31] and used to quantify the organization of triangles in a network [32]. In its simplified form (details about the degree class of the endpoints of edges are ignored here), the multiplicity of an edge $e$, is the number of distinct triangles that passes through $e$ (Fig. 6). *EM* >> 1 means triangles are packed onto shared edges, or alternatively, edges participate in many triangles. $EM \leq 1$ denotes that the triangles are disjoint. Networks with large *EM* values have strong transitivity, while those with small *EM* values have weak transitivity [31]. When the clustering coefficient (which is susceptible to the influence of large node degree variations in scale-free networks) of a network does not reflect the transitivity strength of a network, transitivity strength quantified by *EM* is a better indicator of the percolation properties of a network than its clustering coefficient [33]. *EM* is a measure of edge embeddedness, and its use here is equivalent to the number of direct neighbours common to the endpoints of an edge.

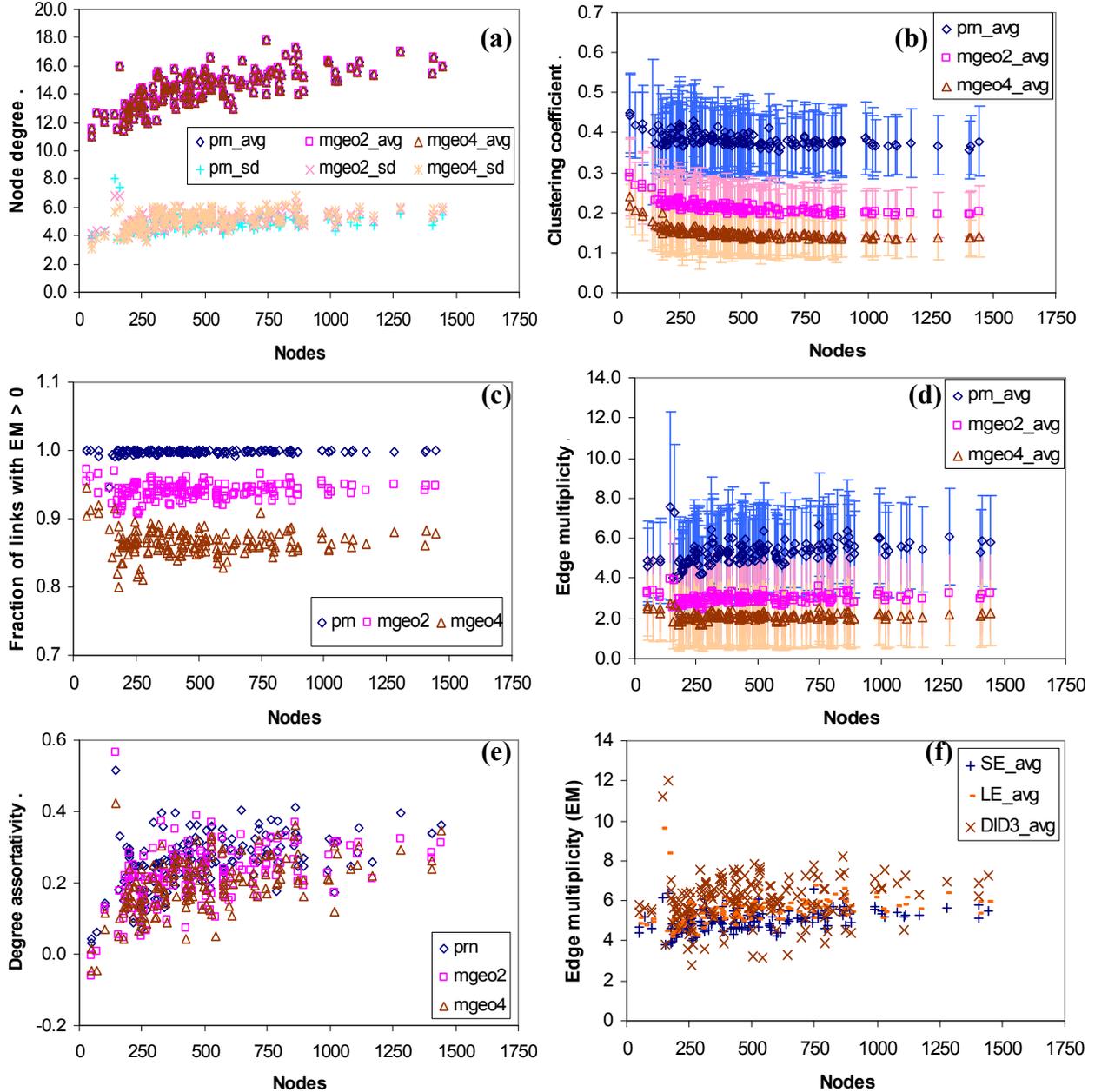

**Fig. 5 Structural characteristics of PRNs and MGEO networks. (a)** PRNs and MGEO networks have Gaussian degree distributions that peak at almost identical average node degrees. **(b)** The clustering coefficients ($C$) of PRNs are significantly larger than the $C$ values of MGEO networks. MGEO2 networks have significantly higher clustering than MGEO 4 networks. **(c)** The fraction of PRNs links with $EM > 0$ is almost 1.0. A link with $EM > 0$ belongs to at least one triangle. The fraction of edges with $EM > 0$ is significantly smaller in the MGEO networks. **(d)** Edge multiplicity averaged over the edges of a PRN is significantly larger edge multiplicity averaged over the edges of a MGEO network. **(e)** $EM$ is also the number of direct neighbors common to the endpoints of an edge. Thus it is influenced by degree-degree assortativity, the propensity for nodes with similar degrees to link with each other. Both PRNs and MGEO networks are mildly positively associative by node degree. **(f)** Edge multiplicity by edge type. In the PRN networks, both long-range links ($LE\_avg$) and 3did links ($DID3\_avg$) have significantly larger average edge multiplicity than short-range links ($SE\_avg$). Error bars in (b) and (d) denote the standard deviation.

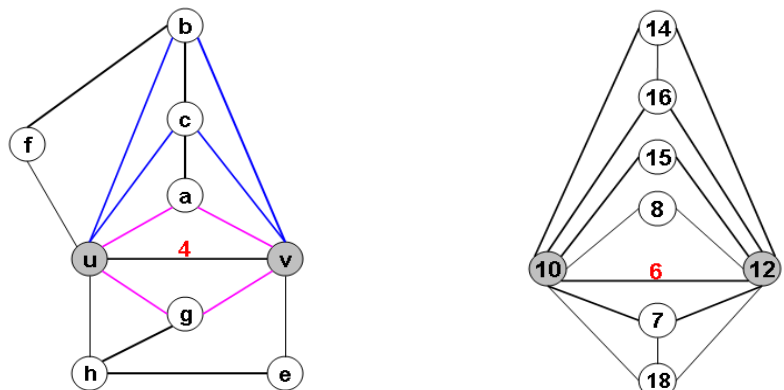

**Fig. 6 Edge multiplicity (*EM*).** *EM* of an edge is the number of triangles the edge completes [31]. In the diagram on the left, *EM* of edge (*u*, *v*) is 4 since nodes *u* and *v* have four direct or first neighbors in common: *b*, *c*, *a* and *g*. The blue and pink edges trace out two diamonds (cycles of length four) that are the result of triangles sharing (*u*, *v*). The diagram on the right is extracted from the 2FAC PRN (only the common first neighbors of nodes 10 and 12, and the links between them are shown). *EM* of edge (10, 12) is 6.

Almost all links in a PRN make up a leg of at least one triangle; the fraction is lower in MGEO networks but still above 80% (Fig. 5c). PRNs also have significantly larger *EM* values than MGEO networks (Fig. 5d). A large clustering coefficient signals an abundance of triangles and when triangles share edges, they stick together and have the potential to form longer cycles. Large *EM* values signal an abundance of diamond motifs. The abundance of triangle and diamond motifs induced by the strong transitivity in PRNs makes it feasible as done in [25] to utilize higher order concepts of local organization such as k-cliques and communities (overlapping cliques) to distinguish decoy structures from native ones. Ref. [34] reports a linear relationship between triangle and diamond motifs in their RINs, and interprets the presence of these local motifs as providing alternative pathways to ensure connectivity as links are made and destroyed when proteins undergo fluctuations at equilibrium. However, their RINs are constructed differently and the diamond motif is also defined differently. A diamond motif in [34] is a chordless four node cycle. By this definition, nodes *b*, *u*, *g*, and *v* in Fig. 6 (left) does not form a diamond motif because of the edge between *u* and *v*.

The multiplicity of an edge is limited by the degree of its endpoint nodes. Thus, networks whose nodes selectively link by degree, i.e. links are more likely between nodes with similar degrees than between nodes with dissimilar degrees, are more conducive to strong transitivity. Nodes in PRNs are mildly positively correlated by degree (Fig. 5e). Assortative mixing of nodes by degree has been reported in other RINs as well [35, 36].

RINs comprising only long-range links have significantly smaller *C* values than complete RINs [28]. In fact, much of the clustering in RINs has been attributed to short-range links [37]. But even though long-range links (*LE*) are less transitive than short-range links (*SE*) when considered separately, when *LE* and *SE* are combine in a network, their interaction results in long-range links having significantly larger EM values than short-range links (Fig. 5f). Nonetheless, a strong linear correlation between sequence distance and EM was not observed. Edge transitivity gains significance when the properties of short-cut edges are investigated in section 3.6 and section 4.0. The 3did edges contribute very little to clustering by themselves, but interact with the other links in a network to form triangles. In fact, 3did edges have significantly larger edge multiplicity than non-3did edges (Fig. 5f).

## 2.4 Expansion property of PRNs and MGEO4 networks

The PRNs have maximum inter-residue Euclidean distances averaged at 15.4Å (std. dev. = 1.3224), which is twice that normally considered in pure (no side-chain consideration) $C_\alpha$-$C_\alpha$ or $C_\beta$-$C_\beta$ contact maps (Fig. 7a). This raises the concern that the PRNs are not preserving topological cavities which are proxies for protein binding sites. Ref [38] examined the expansion factor of RINs and observed that most RINs, especially those of multi-domain proteins with more than 240 nodes, are not homogeneous networks (class I) but belong to the class of networks that exhibits modularity (class II). Class I networks are good expanders. An expander graph is sparse (bounded mean node degree) and its nodes are well-connected such that it is difficult to disconnect the graph, i.e. the cuts need to be large (see Appendix D for a more formal treatment). We can infer from this description that Class I networks will have poor modularity. Modularity describes an organizational structure where clusters of nodes that are more interconnected with each other than with other network nodes outside the cluster exist. Modular structure is an important feature for RINs to have for two complementary reasons: (i) modularity is a natural consequence of the architecture of larger multi-domain proteins, and (ii) the topological cavities or regions of lower connectivity between the clusters correspond to protein binding sites [38]. Using the spectral scaling method in [38], we determined that all but two of our PRNs belong to class II and are therefore modular and cavity preserving in principle (Fig. 7b). The other two PRNs belong to class IV (Figs. 7d & 7f). For comparison, we constructed pure $C_\alpha$-$C_\alpha$ contact maps with a 7.5 Å cutoff. Twelve of these networks belong to Class IV. This means that the PRN construction method presented in this paper allows much larger contact cutoffs, whilst maintaining the essential features of protein structure. Due to their more random construction, the MGEO4 networks are expected to have better expansion which they do. Fourteen MGEO4 networks belong to Class IV, and those MGEO4 networks that belong to Class II show smaller deviation from Class I behavior (Fig. 7e).

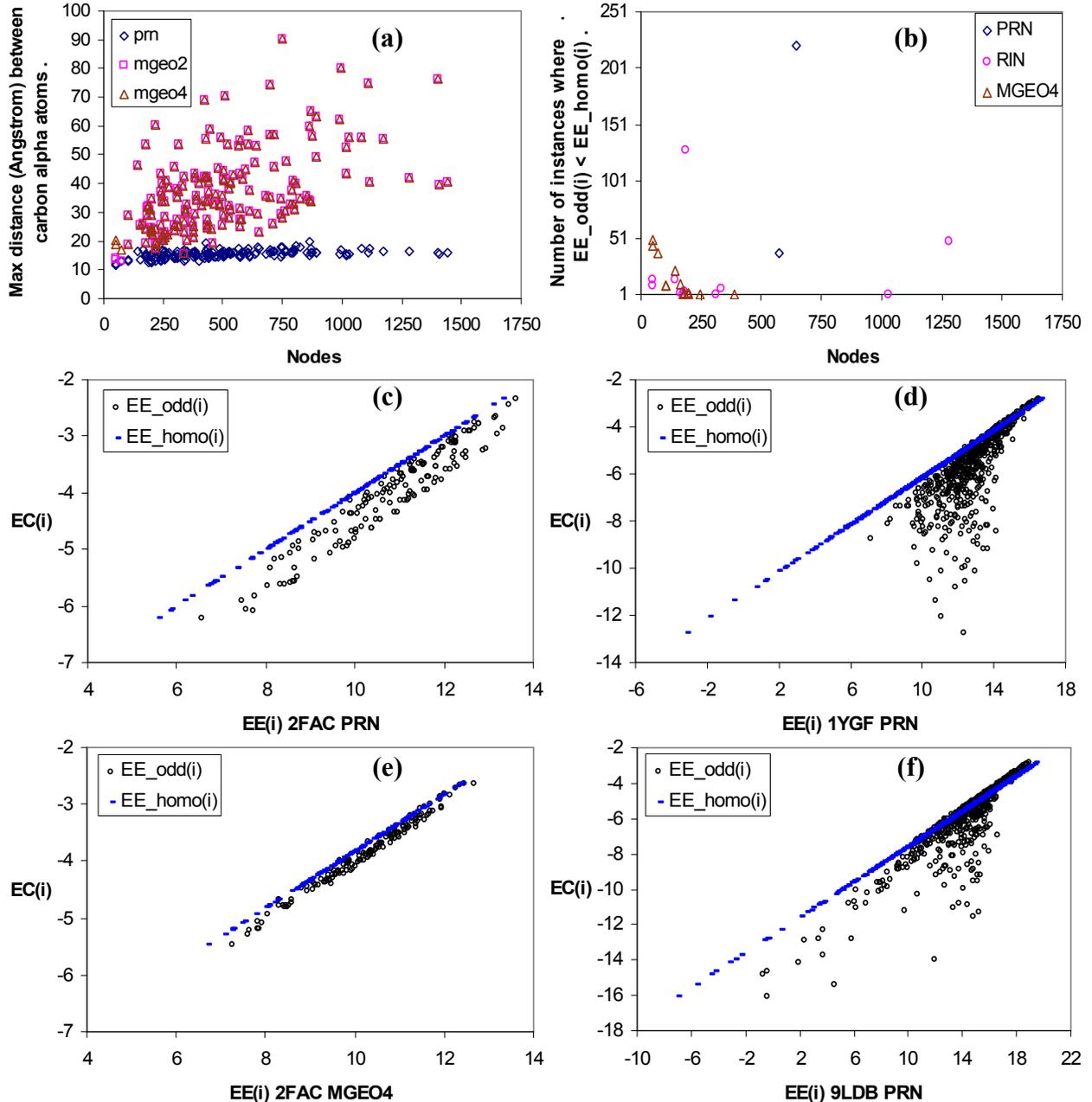

**Fig. 7 Spectral scaling [34] results for the 166 PRNs.** EC($i$) denotes the $i$th component of the principal eigenvector. EE($i$) denotes the subgraph centrality measure for node $i$. EE_odd($i$) denotes the odd-subgraph centrality measure for node $i$. When $\lambda_1 \gg \lambda_2$, EE_odd($i$) = EE_homo($i$). The x- and y- axes in (c - f) are log base 2. **(a)** The maximum Euclidean distance between nodes ($C_\alpha$) in PRNs and between nodes in MGEO networks. **(b)** The number of nodes in a network where EE_odd($i$) < EE_homo($i$). A network with one or more such nodes (but not all nodes), belong to Class IV. The number of PRNs that belong to class IV is 2, compared with 12 for RINs, and 14 for MGEO4 networks. For this plot, RINs are protein contact maps built as pure $C_\alpha$-$C_\alpha$ with 7.5 Å cutoff. **(c)** Spectral scaling results for a Class II network: the 2FAC PRN. All the EE_odd($i$) points are to the right (>) of the EE_homo($i$) points. **(d & f)** Spectral scaling results for the two PRNs (1YGF and 9LDB) that belong to class IV. There are EE_odd($i$) points to the left and right of the line passing through the EE_homo($i$) points. **(e)** The 2FAC MGEO4 network is also Class II, but compared with the 2FAC PRN (c), its EE_odd($i$) points lie closer to the EE_homo($i$) points.

## 2.5 A Euclidean distance directed local search algorithm (EDS)

In [5, 6], Kleinberg describes a local search algorithm that does greedy routing. In this greedy local search algorithm, information used to decide the next move is confined to the neighbourhood within a small radius of the current node, and the search moves to a neighbouring node that is closest to the target node. As such a greedy local search has *directionality*, i.e. is anisotropic.

Similar to Kleinberg's algorithm, the EDS algorithm does greedy routing based on proximity (Euclidean distance) to a target node. However, EDS differs in two main ways: it keeps a memory of all the nodes visited and enquired so far in the current search, and can therefore backtrack and re-route itself to another more promising path midway through the search. The information used is still local, but this information expands over time as more nodes are visited and their direct neighbours queried for their proximity to the target node; in other words, the search radius increases as the search progresses. To keep the search local, information gathered during the search for a path is completely forgotten once the path is found. Technically an EDS path is a walk since an EDS path may retrace edges and in the process revisit nodes.

Starting with the source node, EDS performs the steps outlined below for each node appended to a path until the target node is found. The main steps of the EDS algorithm to find a path from $s$ to $d$ are:

1. Get $N(x)$, the direct neighbors of the most recently visited node $x$. Initially, $x$ is the source node $s$.

2. For each $n$ in $N(x)$, compute the Euclidean distance between $n$ and the target node $d$. (Proximity information for an $n$ to $d$ may already be computed in a previous iteration due to network clustering.)

3. If $n$ is the target node $d$, stop searching and return path.

4. Otherwise, add the new proximity information to memory, i.e. all distance information gathered so far for this search.

5. Sort nodes in memory by proximity to find the next node to visit, $y$.

6. Move to y, which is an as yet unvisited node closest to the target. If necessary (when y is not directly reachable from the current node), backtrack (retrace the current path but also look at the immediate neighborhood of nodes retraced to find a bridge to $y$, i.e. a node neighboring $y$).

7. $x := y$. Go to 1.

The backtrack strategy in Step 6 can lead to unnecessary backtracking. However this apparent inefficiency ensures that the graph an EDS path traces is a tree, and increases the search space for an edge to $y$. A backtrack stops as soon as an edge to $y$ is found. The search tree begins with a single node, $s$. In each iteration, the EDS algorithm adds to the existing search tree a single edge with a node not already in the search tree. The only way for EDS to revisit a node is by tracing the edges of the existing search tree.

The EDS search is conducted on a finite connected graph and EDS terminates at *d* (The efficiency of EDS search depends on the structure of the network).

EDS paths run along the edges of the network. Edges may appear more than once in an EDS path if backtracking occurs. Edges in an EDS path may be classified as short-cut, backtrack, short-cut and backtrack or neither short-cut nor backtrack. An EDS path may have zero or more short-cut and/or backtrack edges. The number of edges in an EDS path is its *length*. Fig. 8 depicts three EDS paths of length four from the 2FAC PRN. The *cost* of an EDS path is the number of unique nodes stored in memory for the search, and is bounded from above by the union of the direct neighbors of all nodes that appear on the path. The sequence of events to construct the rightmost EDS path in Fig. 8 is worked out in Appendix E.

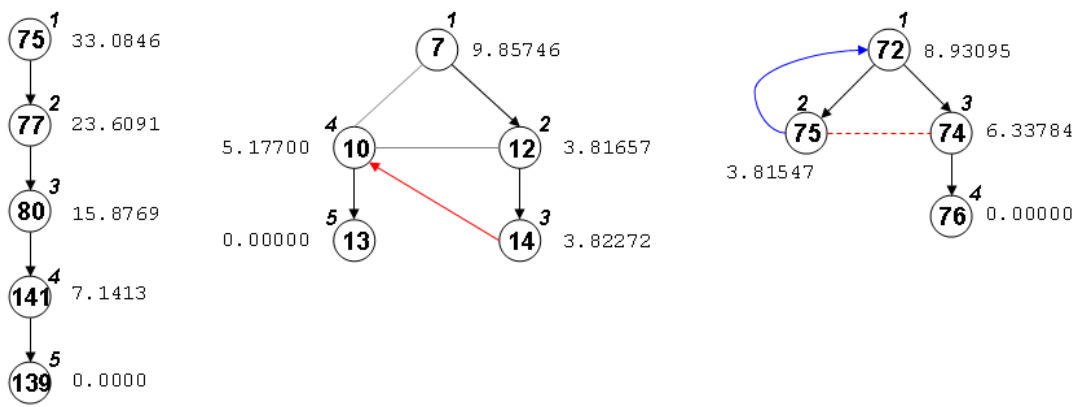

**Fig. 8 Three EDS paths of length four in the 2FAC PRN.** PRN edges are undirected. The arrowheads are used to show the direction the edges are traversed in the respective paths. The leftmost path is ⟨75, 77, 80, 141, 139⟩ which is a straight-forward path. The EDS path in the middle is ⟨7, 12, 14, 10, 13⟩ and it has a *short-cut edge* indicated in red. Without edge (14, 10), EDS would have to backtrack to node 12 to get to node 10. A short-cut edge completes a *navigational cycle*, which in this instance is the triangle comprising nodes 14, 12, and 10. The rightmost EDS path is ⟨72, 75, 72, 74, 76⟩ and it has a *backtrack edge* indicated in blue. If edge (75, 74) existed, it would be a short-cut edge and the backtrack would be avoided. A non-existent short-cut edge marks a *navigational hole*, which in this instance is the two-edge path comprising nodes 75, 72 and 74. Short-cut and backtrack edges are defined more formally in the text. The real number besides each node is the node's Euclidean distance to the target node, which need not decrease monotonically as an EDS progresses. For example, nodes 12 and 14 are much closer to the target node 13 than node 10. The italicized integer besides each node *x* is the node's first visit order $L^T(x)$ (see text for details). The construction of the rightmost path is described in detail in Appendix E.

Unlike BFS which guarantees $\lambda(i, j) = \lambda(j, i)$ by definition (undirected graph), an average of 53.50% (std. dev. 6.28%) of EDS path-pairs are not length invariant, i.e. $\lambda(i, j) \neq \lambda(j, i)$. We do not consider this a disadvantage. In fact, if network topology plays a role in determining paths and their lengths, and the topology is not homogeneous everywhere, then $\lambda(i, j) \neq \lambda(j, i)$ is to be expected. For results in this paper, BFS and EDS searches are run in both directions for each unique ordered node-pair. Average path length of a network is then redefined as: $L = \frac{1}{N(N-1)} \sum_{i \neq j}^{N} \lambda(i, j)$ where $\lambda(i, j)$ is the length (number of edges) of a path) from node *i* to node *j*. The total number of paths sampled by both BFS and EDS is $N(N-1)$. For BFS sampling two possibly different paths per node-pair increases the data available to compute path related statistics such as betweenness centrality. EDS paths that are identical either way, and/or identical to BFS paths exists (Table 1). On average, the set of nodes along a BFS path from $u$ to $v$ is only 44% (std. dev. 6.03%) similar with the set of nodes along an EDS path from $u$ to $v$. This similarity drops significantly to 41% (std. dev. 4.79%) for MGEO4 networks.

**Table 1 Comparing three pairs of EDS paths through the 2FAC PRN with their BFS counterparts.** Paths connecting two nodes need not be symmetrical. Unlike BFS paths, EDS paths connecting a node pair also need not be the same length. An EDS path can be identical to a BFS path.

| Source | | | | Target | | | | EDS path | BFS path |
|---|---|---|---|---|---|---|---|---|---|
| **Node** | Ch | Id | AA | **Node** | Ch | Id | AA | | |
| **75** | A | 76 | GLN | **139** | B | 63 | THR | 75,77,80,141,139 | 75,77,80,141,139 |
| **139** | B | 63 | THR | **75** | A | 76 | GLN | 139,144,83,78,48,75 | 139,141,80,77,75 |
| **72** | A | 73 | ASN | **76** | A | 77 | ALA | 72,75,72,74,76 | 72,74,76 |
| **76** | A | 77 | ALA | **72** | A | 73 | ASN | 76,74,72 | 76,49,72 |
| **7** | A | 8 | LYS | **13** | A | 14 | GLN | 7,12,14,10,13 | 7,11,13 |
| **13** | A | 14 | GLN | **7** | A | 8 | LYS | 13,8,6,10,7 | 13,10,7 |

*Short-cuts (SC) and navigational cycles:* Define $G_{s,d} = (V', E')$ as the sub-graph of a PRN induced by $V'$, the set of nodes on the EDS path from $s$ to $d$ in the PRN. $E'$ is the subset of edges in the PRN that have both their endpoints in $V'$. While PRNs are not directed graphs, the edges in $G_{s,d}$ are oriented in the direction they are traversed by the EDS path from $s$ to $d$. A bi-directional link denotes an edge that has been re-traced in a backtrack. The search tree $T_{s,d}$ of an EDS path from $s$ to $d$ is induced by all oriented edges in $G_{s,d}$ and is rooted on $s$. $T_{s,d}$ spans the nodes of $G_{s,d}$. Let $L^T(x)$ be a positive integer denoting the *first visit order* of node $x$ in the EDS path represented by the search tree $T$. $L^T(s) = 1$. $L^T(d) =$ the number of unique nodes in the EDS path. Nodes are visited for the first time as they appear in an EDS path. Nodes revisited via backtracking are skipped over. See Figs 8 and 9 for examples.

Given the sub-graph $G'$ and the search tree $T$ for an EDS path from $s$ to $d$: an edge $(u, v)$ in $T$ is a short-cut if and only if $L^T(v) = L^T(u) + 1$, and $v$ is adjacent (in $G'$) to a node $w$ such that $L^T(w) < L^T(u)$. There may be more than one such $w$ node. Let $W$ be the set of all such $w$ nodes. To identify the specific $w$ node, EDS retraces its step from $u$ until it finds the *first* $(x, v)$ edge where $x \in W$. The path $\langle u, \ldots, x, v \rangle$ together with the edge $(u, v)$ forms a *navigational cycle*. Hence short-cut edges complete navigational cycles and are made possible by cycles in the underlying network. The condition $L^T(v) = L^T(u) + 1$ is necessary so that the process of backtracking itself does not create shortcut edges.

The sub-graph in Fig. 9 contains one navigational cycle comprising nodes 51, 57, 55, 58, 46 which is completed by the short-cut edge (51, 46). The edge (46, 51) is backtracked by this EDS search to get to node 56. Suppose, with everything else unchanged, that edge (55, 46) exists in the sub-graph but not in the search tree in Fig. 9. Then the EDS path is unchanged and (51, 46) still forms a short-cut. But $W$ = {55, 58}, and the navigational cycle associated with (51, 46) in this EDS search is now circumnavigated by a shorter path: ⟨51, 57, 55, 46, 51⟩.

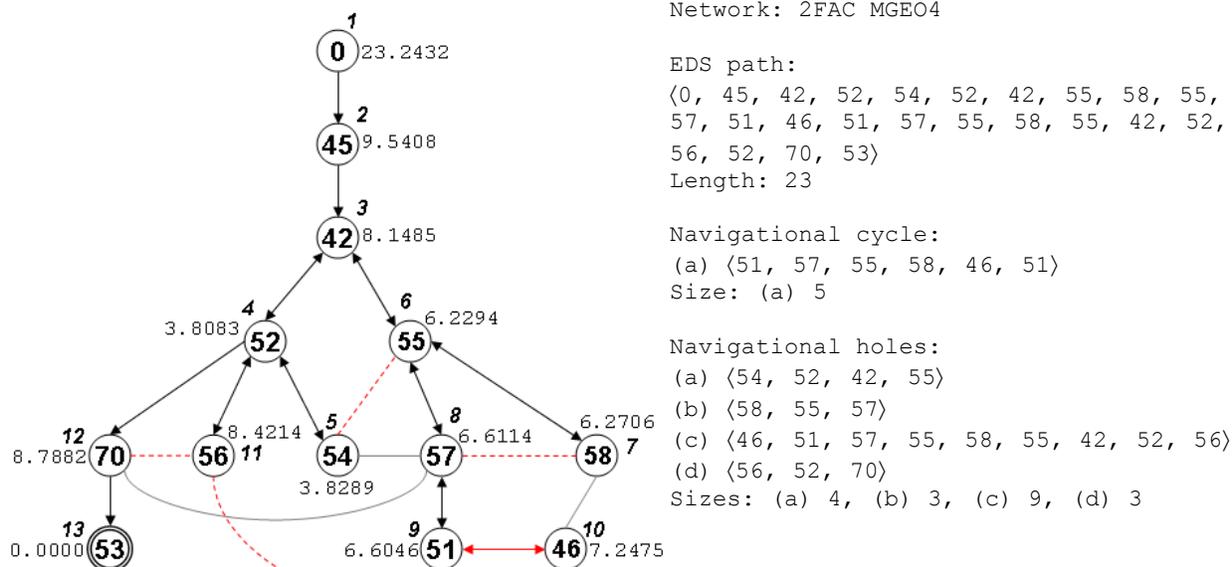

Network: 2FAC MGEO4

EDS path:
⟨0, 45, 42, 52, 54, 52, 42, 55, 58, 55,
57, 51, 46, 51, 57, 55, 58, 55, 42, 52,
56, 52, 70, 53⟩
Length: 23

Navigational cycle:
(a) ⟨51, 57, 55, 58, 46, 51⟩
Size: (a) 5

Navigational holes:
(a) ⟨54, 52, 42, 55⟩
(b) ⟨58, 55, 57⟩
(c) ⟨46, 51, 57, 55, 58, 55, 42, 52, 56⟩
(d) ⟨56, 52, 70⟩
Sizes: (a) 4, (b) 3, (c) 9, (d) 3

**Fig. 9 The sub-graph $G_{0,53}$ for EDS path from 0 to 53 in the 2FAC MGEO4 network.** All lines except the red dashed ones are edges of $G_{0,53}$. Edges are oriented in the direction they are traversed by this EDS path. Bi-directional edges are backtrack edges. Un-oriented edges are not traversed but exist in the underlying MGEO4 network. The search tree for this EDS path is induced by all the oriented edges. The short-cut edge (51, 46) is marked in red. The red dashed lines indicate the non-existent short-cuts (NESC), i.e. edges that would be short-cuts if they existed in the underlying network, with everything else unchanged. The length of a short-cut is the number of edges in the navigational cycle it closes less one. The length of a NESC is the number of edges in the navigational hole it would close less one. The italicized integer beside each node $x$ is the node's first visit order $L^T(x)$. The real number besides each node is the node's Euclidean distance to the target node 53. For comparison, the EDS path from 0 to 53 in 2FAC PRN is a straight-forward path of length four ⟨0, 2, 71, 51, 53⟩.

The size of a navigational cycle is the number of edges it contains. The length of a short-cut is the size of the navigational cycle it closes less one. Since a short-cut edge may be part of one or more navigational cycles in different EDS searches, the length of a short-cut edge is context sensitive. Due to differences in clustering and transitivity, short-cut edges are significantly more abundant and significantly shorter in PRNs than in MGEO networks (Figs. 11a & 11b).

*Non-existent short-cuts (NESC) and navigational holes:* If edge (51, 46) is removed from the sub-graph in Fig. 9, all things being equal, EDS would need to retrace its steps to node 58 to reach node 46 from node 51, and the path ⟨51, 57, 55, 58, 46⟩ would form a *navigational hole*. Essentially, backtracks travel along navigational holes until a direct neighbour of the node EDS needs to move to is met. The sub-graph in Fig. 9 has four navigational holes which could be closed by the red dashed edges if these edges exist in the underlying MGEO network. However, since these edges do not exist, they are called *non-existent short-cuts* (NESC). Two of these (54, 55) and (58, 57), cannot exist by our network definition since their sequence distance is less than 2.

A NESC is found when EDS needs to move from $u$ to $v$ but $(u, v)$ is not an edge in the network. However, if we assume that $(u, v)$ exists, then $(u, v)$ is a short-cut as defined previously, and a special $w$ node $x$ can be located as described previously. But since $(u, v)$ does not actually exist, EDS adds $(x, v)$ to the search tree instead of $(u, v)$, and the path $\langle u, \ldots, x, v \rangle$ forms a navigational hole. Both navigational cycles and holes can contain smaller cycles in them due to the backtracking that was used to find them. E.g.: the largest navigational hole in Fig. 9 contains a cycle $\langle 55, 58, 55 \rangle$.

The size of a navigational hole is the number of edges in an EDS path that circumnavigates it plus one. The length of a non-existent short-cut is the size of the navigational hole it would close if it existed less one. The length of a NESC also gives the search depth for an EDS search. As with short-cut edges, the length of NESCs may be context dependent. *NESC*s are significantly more abundant and significantly longer in MGEO networks than in PRNs (Figs. 11c & 11d).

## 2.6 Random short-cuts

As part of this study, we investigate the properties of short-cut edges and their impact on PRN network statistics. These findings are compared against a set of edges chosen uniformly at random without replacement from all edges of a PRN such that there is a replacement edge for each short-cut. Let *SC* be the set of short-cuts found by EDS for a PRN, and *RSC* be a random short-cut set associated with the PRN. Then *RSC* is a random subset of the PRN's edges such that $|RSC| = |SC|$. Five such random subsets are generated, and they are denoted *rsc1... rsc5* in the figures. Thus, a PRN network without its short-cuts, PRN\SC, has the same number of edges as the PRN without its random short-cuts, PRN\RSC.

For reasons explained in section 4, we use an additional type of random edge set called *RST*. A *RST* is a subset of PRN edges that were selected via random walks to construct two or more trees spanning a PRN such that $|RST| = |SC|$ for the PRN. Five *rst* sets were generated with a different random number seed each time, and they are denoted *rst11…rst15* in the figures. A PRN network without its short-cuts, PRN\SC, has the same number of edges as the PRN without its *RST* edges, PRN\RST.

## 2.7 Molecular Dynamics simulation (MD) dataset

Two datasets are used. The first dataset is taken as a model of protein folding since all runs terminate in the native state (or rather all runs unfold from the native state). The second dataset is only used in section 4, and is of interest because it contains runs which are considered successful and runs which are considered unsuccessful. One of our goals with this research project is to see if it is possible to distinguish between the two using complex network /graph theory and algorithms. We find this is possible in section 4.4.

The first MD dataset is obtained from the Dynameomics project [39, 40, 41]. *il*mm (in lucem Molecular Mechanics) is used to simulate the native and unfolding dynamics of the 2EZN protein which has 101 amino acids and is comprised mainly of beta strands interspersed with helices and loops. There are nine MD runs in this dataset, each with a different number of snapshots (Table 2). The shorter unfolding runs were made to sample early unfolding events more thoroughly. Snapshots are taken at intervals of 1 ps, except for the first 2 ns of the unfolding runs where snapshots are taken at 0.2 ps.

**Table 2** Definition of labels and colors to identify the different 2EZN MD runs.

| Run type | Run label | Color in plots | Number of snapshots |
|---|---|---|---|
| Native dynamics (298K) | 6250 | Black | 51,000 |
| Non-native dynamics (498K) | 6251 | Brown | 50,000 |
| | 6252 | Purple | 68,845 |
| | 6253 | Blue | 2,000 |
| | 6254 | Cyan | 2,000 |
| | 6255 | Green | 2,000 |
| | 6256 | Magenta | 10,000 |
| | 6257 | Orange | 10,000 |
| | 6258 | Yellow | 10,000 |

The second MD dataset is publically available from https://simtk.org/home/foldvillin and it was generated with GROMACS simulating the folding of a subdomain of the mainly alpha-helix 2F4K protein. The subdomain (HP-35 Nle Nle) has 35 amino acids and is modified for fast folding. Folding starts with a different denatured configuration in each run. The nine MD runs in this dataset are partitioned into two sets for reasons stated in [62], and we follow this delineation. In the "successful" set are runs 4, 7, and 8. The starting structures of these runs either folded much faster or folded to a significant extent [62]. In the "unsuccessful" set are runs 0, 1, 2, 3, 5, and 6. The starting structures of these runs generated trajectories that either only briefly visited a folded configuration, or did not fold at all according to the criteria in [62]. We chose at least two trajectories (clones) at random from each run, with preference for clones with more frames (a frame comprises many shapshots). Duplicate snapshots used to splice frames were removed.

For both MD datasets, a PRN is constructed for every snapshot following the method described in section 2.1. The snapshots of a run capture the positions of all protein atoms and the solvent (water) atoms in the MD box at a point in time. Excluding the positions of the solvent atoms, gives one a file of *x*, *y* and *z* coordinates for the protein atoms from which PRNs can be constructed. The results we report are produced by averaging over all snapshots in a run (for the first MD dataset) or a clone (for the second MD dataset). Network statistics for PRNs from the first MD dataset are plotted against the *Radius of Gyration* (also obtained from the MD dataset), which is a measure of the compactness of a protein. Radius of Gyration (Rg) is a commonly used reaction coordinate to monitor progress during protein folding simulations. Protein folding is associated with more order and tighter packing, i.e. smaller Rg values.

## 3. Results

### 3.1 Path-length

Both BFS and EDS search strategies found short paths on both PRNs and MGEO networks. The average length of both BFS and EDS paths increased logarithmically with increase in $N$ (Fig. 10a), but BFS paths are significantly shorter than EDS paths. This is expected given that backtracking occurs in EDS paths and that BFS by definition should produce the shortest paths. The EDS paths are significantly more varied in length than the BFS paths (Fig. 10b). Large variation in path lengths is more congruent with the notion described in section 1 that intra-protein communication pathways are also varied in length. Fast, specific and reliable communication between some sites is crucial, but also some pathways exist to slow down communication as a way to absorb or localize the after effects of undesirable perturbations to maintain protein stability [15].

MGEO networks have significantly shorter BFS paths than PRNs (Figs. 10c & 10d). In contrast, MGEO networks have significantly longer the EDS paths than PRNs (Figs. 10e & 10f). Thus, at the link density levels of the PRNs (Fig. 3a), the significantly weaker clustering of the MGEO networks reduces global path length ($L_G$) but increases local path length ($L_W$). This result suggests a topological reason for the high levels of clustering in PRNs. Clustering is a barrier to short paths found through global search (BFS), but becomes a facilitator for short paths found with local search (EDS). Given the importance of short intra-protein communication pathways, striving towards (folding) and maintaining a configuration with high clustering makes sense, but only from a local search perspective. The benefit of high clustering in a protein's native state, in terms of short intra-protein paths, may also act as a topological barrier preventing its unfolding. Thus we observe a significant gap in $C$ and $L$ values between native protein dynamics and protein unfolding (Fig. 1).

The effect clustering has on BFS path length follows from the relationship: $L_{ER} \sim \ln(N) \, / \, \ln(K)$. Random graphs have little to no clustering and are locally tree-like. The average degree $K$ in ER graphs then approximates the branching factor of a search tree. By introducing cycles into the search tree, some of the branches now loop back to a previous tree level. In other words, clustering reduces the effective branching factor, or effective $K$. Consequently, $L_G$ increases. But an increasing average path-length concomitant with an increasing clustering coefficient does not fit the picture in Fig. 2.

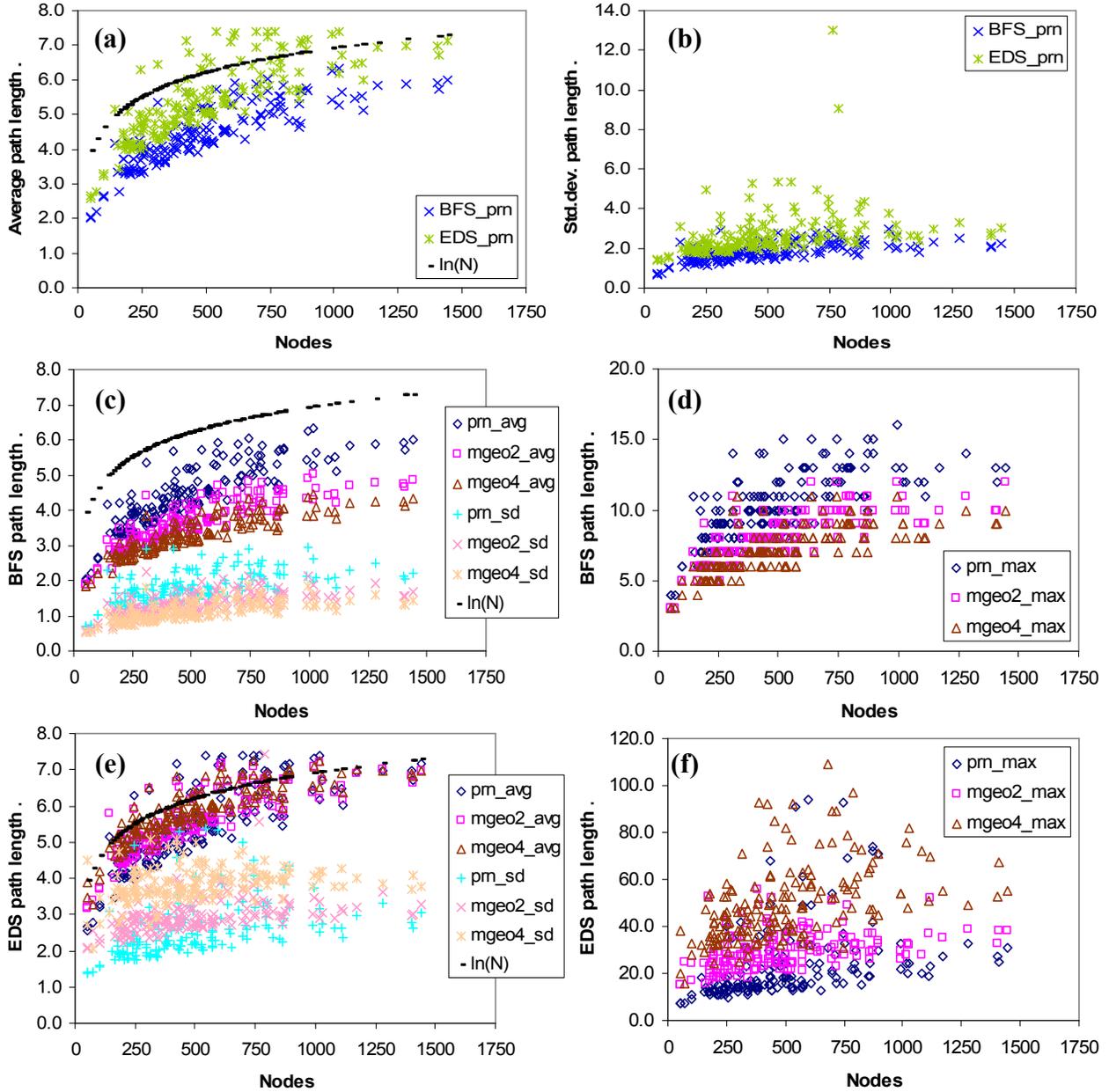

**Fig. 10 BFS and EDS path length.** **(a)** Both the average BFS path length (*BFS_prn*), and the average EDS path length (*EDS_prn*) on PRNs increase logarithmically with the number of nodes *N*. However, *EDS_prn* is significantly longer than *BFS_prn*. **(b)** The standard deviation of EDS path lengths (*EDS_prn*) is significantly larger than the standard deviation of BFS path lengths (*BFS_prn*) on PRNs. Therefore, EDS paths on PRNs are more varied in length than BFS paths on PRNs. **(c)** The more highly clustered PRNs have significantly longer average BFS path length (*prn_avg*) than the less highly clustered MGEO networks. The average BFS path length on MGEO2 and MGEO4 networks, denoted *mgeo2_avg* and *mgeo4_avg* respectively are significantly smaller than *prn_avg*. **(d)** PRNs have significantly longer longest BFS path (*prn_max*) than MGEO networks. Hence from a BFS perspective, PRNs have larger diameters than MGEO networks. **(e)** Average EDS path length on PRNs denoted *prn_avg* is significantly shorter than both average EDS path length on MGEO2 networks (*mgeo2_avg*), and average EDS path length on MGEO4 networks (*mgeo4_avg*). Thus, average EDS path length decreases with increase in clustering. **(f)** PRNs have significantly shorter longest EDS path (*prn_max*) than MGEO networks. Hence from an EDS perspective, PRNs have smaller diameters than MGEO networks.

An explanation for the effect clustering has on EDS path length follows. Unlike the BFS strategy of branch first then prune (when the target node is found), EDS prunes first (or branches conservatively) then if necessary (with the help of backtracking) figures out how to branch further. Define the number of edges retraced to branch onto a more promising path as the *depth* of a search (search depth is also the length of a NESC, section 2.5). The search depth for PRNs is significantly shallower than the search depth for MGEO networks (Fig. 11d), which means that on average, EDS retraces fewer edges when searching PRNs than MGEO networks. EDS also does significantly less backtracking when searching PRNs than MGEO networks (Fig. 11f). Thus, PRNs have significantly shorter average EDS path length than MGEO networks. The shallower search depths and fewer backtracking paths in PRNs are a consequence of their high levels of clustering and strong transitivity (Figs. 5b & 5d), which enrich PRNs with potential short-cut edges. In contrast, the lower levels of clustering and weaker transitivity of MGEO networks create significantly more navigational holes (Fig. 11c).

The short-cut edges which are significantly more abundant in PRNs than in MGEO networks (Fig. 11a) reduce the need for and the extent of backtracking, which in turn keeps the length of EDS paths in check. So while cycles pose a problem for global search (BFS) and increase $L_G$, they help local search (EDS) to decrease $L_W$. In the MD simulations on 2EZN, as the protein becomes more compact (Fig. 12), generally: (a) the number of short-cut edges increases, (b) the number of non-existent short-cuts decreases, (c) the fraction of EDS paths containing at least one short-cut increases, and (d) the fraction of EDS paths doing at least one backtrack decreases. From Fig. 2, we know that clustering and edge multiplicity increases while average path length decreases as the protein becomes more compact in these runs. Taken together, the observations from the MD simulations on 2EZN are aligned with the results of this section which are made for the 166 native PRNs – that clustering (and strong transitivity) facilitates navigation by increasing short-cuts and reducing backtracking.

The linear relationship between the number of short-cuts and the number of nodes in a PRN as ~2N reported in Fig. 11a is not an artifact of finding EDS paths both ways. When the analysis is repeated with only one path chosen at random between every node-pair, the number of short-cuts still approximates 2N (prn_1 in Fig. 11a). The number of short-cuts found by EDS is however sensitive to how the PRNs are constructed. In a less dense PRN, say one constructed as pure $C_\alpha$-$C_\alpha$ with 7.5 Å cutoff, the number of short-cuts are much fewer than 2N.

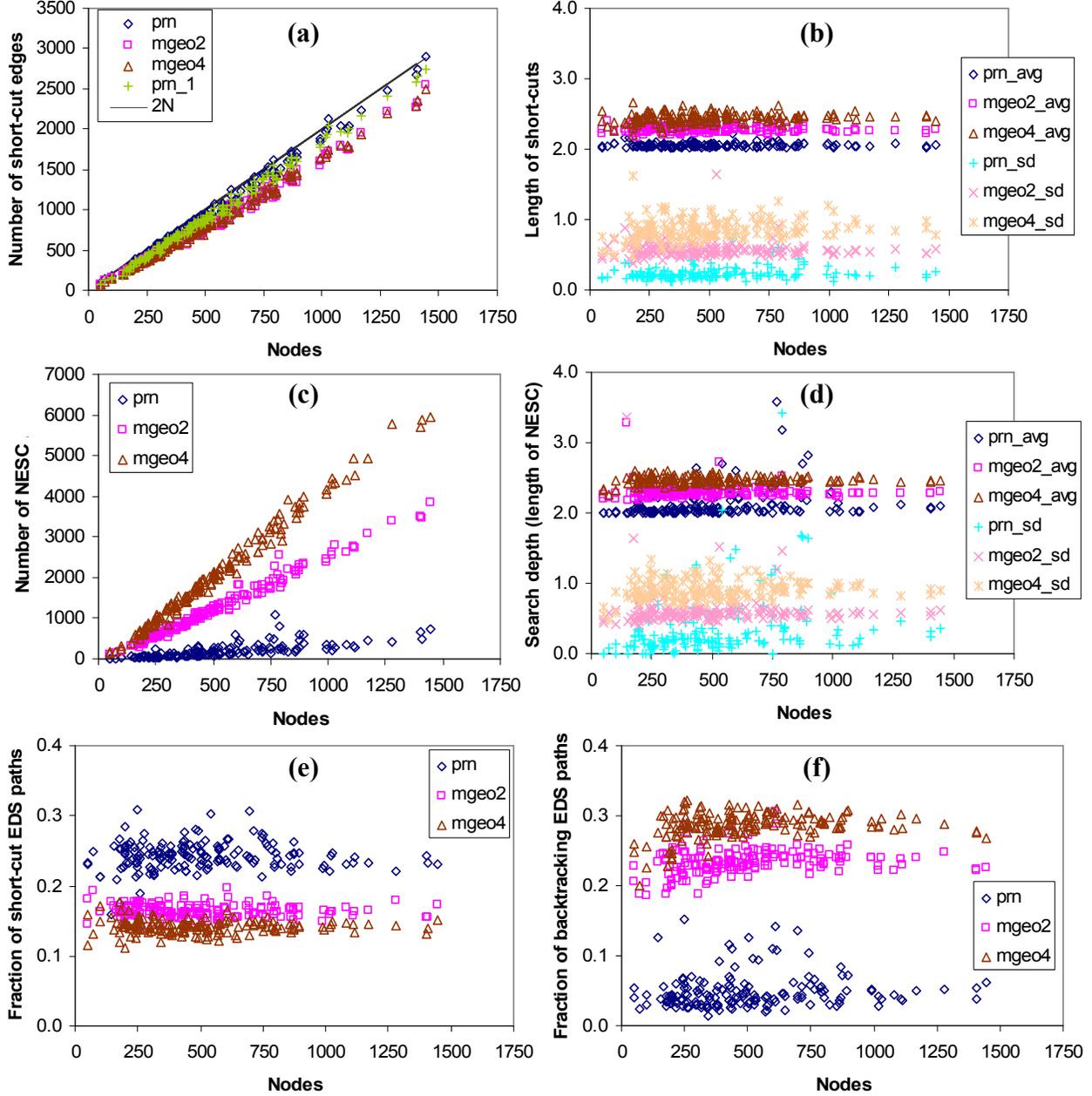

**Fig. 11 The more clustered PRNs have more short-cuts and do less backtracking than the less clustered MGEO networks. (a)** PRNs have significantly more short-cuts than MGEO networks. The number of short-cuts in PRNs is about twice the number of nodes in a network, $|SC| \sim 2N$. This approximation remains valid even when only one EDS path between each node-pair is used in the calculation (*prn_1*). **(b)** PRNs have significantly shorter short-cuts than MGEO networks. The length of a short-cut is the size of the navigational cycle it closes less one. A short-cut may close more than one navigational cycle of different size. We recorded all lengths for each short-cut, and report the average length over all short-cuts found by EDS in a network. **(c)** The MGEO networks have significantly more non-existent short-cut edges (*NESC*) than PRNs. **(d)** MGEO networks have significantly longer *NESC*s than PRNs. The length of an *NESC* is also the number of edges EDS needs to retrace when backtracking, i.e. search depth. Since an NESC can be part of more than one navigational hole of different size, we recorded all lengths for each *NESC*, and report the average length over all *NESC*s found by EDS in a network. **(e)** The fraction of EDS paths that traverses at least one short-cut edge is significantly larger for PRNs than MGEO networks. **(f)** The fraction of EDS paths that does backtracking is significantly smaller for PRNs than MGEO networks. EDS is more likely to backtrack in MGEO networks than in PRNs.

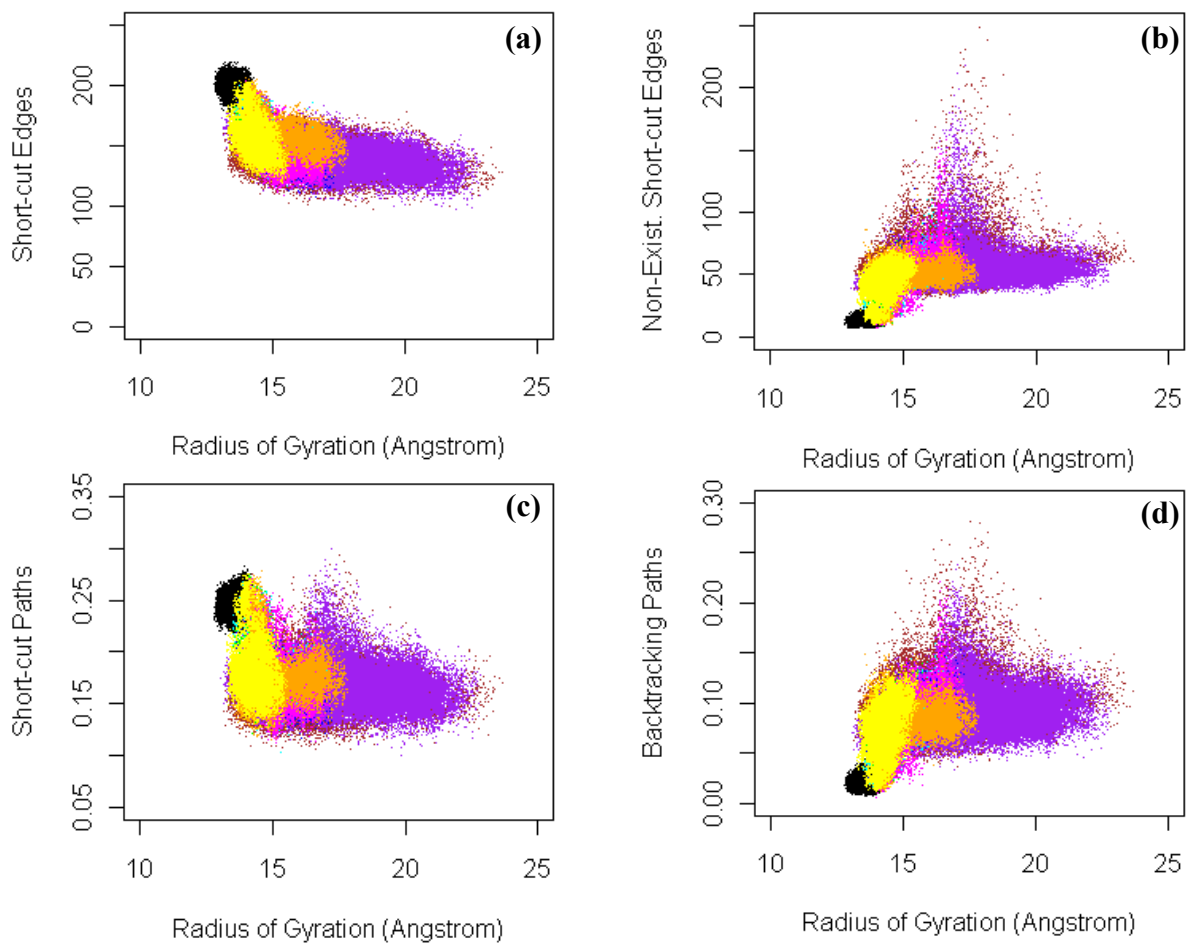

**Fig. 12 Short-cuts and backtracks in PRNs from the 2EZN MD simulation.** As the protein folds, generally: **(a)** the number of short-cut edges increases, **(b)** the number of non-existent short-cuts decreases, **(c)** the fraction of EDS paths containing at least one short-cut increases, and **(d)** the fraction of EDS paths doing at least one backtrack decreases. In the four plots, the points in black denote data from the native dynamics run 6250, while the non-black points come from the rest of the eight runs simulating non-native dynamics (section 2.7).

When short-cut edges are removed from PRNs, there is a significant increase in both BFS and EDS average path length (Figs. 13a & 13b). This is understandable since the reduced network (PRN \ SC) has fewer links. However, the increase in average EDS path length is significantly larger than when a corresponding number of random edges are removed from PRNs (Fig. 13d). Hence, short-cut edges have a significant impact on average EDS path length and are a non-random subset of PRN links. In contrast, the increase in average BFS path length is significantly smaller than when a corresponding number of random edges are removed from PRNs (Fig. 13c). We revisit the effect of short-cuts on EDS path lengths in section 3.6.

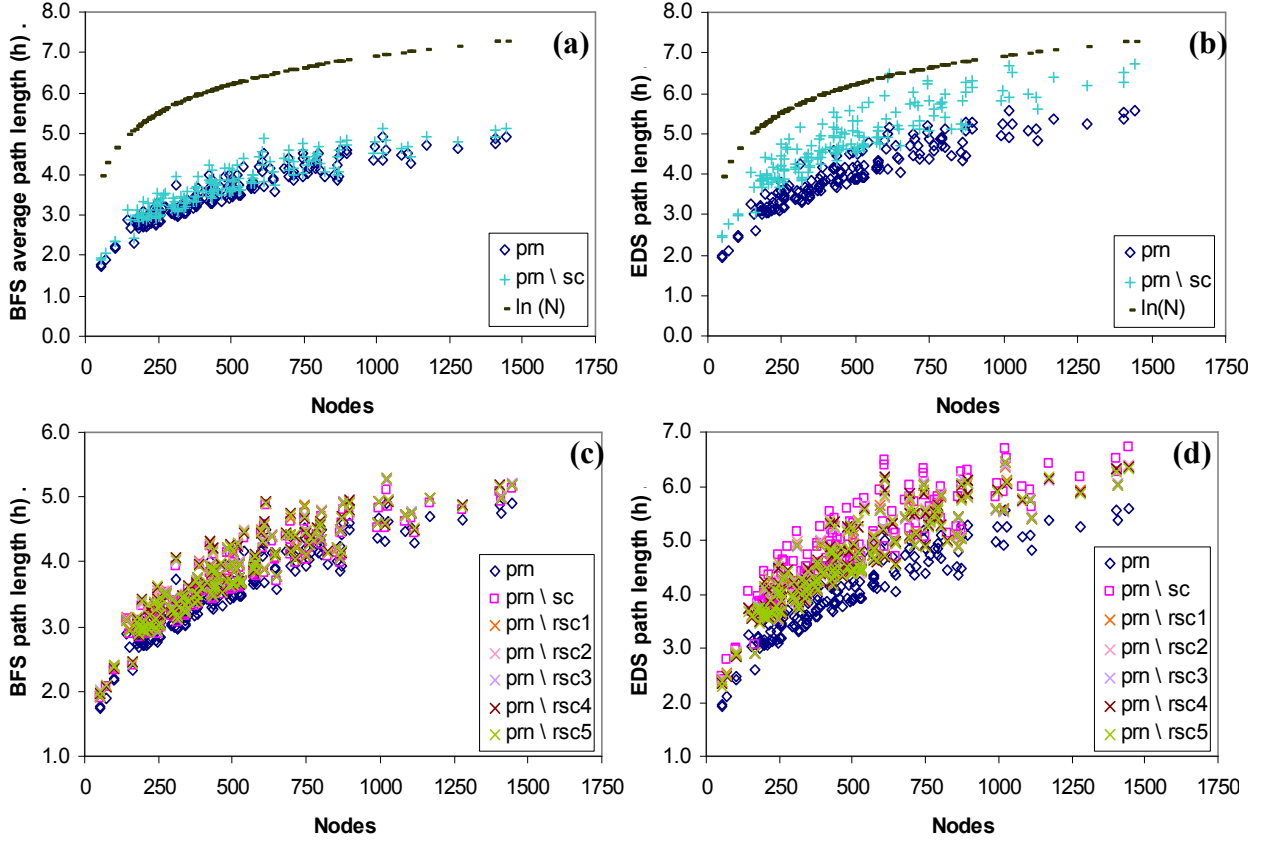

**Fig. 13 Effect of removing short-cut edges on average path length. (a & b)** Removal of short-cut edges from PRNs (denoted *prn\sc*) significantly increases the lengths of both BFS and EDS paths, but the effect is greater with EDS paths. **(c & d)** Removal of *rsc* (random short-cuts are described in section 2.6) from PRNs increased the average length of both BFS and EDS paths. However, *prn\sc* produces significantly longer EDS paths than *prn\rsc*. We revisit the difference in (d) in section 3.6 when more is known about short-cut edges. Since removing edges could potentially disconnect a PRN, the harmonic mean method of calculating average path length [8] (indicated by *h*) is used in these figures.

## 3.2 Search cost

Define the cost of finding a path from node *s* to node *d* as the number of unique nodes in the network that is touched by an algorithm when searching for the path. The search cost of a BFS path is the number of unique nodes visited by the BFS algorithm. The search cost of an EDS path is the number of unique nodes stored in memory (visited and enquired) by the EDS algorithm (see Appendix E for example). Due to the more diffusive nature of BFS, BFS paths rapidly become more expensive than EDS paths. BFS search cost increases linearly with network size, while EDS search cost scales logarithmically with *N* (Fig. 14).

Unlike BFS, EDS defers branching into other paths and only develops the path which appears most promising at the time. EDS touches only the direct neighbours of a node at each step of the way on a path. Thus, the cost of an EDS path *p* is at most the cardinality of the union of the direct neighbors of all nodes on $p$, and is unaffected by backtracking. Due to heavy clustering, average EDS cost of a path is much less than $K\,ln(N)$ and approximates $(1-C)\,K\,ln(N)$ (Fig. 14b). Both mean node degree $K$ and the clustering coefficient $C$ are almost constant for PRNs and MGEO networks (Figs. 5a & 5b). BFS search cost is unaffected by changes in network clustering introduced via the MGEO networks (Fig. 14a). However, reduced clustering has an inflationary effect on EDS search cost. EDS paths on PRNs are significantly less costly to find than EDS paths on MGEO networks (Fig. 14b). Lower search cost indicates a less diffusive type of search, which may be preferable to reduce cross-talk when multiple searches are conducted simultaneously. It is also more reflective of how energy flows in proteins, which is anisotropic and sub-diffusive [13].

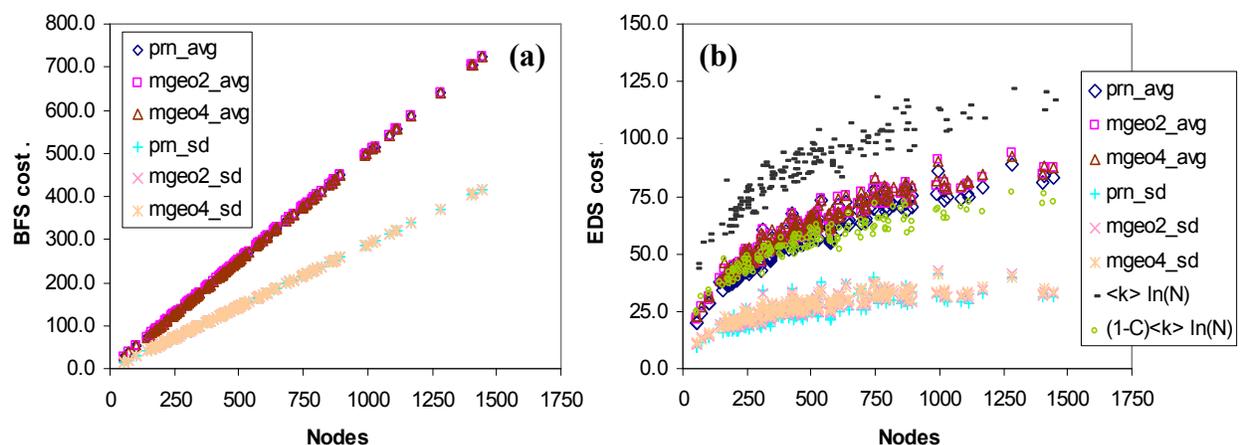

**Fig. 14 Average cost of finding paths.** The cost of finding a path from node *s* to node *d* is the number of unique nodes in the network that is touched by an algorithm when searching for the path. **(a)** BFS search cost increases linearly with the number of nodes in a network, *N*. and it is unaffected by changes in clustering. **(b)** EDS search cost scales logarithmically with *N*. The cost of an EDS path *p* is at most the cardinality of the union of the direct neighbors of all nodes on *p*, which due to heavy clustering is much less than $<k>$ $ln(N)$ and approximates $(1-C)<k>ln(N)$. EDS search cost is affected by changes in clustering. Reduced clustering increases EDS search cost. EDS paths on PRNs are significantly less costly to find than EDS paths on MGEO networks.

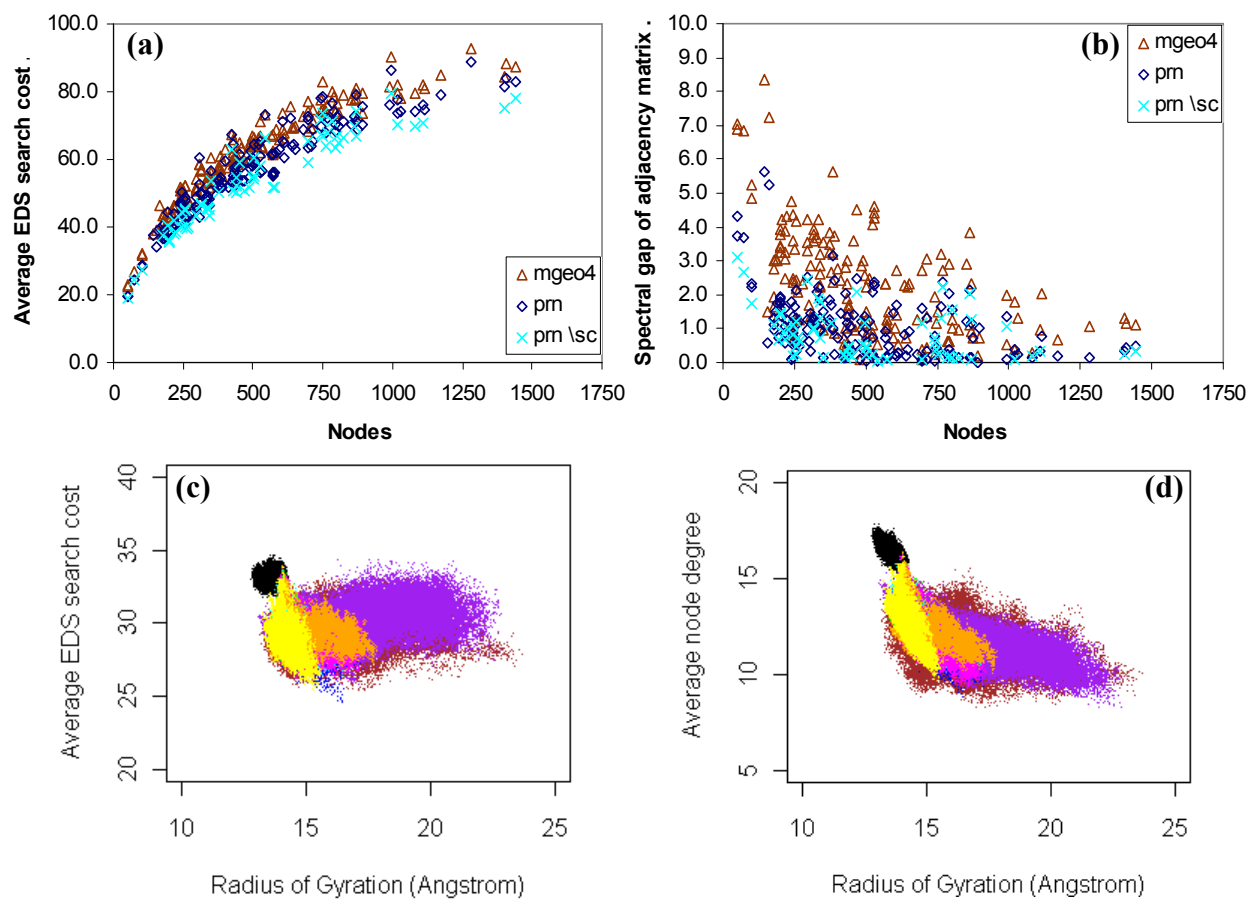

**Fig. 15 Average EDS search cost and expansion of PRNs are positively related. (a & b)** Networks with higher average EDS search cost also have larger spectral gaps. Good expander graphs are associated with large adjacency matrix spectral gaps. **(c)** Average EDS search cost increases as the 2EZN protein folds. **(d)** Average node degree increases as the 2EZN protein folds.

The low EDS search cost is hinted at by the expansion property of PRNs (section 2.4). The expansion factor of a graph with $V$ nodes is determined by the smallest $|\mathcal{N}\,(S)|\,/\,|S|$ found over all sufficiently small ($< |V|/2$) node subsets of $V$ (see Appendix D for more details). By definition, EDS search cost for a path $p$ is also the size of the boundary for nodes in $p$, i.e. $|\mathcal{N}\,(S)|$ with $S$ as the set of unique nodes in $p$. $|S|$ is $\leq N/2$ since EDS paths are short ($< ln(N)$). Thus, by computing the EDS search cost for all paths, we are essentially calculating $|\mathcal{N}(S)|$ for a sample of node subsets which are sufficiently small. A direct relationship is observed between average EDS search cost and the spectral gap of the adjacency matrix. Networks with significantly higher average EDS search cost have significantly larger adjacency matrix spectral gaps (Figs. 15a & 15b).

As with average EDS search cost, the expansion factor (as indicated by the adjacency matrix spectral gap) for PRNs is sensitive to both clustering and average node degree. Both MGEO4 and PRN\SC networks have significantly weaker clustering than PRNs, but MGEO4 networks have significantly larger adjacency matrix spectral gaps than PRNs, while PRN\SC networks have significantly smaller adjacency matrix spectral gaps than PRNs. PRN\SC are worse expanders than PRNs because they have smaller average node degree. Average EDS search cost increases as the 2EZN protein folds, and is highest in the native state (Fig. 15c). This means that the increase in average EDS search cost due to the increase in average node degree as 2EZN folds (Fig. 15d), more than offsets the decrease in average EDS search cost due to the increase in average clustering (Fig. 2). Native PRNs may not be good expanders, but the process of protein folding improves their expansion property and this improvement, along with the increase in clustering, contributes to the navigability of PRNs.

### 3.3 Hierarchy

Kleinberg describes his decentralized (local search) algorithm as homing in on a target node [6]. For PRNs however, even on EDS paths with no backtracking, Euclidean distance to target need not decrease monotonically. But for a fraction of EDS paths (and also BFS paths), what does change monotonically is node degree. Paths with monotonically increasing (type S1), monotonically decreasing (type S2) or monotonically increasing then decreasing node degrees (type S3) are hierarchical paths [42]. Implicit in this definition are paths with constant node degree (type S0). BFS and EDS hierarchical paths exhibit very similar hierarchical path decomposition (Fig. 16). The existence of both BFS and EDS hierarchical paths supports the notion that some hierarchical organization is present in PRNs. The fraction of hierarchical paths of a type is almost constant for $N > 250$ (larger multi-domain proteins). The majority of hierarchical paths are of type S3, which echoes the zoom-in zoom-out navigational pattern reported in other real-world networks [43], and is further evidence of the navigability of PRNs. Zoom-in correlates with increase in node degree and zoom-out with decrease in node degree.

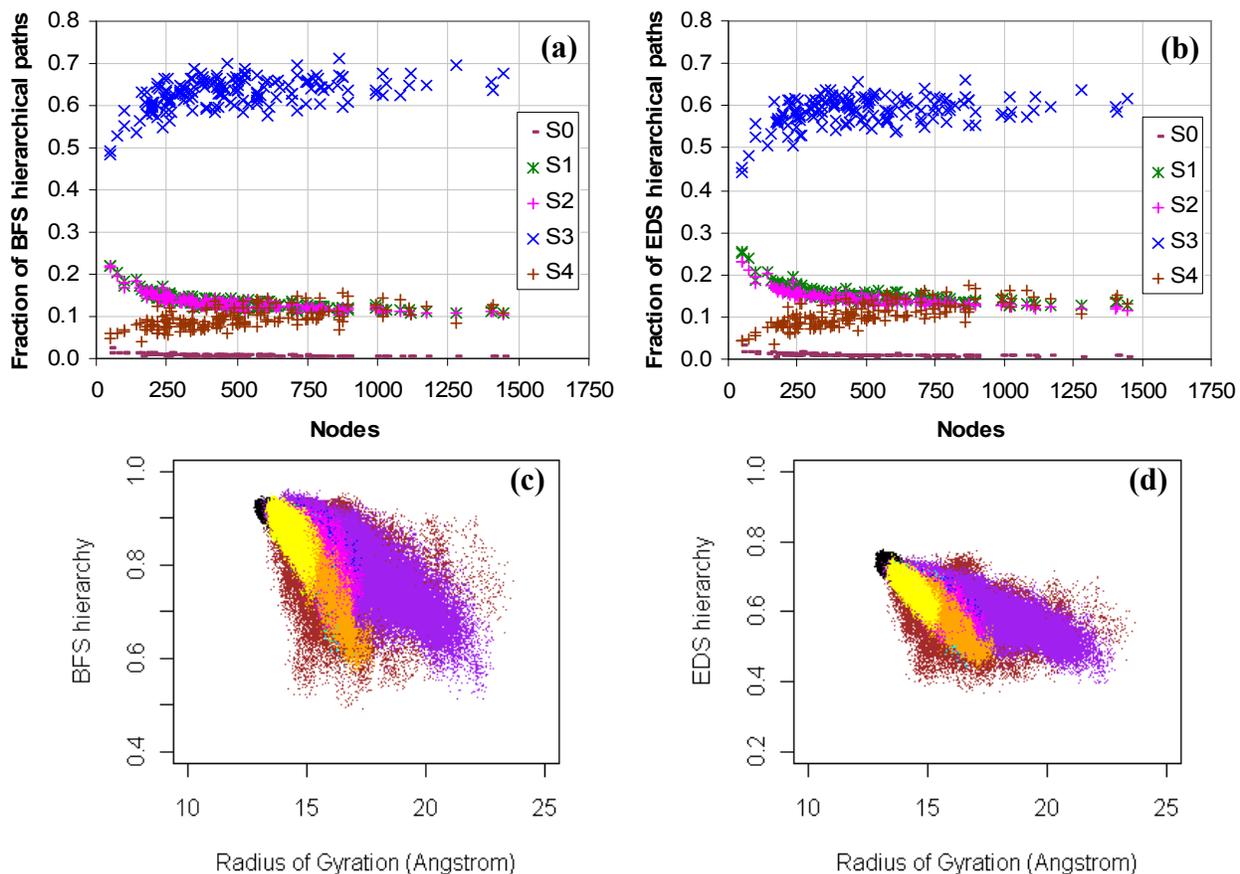

**Fig. 16 Hierarchical paths. (a & b)** Decomposition of BFS and EDS hierarchical paths by type. Hierarchical path types *S0…S3* are explained in the text. *S4* denotes paths with monotonically decreasing then monotonically increasing node degrees. *S4* paths are not typical hierarchical paths; they may even be considered *anti-hierarchy*. But we include them since *S4* paths also demonstrate structure. **(c & d)** The fraction of BFS paths and EDS paths which are hierarchical increases as the 2EZN protein folds.

### 3.4 Centrality

One of the first centrality measures to appear in the protein literature is *betweenness centrality*, defined as the number of shortest paths passing through a node [2]. Ref. [2] observed that betweenness centrality changes as a protein folds to its unique native state and that residues with high (large) betweenness centrality during the transition state play a critical role in the folding process.

Given a set of paths $P$, define the betweenness centrality of a node $v$ as the number of times paths in $P$ passes through it, i.e. enter and exit, $v$. To observe the effect of local search on betweenness centrality in PRNs, we ranked the nodes according to their respective BFS and EDS betweenness centrality values and compared the two node rankings. We found a strong positive correlation between BFS betweenness centrality rank and EDS betweenness centrality rank, and this correlation is significantly stronger for PRNs than the MGEO networks. The average Pearson correlation for PRN, MGEO2 and MGEO4 networks are 0.9018 (std. dev. = 0.0341), 0.8425 (std. dev. = 0.0431) and 0.8117 (std. dev. = 0.0580) respectively. A strong positive correlation between BFS betweenness centrality rank and EDS betweenness centrality rank means that nodes that are ranked highly by BFS betweenness centrality are also likely to be ranked highly by EDS betweenness centrality.

Another centrality measure that has proved insightful in protein research is *closeness centrality*, which measures how close nodes are to each other in a network. A node with high (large) closeness centrality has a low average distance to other nodes in the network. Closeness centrality of a node $i$ is then inversely related to the length of the paths starting from node $i$ to all the other $N$-1 nodes in the network: $CL_i = (N-1) \bigg/ \sum_{j \neq i}^{N} \lambda(i,j)$. The average closeness centrality for a network with $N$ nodes is: $CL = \frac{1}{N} \sum_{i}^{N} CL_i$. Since closeness centrality is a structural measure based on path-length, it too is influenced by clustering. Closeness centrality has been applied to detect functional protein sites [44] and to identify possible protein folding nucleation sites [45].

The decreased levels of clustering in the MGEO networks decreases the length of BFS paths and thus average BFS closeness is significantly larger in MGEO networks than PRNs. In contrast, EDS closeness centrality decreases significantly when clustering decreases. EDS closeness centrality responds to changes in clustering in a way that is more in keeping with the notion that as a protein becomes more compact and clustering increases, nodes or amino acids get closer to each other both in Euclidean distance and in graph distance. We found a strong positive correlation between BFS closeness centrality rank and EDS closeness centrality rank, and this correlation is significantly stronger for PRNs than the MGEO networks. The average Pearson correlation for PRN, MGEO2 and MGEO4 networks are 0.9716 (std. dev. = 0.0470), 0.9541 (std. dev. = 0.0435) and 0.9329 (std. dev. = 0.0573) respectively. This means that nodes that are ranked highly by BFS closeness centrality are also likely to be ranked highly by EDS closeness centrality.

The strong and positive correlations between BFS and EDS betweenness centrality rank, and between BFS and EDS closeness centrality rank, are encouraging from a utilitarian perspective because it implies the transferability of existing research based on BFS, e.g. [44, 45, 46], and also the possibility of improving upon existing results with EDS.

### 3.5 Link usage

Short-range links are those with sequence distance $|u - v| \leq 10$ (section 2.1). Protein sequence distance is a familiar and meaningful metric in protein research. There is much discussion about the role of short- and long-range contacts in proteins [11]. There is a behavioral cost to the presence of long-range links in PRNs. Larger proteins with more long-range links tend to fold more slowly [47, 48, 49]. There is also an entropic cost to the formation of long-range links early in protein folding as the conformational possibilities of a sequence segment book-ended by a long-range contact is greatly reduced [50]. And indeed, the ratio of long-range to short-range links (*LE*/*SE*) increases for PRN edges as the 2EZN protein folds under MD simulation, and so does the *LE*/*SE* ratio for short-cut edges (Fig. 17). The bias towards the appearance of short-range links before long-range links accords with the framework model of protein folding which envisions protein folding as ascending a hierarchy of more complex forms starting from the secondary structure elements and progressing to higher order folds that comprise the organization of lower order elements. This process need not however exclude the necessity for feedback between the levels of organization [51]. In spatial networks, where nodes are located in a metric space e.g.: transportation and neural networks, links cover actual physical distances and as such longer links have a higher wiring cost [52]. For these reasons, relying more heavily on shorter links is deemed preferable.

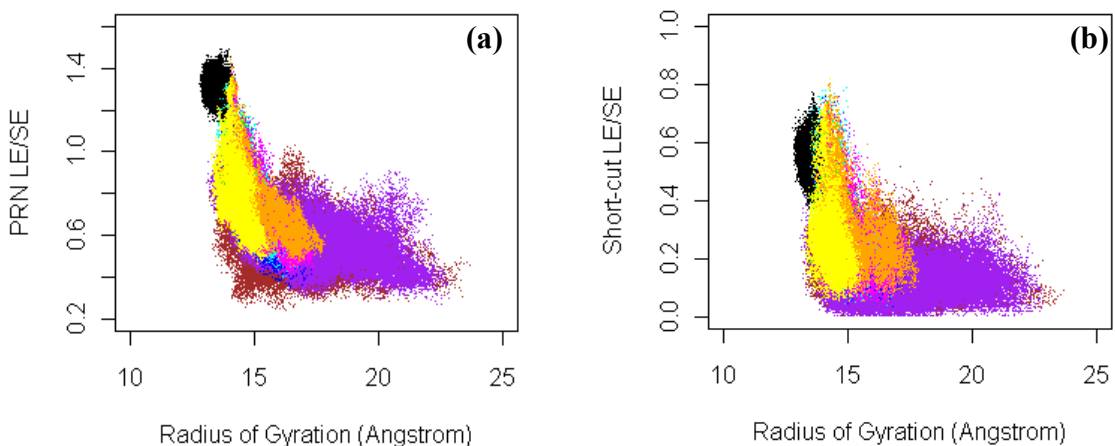

**Fig. 17 The ratio of long-range to short-range links (*LE*/*SE*)** for **(a)** PRN edges and **(b)** short-cut edges increases as the 2EZN protein folds under MD simulation. The *LE*/*SE* ratio for short-cut edges remains less than 1.0.

Usage of an edge $e$ is incremented by one with each traversal of $e$ by a path. We observe that BFS paths are more biased towards long-range links than EDS paths (Fig. 18a). Since BFS paths are significantly shorter on average than EDS paths, link usage is normalized by average path length prior to statistical testing. After normalization, we found that: (i) BFS paths make significantly less use of short-range links than EDS paths; (ii) BFS paths make significantly greater use of long-range links than EDS paths; and (iii) BFS paths make significantly less use of 3did links than EDS paths (Fig. 18b).

Short-cut edges identified by EDS paths are predominantly short-range (Fig. 19). An average of 63.91% (std. dev. = 12.84%) of a short-cut edge set is composed of short-range links, while only 43.42% (std. dev=6.95%) of a PRN's links is short-range. Accordingly, the LE/SE ratio for short-cut edges is less than 1.0 in Fig. 17b. The cutoff of 10 residues was chosen so that links within secondary structure elements would be short-range (on average an α-helix is 11 residues in length and a β-strand, 6 residues) [28]. There is some experimental evidence that energy in proteins is transported via secondary structures [53, 54]. Energy transfer is faster in helixes (reaching the speed of sound if the helixes are rigid) and slower in the hydrogen bond network of beta sheets. And indeed, EDS paths in the 166 PRNs make heavier use of short-cut edges than non-short-cut edges, not only in terms of the number of paths (Fig. 20a) but also the length of paths (Fig. 20b). Thus, short-cuts are not only more central, they also carry heavier loads on average. The load of an edge is defined in [55] as the sum of the length of paths traversing an edge.

The more central position of short-cut edges is so not only for PRNs of configurations at equilibrium, but also for PRNs of configurations not at equilibrium (Fig. 20c). Average EDS edge usage, of both short-cuts and all edges, is significantly lighter (smaller) for PRNs in equilibrium which indicates that EDS edge usage becomes more balanced as a protein folds (Figs. 20d & 20e). A more detailed study of the EDS paths between allosterically linked protein binding sites may be fruitful. In section 3.7, we use the idea of "communication pathways" to show that EDS paths are more plausible channels for allosteric intra-protein communication than BFS paths, and this plausibility is tied to EDS preference for short-range links.

### 3.6 Statistical signature of short-cut edges

As a consequence of being dominated by short-range links, short-cuts do not have significantly higher average edge multiplicity than non-short-cut edges. This follows from Fig. 5f which reports that short-range edges have significantly smaller average EM. However, against the backdrop of all edges, short-cut edges carry a distinctive statistical signature: they are situated in highly clustered areas, but have significantly weaker local community structure [56] themselves. Quantitatively this means that clustering amongst the first neighbors of short-cut links is significantly higher than clustering amongst the first neighbors of all edges, and clustering amongst the common first neighbors of short-cut links is significantly lower than clustering amongst the common first neighbors of all edges (Figs. 21a & 21b).

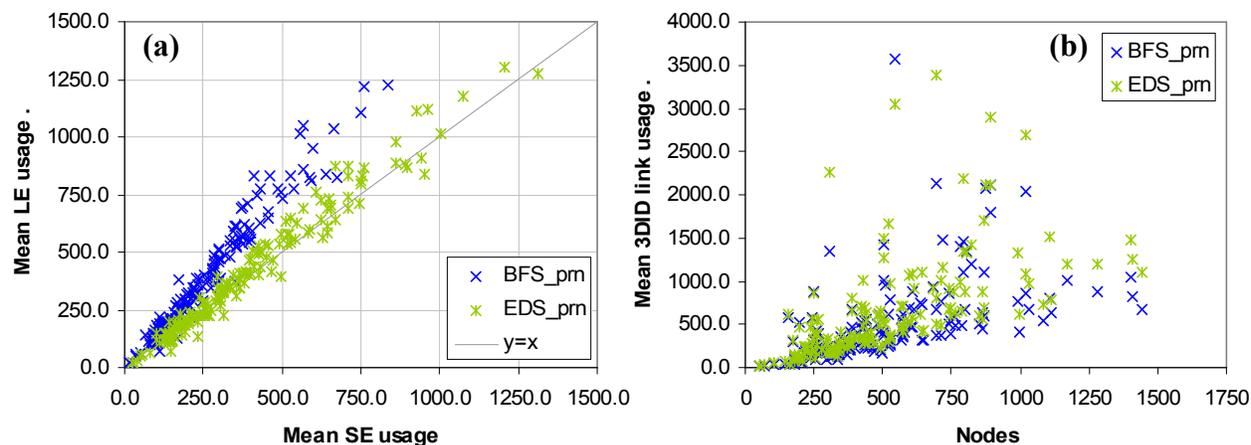

**Fig. 18 EDS paths make significantly more use of short-range links than long-range links, and significantly more use of 3did links than BFS paths.** Usage of an edge $e$ is incremented by one with each traversal of $e$ by a path. Each point in the plot reports the results for one of the 166 PRNs. **(a)** BFS paths are more biased towards long-range links than EDS paths. **(b)** BFS paths make significantly less use of 3did links than EDS paths.

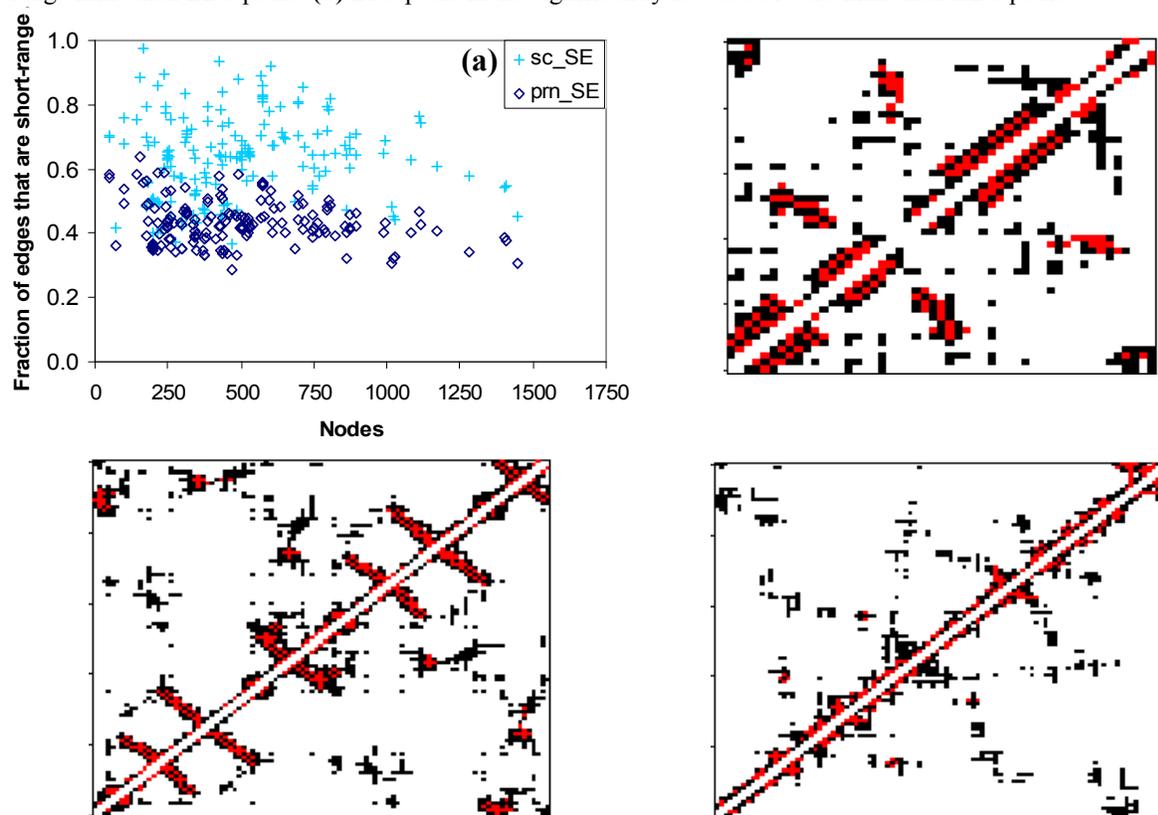

**Fig. 19 Short-cut edges are significantly enriched with short-range links (SE). Top left (a):** On average, 63.91% (std. dev. = 12.84%) of short-cut edge sets are comprised of short-range links, while only 43.42% (std. dev=6.95%) of PRN links are short-range. The cutoff of 10 residues was chosen so that links within secondary structure elements would be short-range (the average length of an α-helix is 11 residues and for a β-strand is 6 residues) [28]. **Top right:** The 1B19 PRN in contact map form as in Fig 5 but with cells representing short-cut edges in red. The majority of the red cells hug the main diagonal where the protein's three alpha helix structures are located. In contact maps, alpha helix contacts are located along the main diagonal, and beta sheet contacts are situated either parallel or perpendicular to the main diagonal. **Bottom left:** Contact map for a 2EZN (beta sheet rich) native configuration with short-cuts indicated by red cells. **Bottom right:** Contact map for a 2EZN non-native configuration – the beta sheets are not discernable yet at this stage.

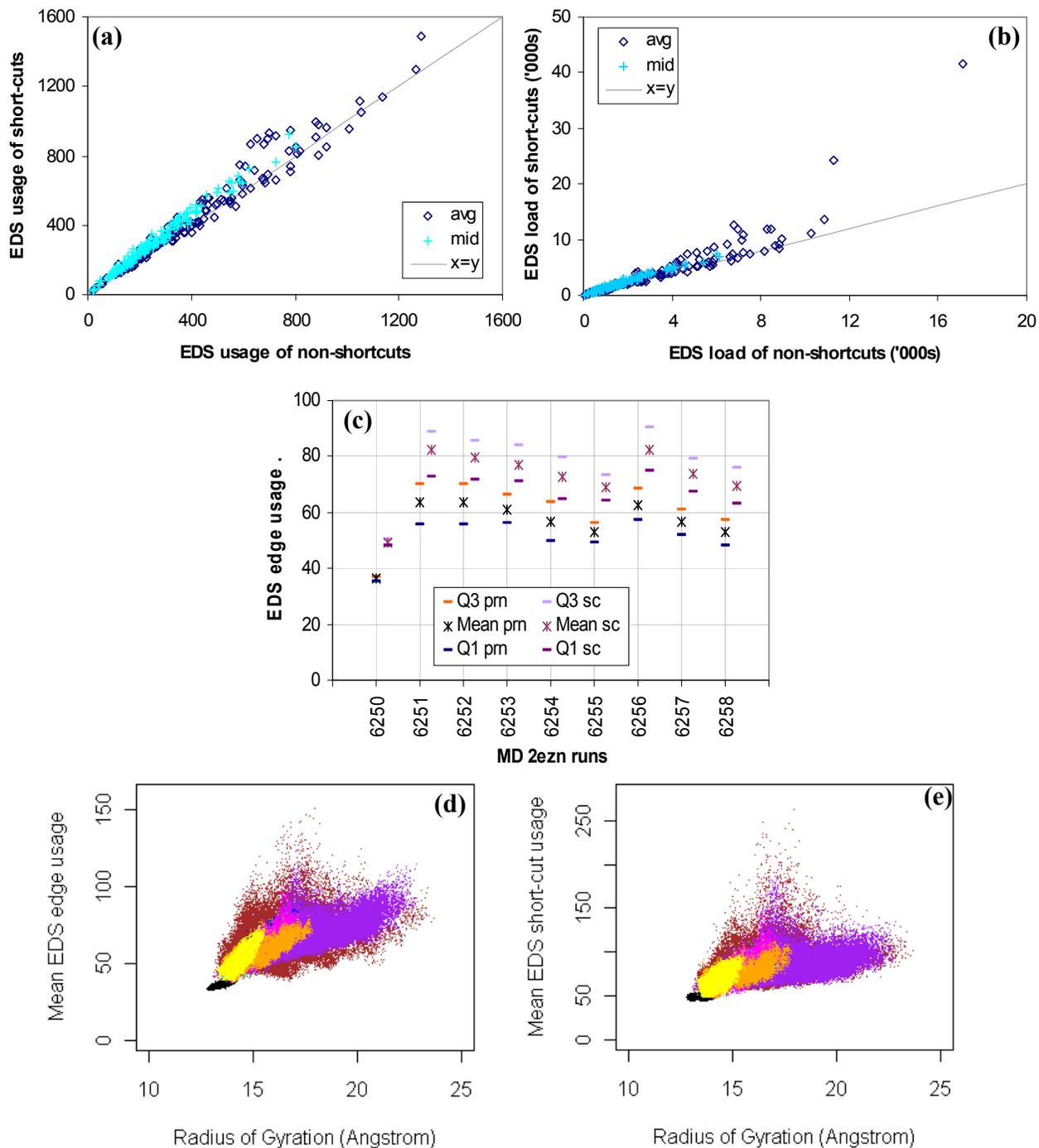

**Fig. 20 EDS paths make heavier use of short-cuts than non-short-cut edges. (a)** Short-cut edges see significantly more use than non-short-cut edges on average. **(b)** Significantly longer EDS paths traverse short-cuts than non-short-cut edges. **(c)** EDS usage of short-cuts is compared with EDS usage of all edges in PRNs from the 2EZN MD dataset (section 2.7). For all MD runs (equilibrium and non-equilibrium), average EDS usage of short-cuts (*Mean sc*) is significantly heavier (larger) than average EDS usage of all edges (*Mean prn*). Average EDS edge usage, of both short-cuts and all edges, is significantly lighter (smaller) for PRNs in equilibrium which indicates that EDS edge usage becomes more evenly balanced as a protein folds. **(d)** EDS usage of PRN edges decreases on average as links increase and paths become more diverse during protein folding (Fig. 27a). **(e)** Average EDS usage of short-cut edges also decreases as short-cut links increase during protein folding (Fig. 12a).

Define clustering amongst the first neighbors of edge $e$ as $C_{FN}(e)$, and clustering amongst the common (first) neighbors of edge $e$ as $C_{CN}(e)$. The first neighbors of edge $e = (u, v)$ in Fig. 6(left) is the nine node set $\{a, b, c, e, f, g, h, u, v\}$. Since there are 17 links amongst the nodes of this set, $C_{FN}(e) = (2 \times 17)/(9 \times 8) = 0.47222$. The common neighbors edge $(u, v)$ in Fig. 6(left) is the four node set $\{a, b, c, g\}$. As there are two links amongst the nodes of this set, $C_{CN}(e) = (2 \times 2)/(4 \times 3) = 0.33333$. The $FC$ ratio for a given set of edges $E$ is the number of edges in $E$ where $C_{FN}(e) > C_{CN}(e)$ divided by the number of edges in $E$ where $C_{FN}(e) \leq C_{CN}(e)$. For each of the 166 PRNs, the $FC$ ratio when $E$ is the set of short-cuts ($FC_{SC}$) is significantly larger than the $FC$ ratio when $E$ is the set of all PRN edges ($FC_{PRN}$) (Fig. 21c). $FC_{SC}$ is also significantly larger than the $FC$ ratio when $E$ is a set of random short-cuts. The ratio of $FC_{SC}$ to $FC_{PRN}$ increases as the 2EZN protein folds under MD simulation, signifying that the statistical signature of short-cut edges becomes stronger as PRNs approach equilibrium (Fig. 21d).

Local community structure around a link serves as a reservoir of alternate short routes should the link fail. This utility is evident from a global search or BFS perspective. But it applies as well for EDS on PRNs. Due to the relatively weaker local community structure surrounding short-cut links, EDS paths that connect node-pairs previously linked by short-cuts suffer greater positive dilation when short-cut links are removed, compared with EDS paths that do not connect node-pairs previously linked by short-cuts (Fig. 22a). This explains why the average EDS path length for *prn\sc* is a significantly longer than the average EDS path length for *prn\rsc* in Fig. 13d. *Dilation* refers to the change in path length between a node-pair under two different circumstances. It is positive when the path is elongated and negative when the path is shortened. Some negative dilation did occur when short-cuts were removed from PRNs. Negative dilation applies only to paths between node-pairs that are not originally endpoints of edges. Since $|SC|\sim 2N$, a node loses an average of four edges when short-cuts are removed from PRNs. This further weakens the local community structure around short-cut edges.

Average EDS path dilation for short-cuts increases as the 2EZN protein folds, and is significantly larger for native state PRNs (Fig. 22b). This behavior reflects the strengthening statistical signature of short-cut edges depicted in Fig. 21d. For Fig. 22, EDS path dilation between nodes $u$ and $v$ is $[\lambda_{PRN \setminus SC}(u, v) - \lambda_{PRN}(u, v)]$ or $[\lambda_{PRN \setminus RSC}(u, v) - \lambda_{PRN}(u, v)]$ where $\lambda_{PRN}(u, v)$, $\lambda_{PRN \setminus SC}(u, v)$ and $\lambda_{PRN \setminus RSC}(u, v)$ are respectively, the EDS path length between nodes $u$ and $v$ on a PRN, on a PRN without its short-cut edges, i.e. PRN\SC, and on a PRN without its random short-cuts, i.e. PRN\RSC. Average EDS path dilation for a set of edges $E$ is $\frac{1}{|E|} \sum_{i=1}^{|E|} \left[ \lambda_{PRN \setminus E}(u, v)_i - \lambda_{PRN}(u, v)_i \right] \forall (u, v) \in E$ and $(x, y) \in E \Leftrightarrow (y, x) \in E$. Average SC EDS dilation for a MD run in Fig. 22b, is the average EDS dilation over the set of short-cut edges in a PRN (snapshot), averaged over all snapshots for that run.

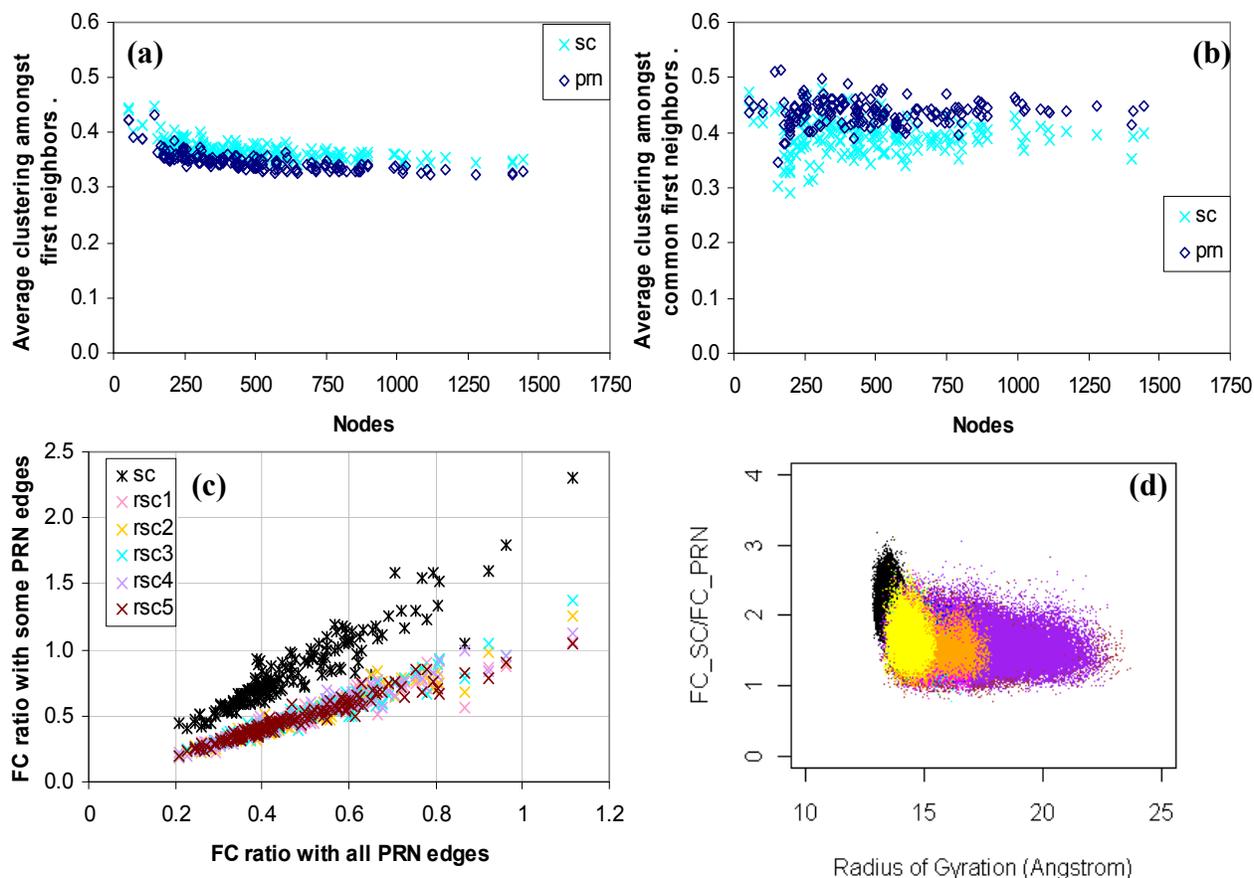

**Fig. 21 Statistical signature of short-cuts.** **(a)** Average clustering amongst first neighbors for short-cuts is significantly larger than average clustering amongst first neighbors for all edges. Short-cuts are more likely to be situated in highly clustered or dense areas in the network. **(b)** Average clustering amongst common first neighbors for short-cuts is significantly smaller than average clustering amongst common first neighbors for all edges. Short-cuts have weaker local community structure than non-shortcuts. **(c)** Short-cuts have significantly larger $FC$ ratio than random short-cuts and than all edges. An edge set with a larger $FC$ ratio has more edges whose first neighbor clustering is larger than its common first neighbor clustering, i.e. $C_{FN}(e) > C_{CN}(e)$. **(d)** The statistical signature of short-cuts strengthens as the 2EZN protein folds. FC_SC and FC_PRN are respectively the $FC$ ratio for the set of short-cuts, and the set of all edges in a PRN.

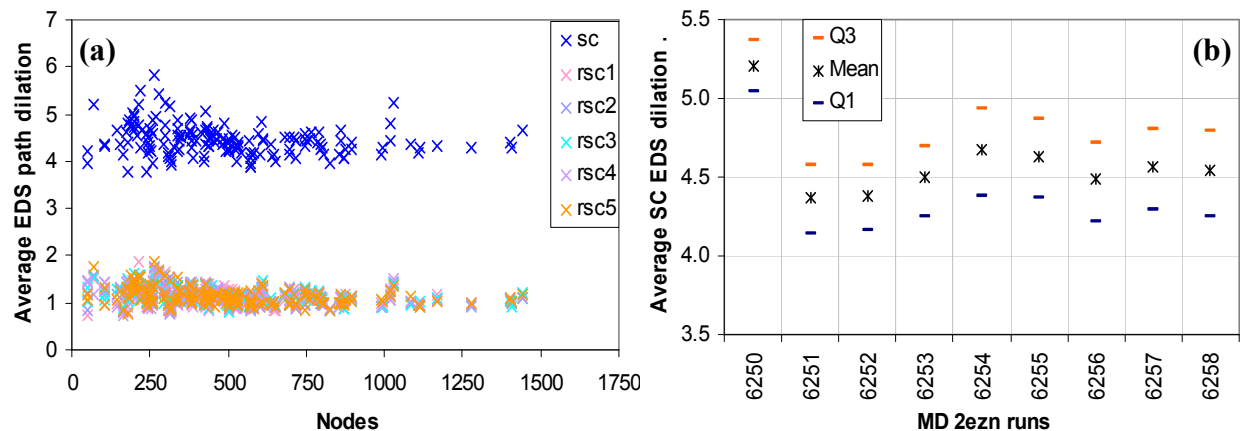

**Fig. 22 EDS path dilation.** **(a)** Short-cuts effect a significantly larger average EDS path length dilation than random short-cuts. **(b)** Average EDS path dilation for short-cuts is significantly larger for native state PRNs.

### 3.7 Path stability and communication propensity

The allosteric communication routes between the activation loop (A-loop) and the distant juxtamembrane region (JMR) of the receptor tyrosine kinase KIT was represented in [65] as a modular network of *independent dynamic segments* connected by *communication pathways*. An independent dynamic segment (IDS) is a cluster of residues whose atomic fluctuations are correlated with each other, but whose dynamical behavior is independent from other IDSs [66]. A communication pathway (CP) connects two IDSs and is composed of a chain of residues such that each link in the chain is stable and the commute time between any pair of residues in the chain is small. A link between residues of a protein is stable if the link is present in a large fraction (above a threshold e.g. $\geq 50\%$) of the protein's native ensemble (conformations generated in a MD simulation of the protein's native dynamics). The commute time between a pair of residues $(i, j)$ is the (population) variance of the Euclidean distance between $(i, j)$ in protein's native ensemble. A larger variance increases commute time and decreases communication propensity between a residue pair. Hence, CPs are stable paths with high communication propensity. The combination of IDSs and CPs capture two ways a signal or perturbation may propagate in a protein: via concerted local atomic fluctuations (typically short-ranged), and via well-defined interactions of longer range.

Ref [65] used the stability and commute time of links to construct communication pathways between IDSs. We use the idea of link stability and commute times to compare BFS paths with EDS paths in terms of their plausibility as communication pathways. To conduct this test, the PRNs from MD run 6250 which simulates the native dynamics of 2EZN are used to compute link stability and commute time, and the edges and paths from 2EZN *PRN0*, i.e. the PRN of 2EZN's PDB structure (Model 1), are evaluated. We are mainly interested in long-range paths (*LP*), i.e. paths with more than one edge that connect residues more than 10 sequence distance apart. Define path stability and path commute time as follows: Stability of a link is interpreted as link probability. Assuming links of a PRN are independent of each other (this is not entirely true because of geometric constraints), stability values for links on a path are multiplied to estimate the probability of the path. Path stability is maximum when the probability of the path is 1.0. Path commute time is the average commute time between *all* pairs of nodes on the path. A path of length $\lambda$ has $\lambda(\lambda+1)/2$ node pairs; some of the node pairs on an EDS path with backtrack may not be distinct from each other, and commute time between a residue and itself is zero.

Over all paths, we found that EDS paths are significantly more stable and have significantly lower commute times than BFS paths (Tables 3A & 3B). 72.79% of 2EZN node-pairs have EDS paths that are not less stable than BFS path. The median path stability of BFS paths is 0.5393 while the median path stability of EDS paths is 0.6350. The proportion of paths with high ($\geq 50\%$) path stability is 57.39% for EDS paths and 51.99% for BFS paths. 76.15% of node-pairs have EDS path commute time that is not larger than BFS path commute time. The median path commute time of BFS paths is 0.4200 while the median path commute time of EDS paths is 0.3698. Hence, EDS paths exhibit better stability and communication propensity than BFS paths. This conclusion holds when the analysis is broken down by path type (Tables 4A & 4B), and applies to other protein structures as well (Appendix F).

**Table 3A Path stability and path commute time statistics of BFS and EDS paths in 2EZN PRN0.** EDS paths are significantly more stable and have significantly better communication propensity (smaller commute time) than BFS paths.

| 2EZN protein Native dynamics MD run 6250. Number of paths (source-target node pairs) = 10100 | BFS paths median mean std. dev. | EDS paths median mean std. dev. |
|---|---|---|
| Path stability | 0.5393  0.5245  0.3887 | 0.6350  0.5696  0.3844 |
| Path commute time | 0.4200  0.5606  0.4653 | 0.3698  0.5135  0.4582 |

**Table 3B Distribution of paths in 2EZN PRN0 by path stability.** 51.99% of BFS paths and 57.39% of EDS paths have path stability of at least 0.5.

| Path stability | Proportion of all paths in 2EZN PRN0 | |
|---|---|---|
| | BFS | EDS |
| = 1.0 | 12.67 % | 14.60 % |
| $\geq$ 0.5 | 51.99 % | 57.39% |
| $\geq$ 0.25 | 65.07 % | 70.06 % |

**Table 4A Path stability and path commute time statistics of paths in 2EZN PRN0 by path type.** A long-range path is one whose source and destination nodes are more than ten residues apart on the protein sequence. Short-range paths are significantly more stable and more conducive to signal propagation than long-range paths. EDS paths are significantly more stable and have significantly shorter commute times than BFS paths.

| Path characteristic | Source to target sequence distance | Number of paths > 1 edge | BFS paths median | EDS paths median | Significant difference |
|---|---|---|---|---|---|
| | Any | 8426 | 0.4396 | 0.5446 | < |
| Stability | Short ($\leq$ 10) | 1218 | 0.7927 | 0.9347 | < |
| | Long (> 10) | 7208 | 0.3779 | 0.4627 | < |
| | Any | 8426 | 0.4548 | 0.3976 | > |
| Commute time | Short ($\leq$ 10) | 1218 | 0.2519 | 0.1780 | > |
| | Long (> 10) | 7208 | 0.4864 | 0.4308 | > |

**Table 4B Distribution of long-range (*LP*) paths in 2EZN PRN0 by path stability.** 41.88% of *LP* BFS paths and 47.41% of *LP* EDS paths have path stability of at least 0.5.

| Path stability | Proportion of long-range paths with > 1 edge | |
|---|---|---|
| | BFS | EDS |
| = 1.0 | 4.70 % | 5.66 % |
| $\geq$ 0.5 | 41.88 % | 47.41 % |
| $\geq$ 0.25 | 57.49 % | 62.76 % |

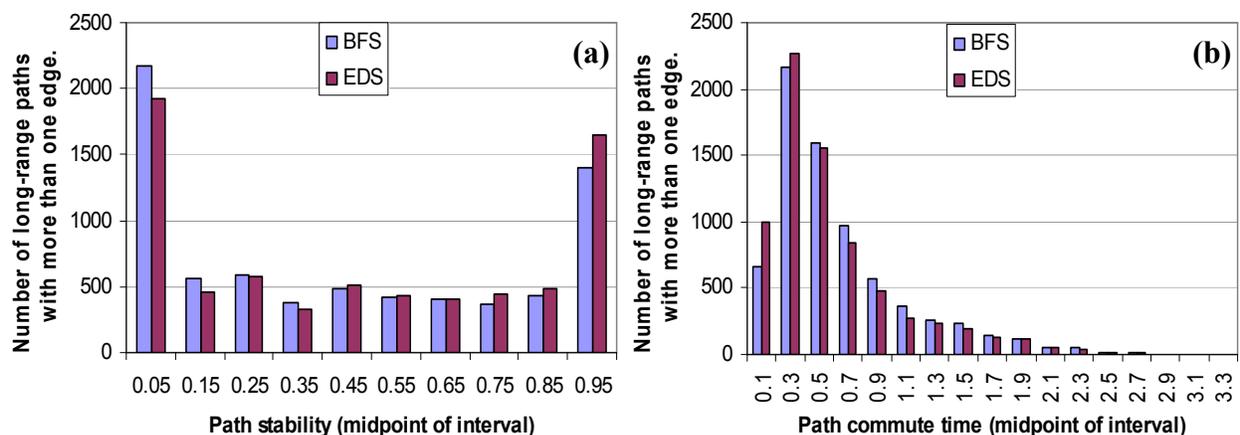

**Fig. 23 Distribution of *LP* paths by path stability (a) and by path commute time (b).** EDS paths tend to outnumber BFS paths as path stability increases and path commute time decreases. EDS paths are more plausible communication pathways for protein allostery than BFS paths.

Over all paths of length greater than one, EDS paths are significantly more stable and have significantly lower commute times than BFS paths (Tables 4A & 4B). Both short- and long-range EDS paths are significantly more stable and have significantly better communication propensity than BFS paths of the same range. The short-range paths of both BFS and EDS exhibit significantly better stability and significantly higher communication propensity than their respective long-range paths. BFS and EDS exhibit similar distribution of paths by path stability or by path commute time (Fig. 23). However, *LP* EDS paths tend to outnumber *LP* BFS paths as path stability increases and path commute time decreases. This suggests that EDS paths are more plausible communication pathways for protein allostery than BFS paths.

The above findings further evince that EDS paths are more plausible allostery channels than BFS paths. These findings follow from the different link usage pattern observed in section 3.5. Averaged over the links in 2EZN PRN0, short-range links (*SE*) are significantly more stable and have significantly smaller commute times than long-range links (*LE*); and short-cut edges are significantly more stable and have significantly smaller commute times than non-short-cut edges (Table 5). Similar observations are made with other protein structures (Appendix F).

**Table 5 Stability and commute time of links in 2EZN PRN0.** Short-range links are significantly more stable and have significantly shorter commute times than long-range links. Short-cut edges, which are predominantly short-range, are significantly more stable and have significantly shorter commute times than non short-cut links.

| 2EZN PRN0 links by type | Number of links | Stability median | mean | std. dev. | Commute time median | mean | std. dev. |
|---|---|---|---|---|---|---|---|
| All | 837 | 0.9995 | 0.8153 | 0.3102 | 0.2081 | 0.3933 | 0.4747 |
| Short-range | 346 | 1.0000 | 0.9381 | 0.1753 | 0.0993 | 0.1693 | 0.1969 |
| Long-range | 491 | 0.9510 | 0.7287 | 0.3526 | 0.3697 | 0.5511 | 0.5449 |
| Short-cut | 201 | 1.0000 | 0.9728 | 0.1358 | 0.0915 | 0.1958 | 0.3948 |
| Non-short-cut | 636 | 0.9739 | 0.7655 | 0.3324 | 0.2679 | 0.4556 | 0.4810 |

# 4. A PRN's short-cut network

The *short-cut network* (*SCN*) of a PRN is a graph whose edge set is *SC*, the set of short-cut edges identified by EDS (section 2.5) for the PRN, and whose vertex set is induced by its edge set. The number of short-cuts in a PRN was estimated in section 3.1 to be twice the number of nodes in the PRN, i.e. $|SC| \sim 2N$ (Fig. 11a). In the 2EZN MD runs, the number of short-cuts in native state configurations also approximates $2N$ (run 6250 in Fig. 24a), while in non-native state configurations, the number of short-cuts is about *1.5N* (runs 6251…6258 Fig. 24a). The number of short-cuts tends to increase as a protein folds (Fig. 12a).

While an SCN spans almost all nodes of its PRN, it may not be a connected graph. We are interested in the largest connected component of a SCN, denoted *gSCN*. The size of gSCNs is largest in PRNs for configurations at equilibrium (6250 in Fig. 24b). The two plots in Fig. 24 suggest a process of SCN growth in terms of both edges and nodes, as the 2EZN protein folds.

In this section, we investigate SCNs to discover their essential characteristics with a view to understand their formation. The linear relationship $|SC| \sim 2N$ prompted us to hypothesize that short-cut edges form a tree-like structure that spans the nodes of a PRN. We could not confirm this hypothesis, but this view point proved fruitful (section 4.2). SCNs are strongly transitive, much more so than expected from the union of two or more random spanning trees (section 4.1), and affect EDS path-pair diversity and stretch differently (section 4.2). From our comparison of "successful" and "unsuccessful" MD trajectories (section 4.4), we posit that strong SCN transitivity is an essential attribute of well-formed SCNs, and by extension well-formed PRNs.

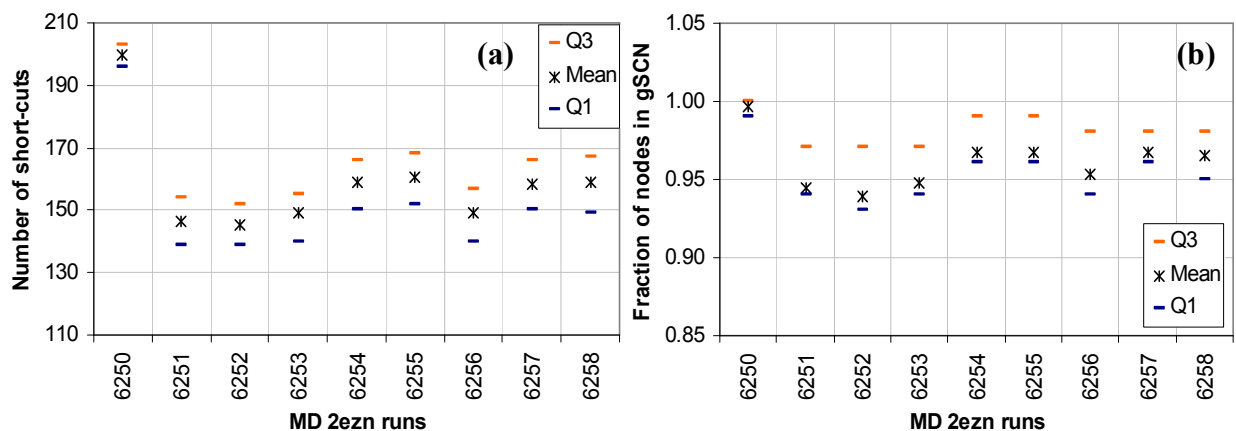

**Fig. 24 (a)** Native state PRNs (6250) have significantly larger short-cut sets than non-native state PRNs. **(b)** gSCNs span almost all nodes of the 2EZN PRN. gSCN node coverage averages around 95% in non-native state PRNs and is almost 100% in native state PRNs (6250).

## 4.1 SCN structure: transitivity

*SCN transitivity* is the average edge multiplicity of the edges of a SCN evaluated solely within the SCN. Edge multiplicity (*EM*) and transitivity was defined in section 2.3. SCN transitivity is stronger when average edge multiplicity is larger. The PRNs (of section 2.1) have significantly stronger SCN transitivity than the MGEO4 networks (Fig. 25a). This seems to follow from the fact that MGEO4 networks have significantly weaker transitivity than PRNs (Fig. 5d). However, it is not necessarily true that the average edge multiplicity or transitivity strength of a sub-graph is limited by the transitivity strength of its parent graph. In particular, graphs where numerous edges have $EM = 0$, e.g. a kite graph with a long tail. On average, 86.49% of links in MGEO4 networks have $EM > 0$ (Fig. 5c). Like PRN transitivity (Fig. 2b), SCN transitivity increases as the 2EZN protein folds in the MD simulation (Fig. 25b).

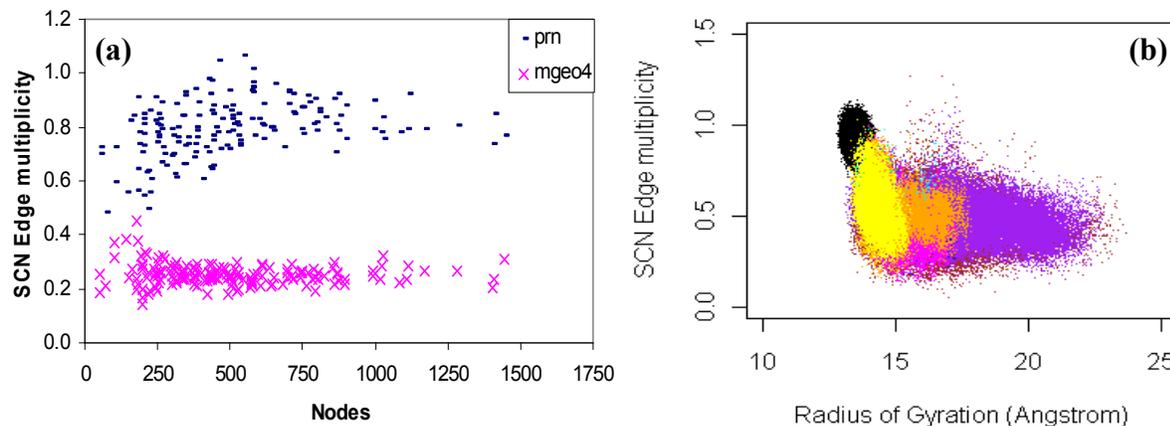

**Fig. 25 SCN transitivity. (a)** PRNs have significantly stronger SCN transitivity than MGEO4 networks. **(b)** SCN transitivity increases as the 2EZN protein folds in the MD simulation.

From the observations in Fig. 24, SCNs face the problem of node percolation on PRNs, i.e. growing their largest connected component (gSCN) so that it spans the nodes of their respective PRNs. Ref [33] found that, with degree distribution and clustering coefficient values being equal, networks with strong transitivity are more resilient to random edge removals than weakly transitive networks. The reduction in size of the giant component as edges are removed uniformly at random actually slows down in strongly transitive networks as the probability of removing a random edge $q$ increases. Conservatively, this effect is pronounced for $q > 0.2$. (Fig. 4 in [33]).

In the 2EZN MD dataset, on average between 20% – 30% of short-cut edges disappear in a step in non-equilibrium runs (Fig. 26a). A short-cut edge disappears when it either no longer exists in the PRN or it becomes a non-short-cut edge and so is no longer a part of a SCN, i.e. the short-cut edges undergoes either a *d20* or *d21* transition (Fig. 30). The fraction of short-cuts removed can be well above 35% in some steps, which calls for a SCN structure that is resilient to aggressive edge deletions. And indeed, the more strongly transitive gSCNs show significantly more resilience to random edge removals than the weakly transitive *grsc* and *grst* networks. The fraction of nodes in the largest component that remain in the largest component after random edge deletion is significantly larger for gSCNs than for both *grsc* and *grst* networks (Fig. 26b). Both *grsc* and *grst* networks fracture into significantly more pieces than gSCNs

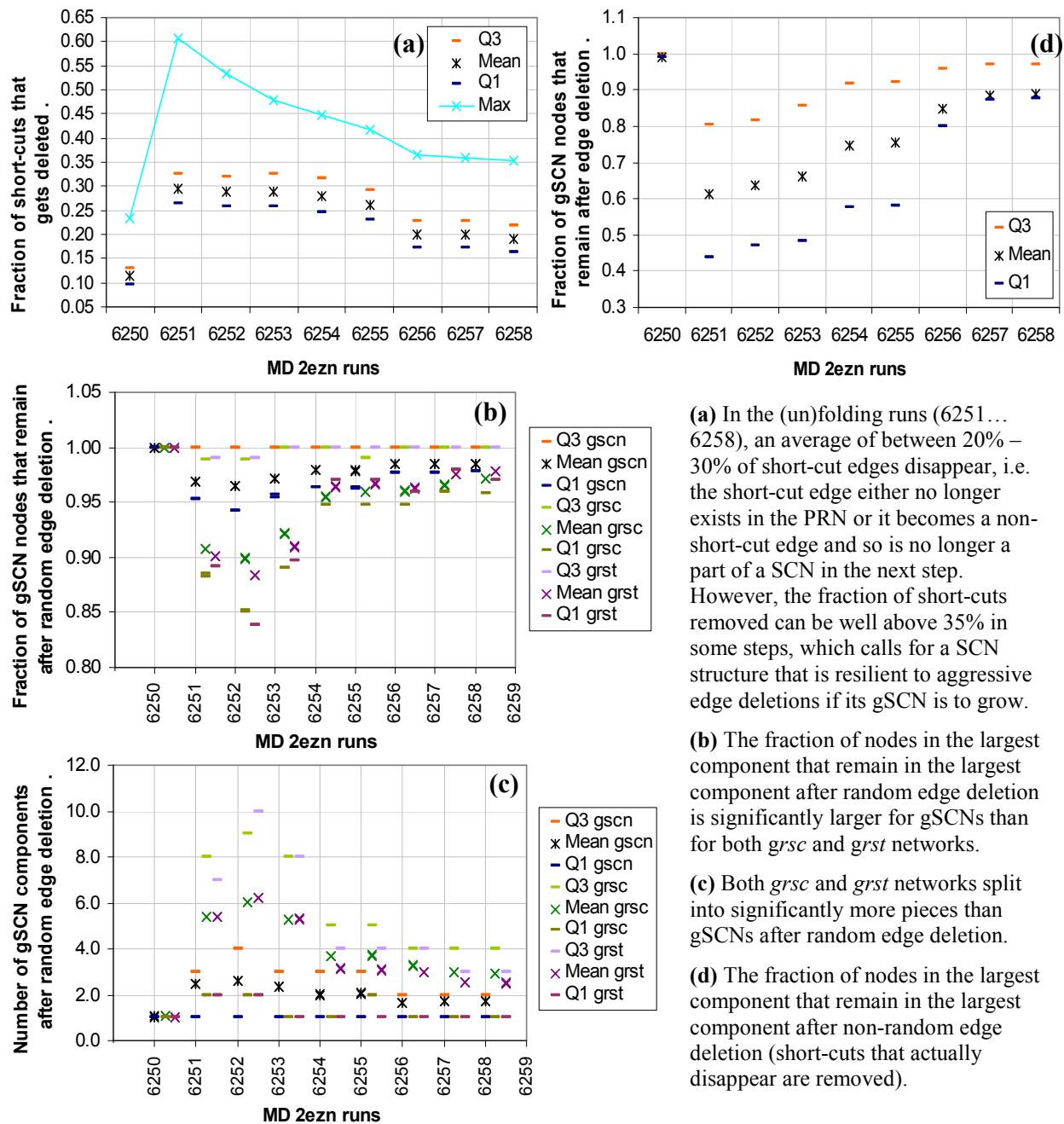

**(a)** In the (un)folding runs (6251...6258), an average of between 20% – 30% of short-cut edges disappear, i.e. the short-cut edge either no longer exists in the PRN or it becomes a non-short-cut edge and so is no longer a part of a SCN in the next step. However, the fraction of short-cuts removed can be well above 35% in some steps, which calls for a SCN structure that is resilient to aggressive edge deletions if its gSCN is to grow.

**(b)** The fraction of nodes in the largest component that remain in the largest component after random edge deletion is significantly larger for gSCNs than for both g*rsc* and g*rst* networks.

**(c)** Both g*rsc* and g*rst* networks split into significantly more pieces than gSCNs after random edge deletion.

**(d)** The fraction of nodes in the largest component that remain in the largest component after non-random edge deletion (short-cuts that actually disappear are removed).

**Fig. 26 Robustness of the largest SCN component (gSCN) to (random) edge deletions relative to *grsc* and *grst*.**

after random edge deletion (Fig. 26c). The native configurations in the MD simulation have significantly stronger SCN transitivity than non-native configurations (Fig. 25b), and this strength is reflected in the results in Fig. 26 where the size of native configuration gSCNs (6250) is almost immune to both random and non-random edge deletions.

*grsc* (*grst*) denotes the largest connected component of a *rsc* (*rst*) network. A *rsc* (*rst*) network is a graph comprising *RSC* (*RST*) edges (section 2.6). The existence of a *grsc* (*grst*) is expected since with |SC|

$\approx 2N$, the average degree of a *rsc* (*rst*) network is approximately $2(2N)/N = 4$, which is larger than the threshold required for a random graph to have a giant component [9, p.406]. gSCN is significantly more strongly transitive than both *grsc* and *grst*. The weak transitivity of *rsc* networks is expected since they are essentially random graphs (edges in *rsc* are selected uniformly at random from PRN edges and so are not expected to be correlated). The weak transitivity of *rst* networks is also expected since they are constructed from the union of trees.

Let $q_t$ be the fraction of edges deleted from gSCN$_t$ (gSCN at step $t$); and let *grsc*$_t$ and *grst*$_t$ be respectively, the largest component of the *rsc* and *rst* networks associated with PRN$_t$. Fig. 26a gives the $q$ value averaged over all steps in a run. For results reported in Figs. 26b & 26c, $q_t$ of the edges in gSCN$_t$, *grsc*$_t$ and *grst*$_t$ were chosen uniformly at random for removal per step $t$, and the results are averaged over all steps in a run. In Fig. 26d, $q_t$ of the edges in gSCN$_t$ is also removed but these edges are the actual short-cut edges that disappear at step $t+1$. The fraction of nodes in the largest component that remain in the largest component after this non-random edge deletion is significantly smaller than when the same number of random edges is removed from gSCN. Thus, the deleted SCN edges are not a random edge set.

## 4.2 SCN function: EDS path-pair stretch and independence

Our spanning tree point of view mentioned at the start of section 4 is motivated by several factors. Ref. [57] showed that the union of two random spanning trees can capture the expansion property of the underlying graph and is an efficient way to design scalable, fault-tolerant and path-diverse routing. Proteins are fairly robust to random attacks according to mutagenesis studies [44], and possess alternative pathways or redundant links between critical sites [15]. These alternate pathways equip a protein with a repertoire of responses for different sets of conditions [65]. Further, path diversity helps to lower a network's congestion threshold [58, 59], and to increase ease of synchronization stability [55]. For these reasons, we consider the existence of alternative pathways (*edge independent* paths) with conservative increase in average distance between alternative pathways (small *stretch* factor) to be an essential feature of well-formed native PRNs.

Let $p$ be a path from node $u$ to node $v$, and $q$ be a path from $v$ to $u$. Paths $p$ and $q$ make a path-pair. $p$ and $q$ is an edge independent or edge disjoint path-pair if and only if $p$ and $q$ have no edges in common. A network that supports the existence of many edge independent path-pairs is *path diverse*. The stretch of a path-pair is the absolute difference between the lengths of its constituent paths. And indeed, the MD simulation of 2EZN shows that PRNs of native state configurations (run 6250 data points are denoted in black in Figs. 27a & 27c) have significantly more edge independent EDS path-pairs (Fig. 27b) and a significantly smaller stretch (Fig. 27d) than the PRNs of non-native state configurations. Further, edge independence generally increases while stretch generally decreases as the 2EZN protein folds.

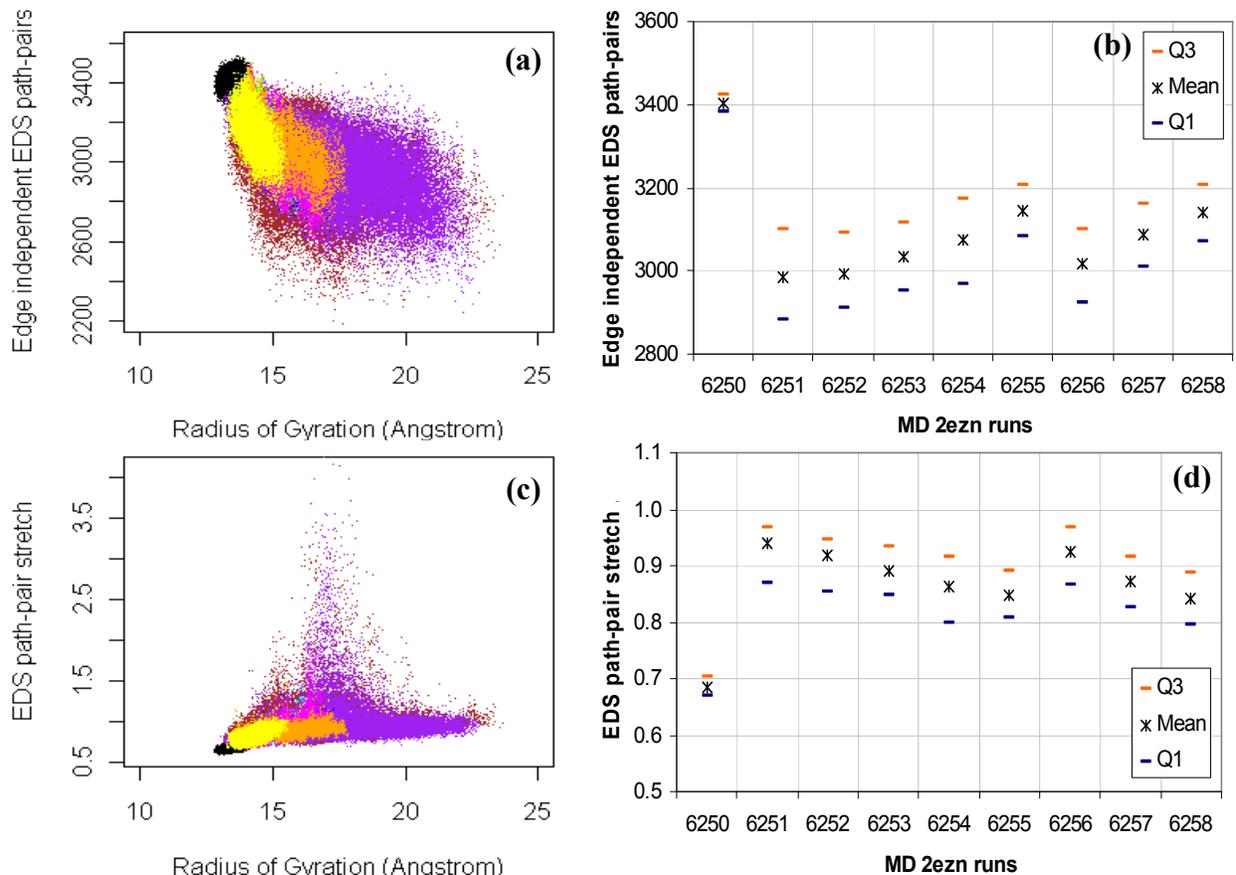

**Fig. 27 Edge-independence and stretch of EDS path-pairs in 2EZN MD runs. (a)** The number of edge independent EDS path-pairs generally increases as the 2EZN protein folds. **(b)** PRNs of native state configurations (6250) have significantly more edge independent EDS path-pairs. **(c)** EDS path-pair stretch generally decreases as the 2EZN protein folds. **(d)** PRNs of native state configurations (6250) have significantly smaller stretch than the PRNs of non-native state configurations.

For the set of 166 native state PRNs (these PRNs were inspected in detail in section 3), the proportion of EDS path-pairs which are edge independent is significantly larger than the proportion of BFS path-pairs which are edge independent (Fig. 28a). An average of 32.55% (std. dev. 1.48%) of EDS path-pairs are edge independent while only 18.89% (std. dev. 1.45%) of BFS path-pairs are edge independent (note that no explicit effort was made, either by BFS or by EDS, to construct independent paths).

An EDS path-pair where either one or both paths traverse at least one short-cut edge is more likely to be edge-independent than an EDS path-pair where neither paths traverse at least one short-cut edge (Fig. 28b). It appears then that short-cut edges play a role in keeping path-pairs edge-wise separate from each other. Investigation by elimination reveals that this conclusion applies to BFS path-pairs more so than to EDS path-pairs (Figs. 28c). Compared with BFS, EDS path diversity is more robust to the removal of short-cut edges and also to the removal of *rst* edges from PRNs (Fig. 28d). Hence, EDS is more able than BFS to exploit the alternative connections already present between nodes in PRNs. Nonetheless the removal of short-cut edges from PRNs does reduce path diversity significantly, but this reduction is significantly smaller in magnitude than the reduction caused by the removal of *rst* or *rsc* edges. The fraction of independent EDS path-pairs in 111 of the 166 PRNs decreased when short-cuts were removed. In contrast, the removal of *rst* (*rsc*) edges negatively affected path diversity in at least 156 (138) PRNs.

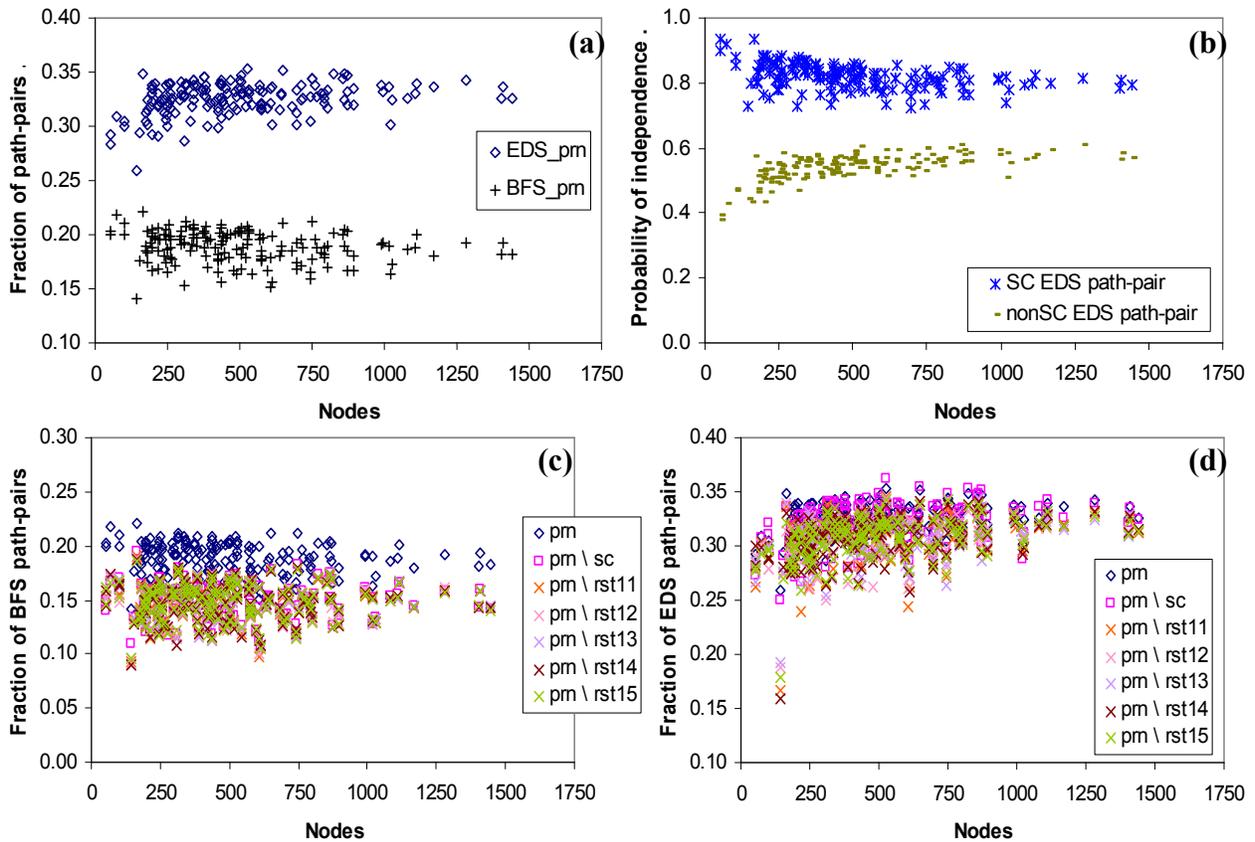

**Fig. 28 Effect of SCN on edge independence of path-pairs in the set of 166 native state PRNs. (a)** EDS path-pairs are more edge independent than BFS path-pairs. EDS paths are more diverse than BFS paths. **(b).** An EDS path-pair that traverses at least one short-cut (SC EDS path-pair) has about 80% chance of being edge independent. This chance drops to at least 60% for an EDS path-pair that does not traverse any short-cut (nonSC EDS path-pair). **(c)** BFS path diversity is significantly compromised by the removal of short-cuts (*prn\sc*) and also by the removal of *rst* edges (*prn\rst*). **(d)** EDS path diversity is robust to the removal of short-cuts (*prn\sc*) and also to the removal of *rst* edges (*prn\rst*). EDS is more able than BFS to exploit the alternative connections already present between nodes in PRNs.

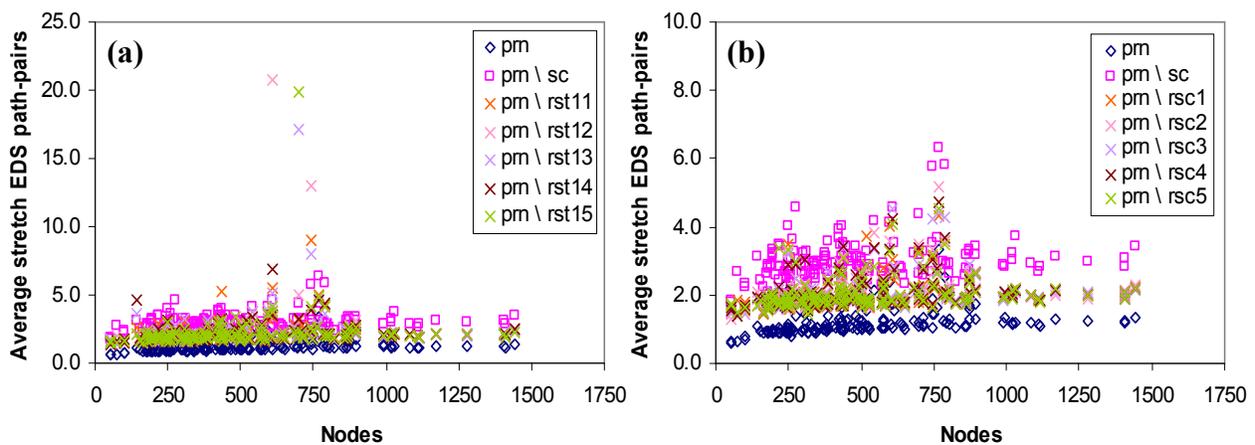

**Fig. 29 Effect of SCN on EDS path-pair stretch in the set of 166 native state PRNs.** EDS path-pairs in *prn\sc* networks experience significantly longer stretch on average than EDS path-pairs in both (a) *prn\rst* networks and (b) *prn\rsc* networks.

The removal of short-cut edges from PRNs (*prn\sc*) increases EDS path stretch significantly (Fig. 29), and this increase is significantly larger compared to the removal of *rst* edges (*prn\rst* in Fig. 29a), and of *rsc* edges (*prn\rsc* in Fig. 29b). EDS path-pairs in the 166 PRNs (*prn* in Fig. 29) experience an average stretch of 1.1650 with a standard deviation of 0.3502. The average stretch for PRNs without short-cuts is 2.9890 with a standard deviation of 0.6182. Greater average stretch contributes to longer average path lengths, and indeed PRNs sans short-cuts have significantly longer EDS average path lengths than PRNs (Figs. 13b & 13d).

The significant differences between *prn\sc* and *prn\rst* reported in this section, coupled with the structural differences between SCNs and *rst* networks identified in section 4.1, weaken the hypothesis that a PRN's set of short-cut edges is the union of two or more *random* trees spanning a PRN. Path stretch is actually not an explicit design consideration in the proposal of splicers [57]. Other graph sparsification strategies designed to preserve different characteristics of the original graph up to a factor exist. It could be insightful to investigate SCNs in terms of these other graph sparsification techniques.

### 4.3 SCN formation: preliminary observations

A consequence of a SCN's role in EDS path diversity and stretch is the volatility of its short-cut edges. The atoms of a protein are in flux, more so when the protein is far from equilibrium. This means changes in EDS paths. To preserve or increase path diversity and to maintain or reduce path stretch, the formation and destruction of short-cut edges need to be responsive to the changes in paths and their lengths. PRNs of native state configurations are more path diverse and do have smaller stretch than the PRNs of non-native state configurations (Fig. 27). Compared to the set of edges that are not short-cuts, the set of short-cut edges is significantly more volatile (Fig. 30). Except for the equilibrium state configurations (6250), a short-cut edge is significantly more likely to transition out of its current state, i.e. become a non-short-cut edge or a non-edge in the next step, than a non-short-cut edge. Non-edges are most likely to remain in their current state as non-edges in the next step.

Despite the volatility of short-cut edge sets, we could discern several patterns in them. Define the set of deleted short-cuts at step $t+1$ as delSC$_{t+1}$. An edge in delSC$_{t+1}$ is a short-cut edge at step $t$ and either a non-short-cut edge or a non-edge at step $t+1$. Per the state transition diagram in Fig. 30, deleted short-cuts are those that make either a *d21* or *d20* transition. Define the set of added short-cuts at step $t+1$ as addSC$_{t+1}$. An edge in addSC$_{t+1}$ is a non-short-cut edge at step $t$ and a short-cut edge at step $t+1$. Per the state transition diagram in Fig. 30, added short-cuts are edges that make a *d12* transition. It is also possible for a non-edge at step *t* to become a short-cut at step *t*+1 (via a *d02* transition) but we exclude such edges for now.

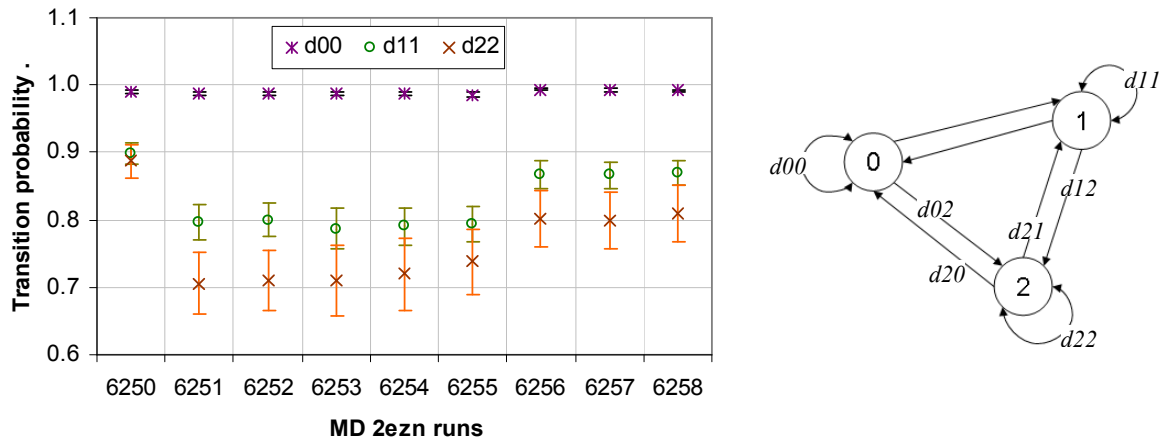

**Fig. 30 Short-cut edges are significantly more volatile than non-short-cut edges** as a result of having to adapt to changes in movement of nodes and changes in edges to support route diversity and to restrain stretch in path-pairs. *d00* denotes non-links that persist from one step to another. Since PRNs are sparse, it is unsurprising that almost all zero entries in the contact map (adjacency matrix of a PRN) at step *t* remain zero at step *t*+1. *d11* denotes non-short-cut links at step *t* that remain a non-short-cut link at step *t*+1. Except for run 6250, about 80% of non-short-cut links remain as such in the next step time. *d22* denotes short-cut links at step *t* that remain a short-cut link at step *t*+1. Except for run 6250, about 70% of short-cut links remain as such in the next step time. This means that about 30% of short-cut links undergo a change in state, i.e. either become non-links or become non-short-cut links in the next step. This significantly higher percentage of state change is why short-cut edges are considered more volatile than non-short-cut edges. Run 6250 is a native dynamics simulation. Expectedly, its *d11* and *d22* values are both significantly larger than the other runs. The state transition diagram on the right summarizes the possible transitions between the three different states a contact map entry can be in: 0 is the non-edge state, 1 is the non-short-cut edge state and 2 is the short-cut state. In an MD run, zero or more of these transitions can happen from one step to the next. Error-bars denote standard deviation about the mean.

In section 4.1, we observed that delSC$_{t+1}$ is a distinct subset of short-cut edges, i.e. their removal affects gSCN in a significantly different way than the removal of a random subset of short-cut edges (Fig. 26). In this section, we observe that (i) deleted short-cuts are more likely to be part of EDS path-pairs with positive stretch (Fig. 31) and thus *SCN function influences its formation*; and (ii) there is a non-random relationship between delSC$_{t+1}$ and addSC$_{t+1}$ (Figs. 32 & 33).

To see how SCN function influences its own formation, we partitioned EDS path-pairs into four types and examined the edges of EDS path-pairs by type. Type *00* denotes edge independent path-pairs with zero stretch. Type *01* denotes edge independent path-pairs with positive stretch. Type *10* denotes path-pairs with zero stretch that are not edge independent, and type *11* denotes path-pairs with positive stretch that are not edge independent. Single-edge paths were excluded from this analysis so that the results for type *10* are not artificially inflated. Irrespective of edge independence, EDS path-pairs with positive stretch (type *01* or *11*) were more likely to traverse at least one short-cut that will disappear in the next step (Fig. 31a). The same affinity is observed with added short-cuts by path-pair type (Fig. 31b). Hence, path-pair stretch may be a more significant factor than edge independence in determining whether a short-cut edge will be deleted.

An average of 73.41% (std. dev. = 5.26%) of deleted short-cuts are recoverable in the first 100 short-cut edges sorted by Euclidean distance in descending order with an average false positive rate (FPR) of 61.08 % (std. dev. = 5.27%) (Fig. 31c). The recovery rate (TPR) is much higher than expected by random chance which on average is 13.03% for 6250, and for the non-native runs 6251…6258, ranges from 19.15% to 32.17%. The FPR is high by conventional standards, but is a consequence of the number of edges required to create a spanning tree with 101 nodes (although some gSCNs may not span the entire PRN, Fig. 24b), and $|SC| \approx 2N$. It remains to be seen whether an even larger FPR is necessary given the transitivity of SCNs.

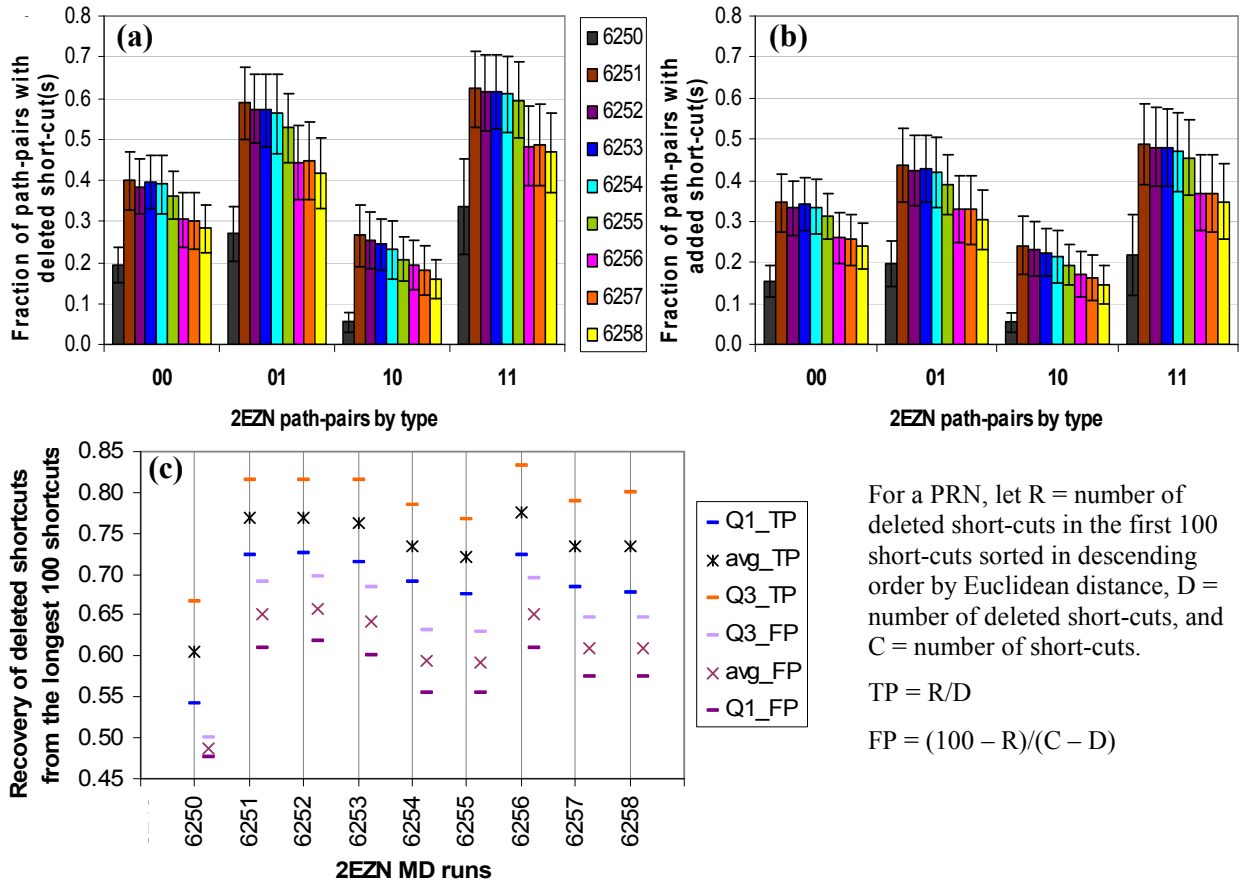

For a PRN, let R = number of deleted short-cuts in the first 100 short-cuts sorted in descending order by Euclidean distance, D = number of deleted short-cuts, and C = number of short-cuts.

TP = R/D

FP = (100 − R)/(C − D)

**Fig. 31 (a)** A significantly larger fraction of EDS path-pairs with positive stretch (*01* and *11*) traverse at least one deleted short-cut. **(b)** A significantly larger fraction of EDS path-pairs with positive stretch traverse at least one added short-cut. In (a) & (b), the fraction of path-pairs by type with at least one deleted (added) short-cut was calculated for each snapshot (PRN) in a run, and averaged over all snapshots for a run. The columns report the average and the error bars report the standard deviation. **(c)** The average (avg) true positive rate (TP) and false positive rate (FP) of recovering deleted short-cuts from a PRN's short-cut set. Q1 and Q3 are the first and third quartiles respectively.

To detect the existence of a relationship between $delSC_{t+1}$ and $addSC_{t+1}$, we assume that SCN connectivity is important, i.e. when short-cut edges disappear at time $t$, they are replaced with other short-cut edges that appear at time $t$ and this edge replacement process, together with strong SCN transitivity, helps to grow the largest SCN component (gSCN). If a relationship exists between a pair of deleted and added short-cut sets, then most deleted short-cuts could be replaced by at least one added short-cut, and most added short-cuts could replace at least one deleted short-cut. An added short-cut could replace a deleted short-cut if the added short-cut is found in the edge cut-set of the deleted short-cut, with respect to a spanning tree. The idea of replacing a deleted edge with another edge from its cut-set to maintain connectivity between all pairs of nodes in a graph is borrowed from dynamic graph theory [e.g.: 60], and we implement it as follows:

1. Identify $delSC_{t+1}$, the set of deleted short-cuts at step $t+1$. An edge in $delSC_{t+1}$ is a short-cut edge at step $t$ and either a non-short-cut edge or a non-edge at step $t+1$. Per the state transition diagram in Fig. 30, deleted short-cuts are those that make either a *d21* or *d20* transition.

2. Create a spanning tree $ST_t$ from $gSCN_t$ (the largest SCN component at step $t$), taking care to pack into $ST_t$ as many of the edges in $delSC_{t+1}$ as possible. We did this by first creating a forest of spanning trees from $delSC_{t+1} \cap gSCN_t$ (the deleted short-cuts in the largest SCN component at step $t$), and then use the other short-cut edges in $gSCN_t$ to grow and join the trees to complete the construction of $ST_t$. Further details in Appendix G.

3. Generate the cut-set for each edge $e$ in $ST_t \cap delSC_{t+1}$. The removal of $e$ from $ST$ splits $ST$ into two sub-trees. All edges with one endpoint in one sub-tree and another endpoint in the other sub-tree complete the cut-set for $e$. $CUTS_t$ is the union of cut-sets for edges in $ST_t \cap delSC_{t+1}$.

4. Identify $addSC_{t+1}$, the set of added short-cuts at step $t+1$. An edge in $addSC_{t+1}$ is a non-short-cut edge at step $t$ and a short-cut edge at step $t+1$. Per the state transition diagram in Fig. 30, added short-cuts are edges that make a *d12* transition. It is also possible for a non-edge at step $t$ to become a short-cut at step $t+1$ (via a *d02* transition) but such an added short-cut edge will not be in any cut-set generated in step 3. Non-edges that become short-cuts in the next step (*d02* edges) are much fewer than *d12* edges (Fig. G1b in Appendix G). Nonetheless, *d02* edges may be pivotal for correct SCN formation and this is a limitation of our current approach.

5. Define $gSCN'_t$ as $gSCN_t$ augmented by all edges in $PRN_t$ that have both endpoints in $gSCN_t$. For each edge in $gSCN'_t \cap addSC_{t+1}$, find a cut-set from step 3 that includes it. If a cut-set is found, then we say that the added short-cut edge is *matched* and the deleted short-cut edge that generated the matching cut-set is *used*.

The implementation here assumes that most of the short-cut changes take place within the largest SCN component, which they do (Fig. G1 in Appendix G). The fraction of unused deleted short-cuts (those without any added short-cuts in their cut-sets) and unmatched added short-cuts (those that do not appear in any cut-set of a deleted short-cut) were calculated for each snapshot (PRN) and the average over all snapshots of a MD run are reported in Figs. 32 & 33 for the two MD datasets. The quality of $CUTS_t$ depends on the edges that make up $ST_t$. Better results (lower unused and unmatched rates) are expected with larger $CUTS_t$. For this reason, we repeated the test on five spanning trees generated with a different random seed each time. The results from these different spanning trees (superimposed in Figs. 32 & 33) are not significantly different from each other.

For the 2EZN MD dataset, an average of 15.45% (standard deviation = 2.77%) of deleted short-cuts were unused (*opt_st* in Fig. 32a), and an average of 11.41% (standard deviation = 4.13%) of added short-cuts were unmatched (*opt_st* in Fig. 32b). The failure rates are much higher for native state PRNs (6250) than non-native state PRNs. This may be because our assumption that edge replacement is important for gSCN formation does not apply as well to already formed SCNs. For the Villin MD dataset, an average of 19.41% (standard deviation = 1.76%) of deleted short-cuts were unused (*opt_st* in Fig. 33a), and an average of 16.92% (standard deviation = 3.22%) of added short-cuts were unmatched (*opt_st* in Fig. 33b). Interestingly, the "successful" runs had a significantly (p-value = 0.0398) lower average unused rate, but a significantly (p-value = 0.0339) higher average unmatched rate than the "unsuccessful" runs (Table 5).

When any edge in gSCN$'_t$ (including non short-cut edges) can be used to grow and join the spanning trees in step 2, failure rates are significantly higher for both MD datasets. For the 2EZN MD dataset, an average of 16.55% (standard deviation = 3.89%) of deleted short-cuts were unused (*ropt_st* in Fig. 32a), and an average of 18.84% (standard deviation = 4.77%) of added short-cuts were unmatched (*ropt_st* in Fig. 32b). For the Villin MD dataset, an average of 21.83% (standard deviation = 1.72%) of deleted short-cuts were unused (*ropt_st* in Fig. 33a), and an average of 24.79% (standard deviation = 2.24%) of added short-cuts were unmatched (*ropt_st* in Fig. 33b). For both MD datasets, *ropt_st* produced significantly smaller $CUTS_t$ than *opt_st* (Figs. 32c & 33c). The differences between "successful" and "unsuccessful" runs in terms of unused and unmatched rates reported with *opt_st* were not significant with *ropt_st*.

The results produced by *opt_st* in this analysis, which is significantly different from the outcomes produced by its randomized counterpart *ropt_st*, support the notions that (i) short-cuts are a set of distinctly placed edges within PRNs (section 3.6. gave an inkling of this), and (ii) the deleted and the added short-cut sets of a step are not random, and that a relationship exists between the two sets. One of the tasks ahead is to decipher the conditions under which short-cuts get deleted and how their replacements are selected.

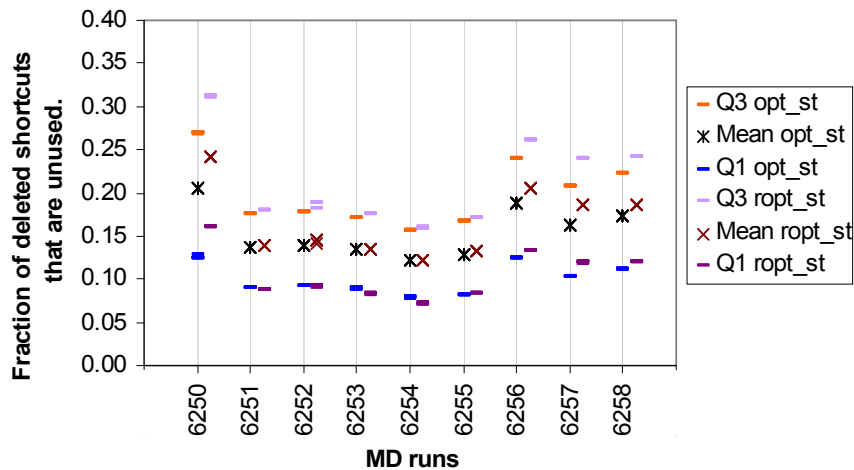

**(a)** A deleted short-cut with an unused cut-set means that a replacement edge was not found from the set of added short-cuts. Averaged over all runs, about 15.45% (std. dev. 2.77%) of deleted short-cuts have unused cut-sets (*opt_st*). This percentage increased significantly to 16.55% (std. dev. 3.89%) when *ST* is supplemented with random edges in step 2 (*ropt_st*).

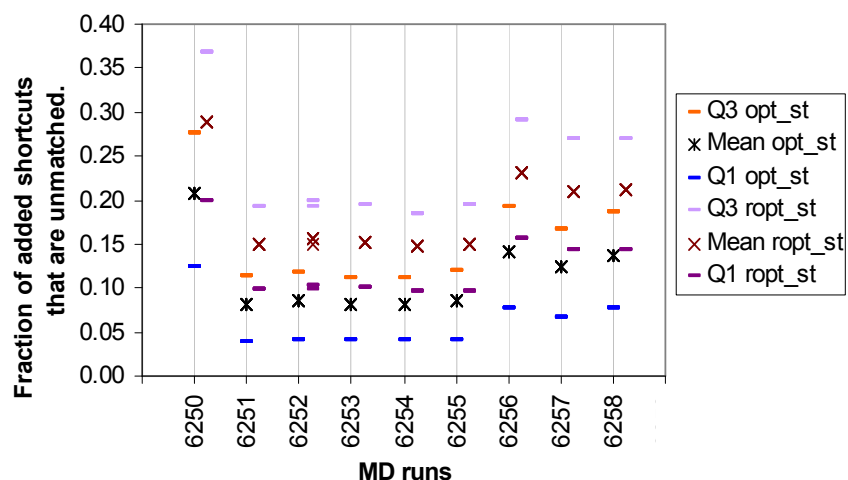

**(b)** An unmatched added short-cut is one whose existence is not explained by it being a replacement for any deleted short-cut. Averaged over all runs, about 11.41% (std. dev. 4.13%) of added short-cuts are unmatched (*opt_st*). This percentage increased significantly to 18.84% (std. dev. 4.77%) when *ST* is supplemented with random edges in step 2 (*ropt_st*).

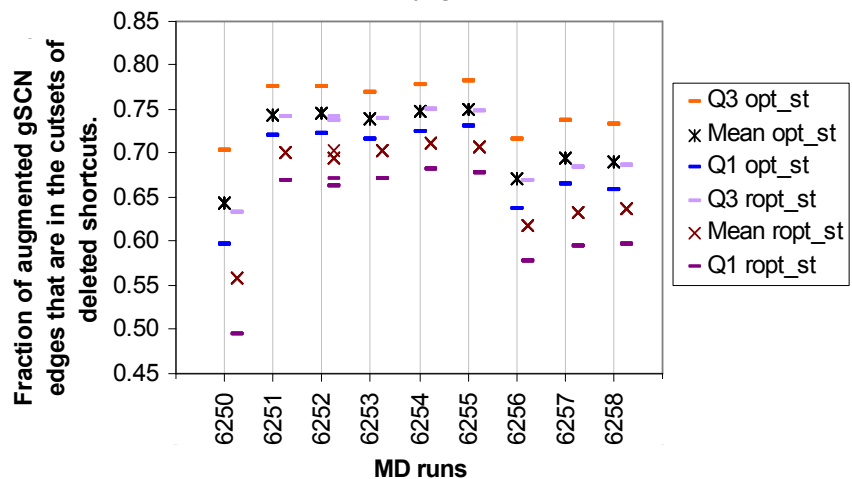

**(c)** The fraction of edges in gSCN′$_t$ that are in CUTS$_t$ is significantly smaller when *ropt_st* is used to produce the edge cut-sets than when *opt_st* is used. CUTS$_t$ is the union of cut-sets for edges in *ST*$_t$ ∩ delSC$_{t+1}$. Accordingly, *ropt_st* has poorer outcomes in (a) & (b) above. That the *opt_st* and *ropt_st* outcomes are significantly different supports the notion that the short-cuts are a distinct set of PRN edges.

**Fig. 32 Results for 2EZN showing the relationship between deleted and added short-cut sets.**

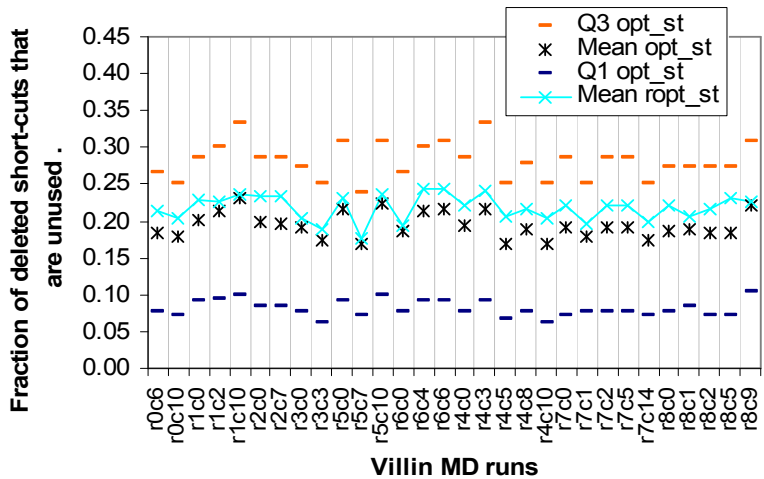

**(a)** Averaged over all clones, about 19.41% (std. dev. 1.76%) of deleted short-cuts have unused cut-sets (*opt_st*). This percentage increased significantly to 21.83% (std. dev. 1.72%) when *ST* is supplemented with random edges in step 2 (*ropt_st*).

Successful runs have a significantly lower average unused rate than unsuccessful runs (Table 5). This difference is not significant with *ropt_st*.

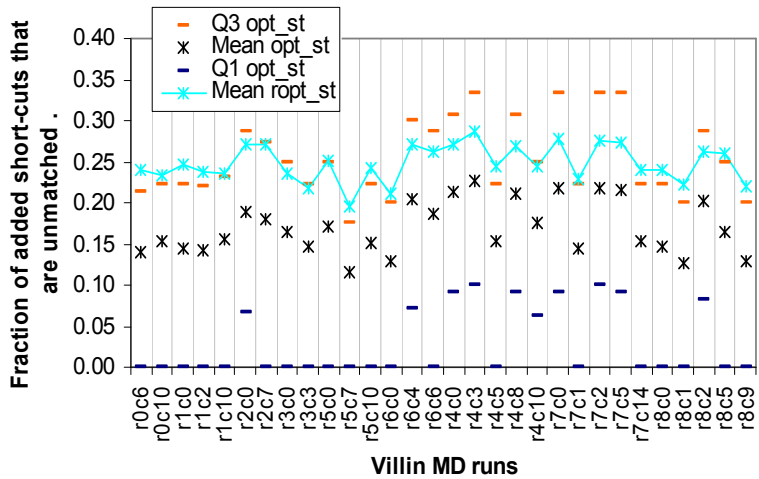

**(b)** Averaged over all clones, about 16.92% (std. dev. 3.22%) of added short-cuts are unmatched (*opt_st*). This percentage increased significantly to 24.79% (std. dev. 2.24%) when *ST* is supplemented with random edges in step 2 (*ropt_st*).

Successful runs have a significantly higher average unmatched rate than unsuccessful runs (Table 5). This difference is not significant with *ropt_st*.

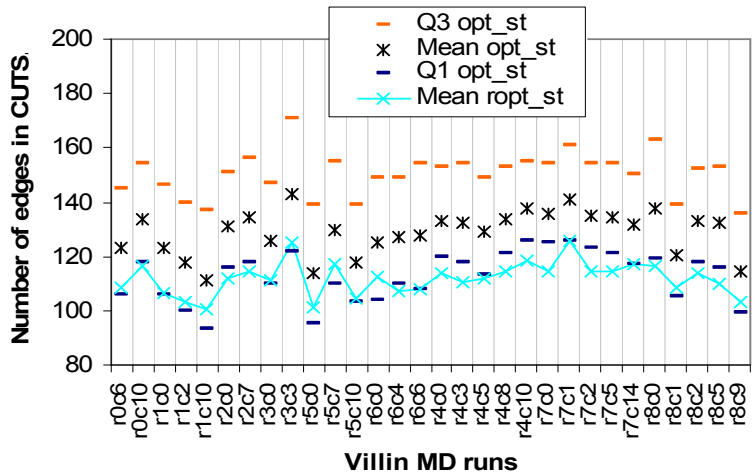

**(c)** $CUTS_t$ is the union of cut-sets for edges in $ST_t \cap delSC_{t+1}$. The number of edges in $gSCN'_t$ that are covered by the cut-sets of deleted short-cuts is significantly smaller when random edges are used in step 2 (*ropt_st*) than when only short-cuts are used (*opt_st*). Accordingly, *ropt_st* produced poorer outcomes in (a) & (b) above.

**Fig. 33 Results for HP-35 NLE NLE showing the relationship between deleted and added short-cut sets.**

## 4.4 Discussion: structure, function and formation

MGEO4 networks have significantly weaker SCN transitivity than PRNs (Fig. 25a). How do MGEO EDS path-pairs fare in terms of the metrics in section 4.2? Compared with PRN EDS path-pairs, MGEO4 EDS path-pairs are significantly less edge independent (Fig. 34a), and suffer significantly larger stretch on average (Fig. 34b).

The artificial MGEO4 example illustrates how the structure of an SCN influences its ability to fulfill its function, namely strong SCN transitivity is conducive towards path diversity with low stretch. This SCN structure-function relationship has been hinted at with the 2EZN MD dataset. As the 2EZN protein folds under MD simulation, SCN transitivity increases (Fig. 25b), EDS path-pair edge independence increases (Fig. 27a), and average EDS path-pair stretch decreases (Fig. 27c). Similar correlations are observed with the second MD dataset. There is a strong positive correlation (r = 0.8865) between SCN transitivity and EDS path-pair edge independence (Fig. 35a), and a strong negative correlation (r = –0.8751) between SCN transitivity and average EDS path-pair stretch (Fig. 35b). Moreover, PRNs in the "successful" Villin MD runs have SCNs that are significantly (p-value=0.0101) more strongly transitive on average than PRNs in the "unsuccessful" Villin MD runs (Fig. 35c). Accordingly, EDS path-pairs in the "successful" runs are significantly (p-value=0.0070) more edge independent and suffer significantly (p-value=0.0073) smaller average stretch than EDS path-pairs in the "unsuccessful" MD runs.

There are two clones (r8c1 and r8c9) classified as "successful" according to the criteria in [62] which have notably weaker SCN transitivity (Fig. 35c). However, in terms of the fraction of native SCN contacts and the number of residues with alpha secondary structure according to DSSP [64], these two clones are laggards compared with the other "successful" clones (Fig. 36a). Native SCN contacts are the SCN edges of the 2F4K PRN0 (PRN generated from the PDB structure). Let the native SCN be $SCN_0$. The fraction of native SCN contacts in a $SCN_t$ is the number of edges $SCN_t$ has in common with $SCN_0$ divided by the number of edges in $SCN_t$. The alpha structure residues for a PRN were obtained by running DSSP [64] on the PRN's frame (coordinates file), and counting the number of residues marked 'H', 'I' or 'G'. Despite having these two "outlier" clones, the "successful" HP-35 NLE NLE MD trajectories register on average a significantly larger fraction of native SCN contacts and significantly more alpha structure residues than the set of "unsuccessful" clones. The significant differences, in terms of SCN structure, function and formation, between "successful" and "unsuccessful" Villin MD runs reported in this paper, are summarized in Table 5.

Native contacts play a dominant role in protein folding and the fraction of native contacts was found to be a suitable folding coordinate even for all-atom MD simulations [63]. The fraction of native SCN contacts increases steadily as the 2EZN protein folds under MD simulation (Fig. 36c). SCN transitivity correlates positively and strongly with both fraction of SCN native contacts (Fig. 36b) and number of alpha residues (maturity of secondary structures) (Fig. 36d). This is expected since short-cuts are dominated by short-range links, and short-range links are found mainly within secondary structures (Fig.

19). These strong correlations, coupled with the significant differences in Table 5, suggest that SCN transitivity could be an easy to compute reaction coordinate for protein folding without requiring *a priori* knowledge of the native configuration.

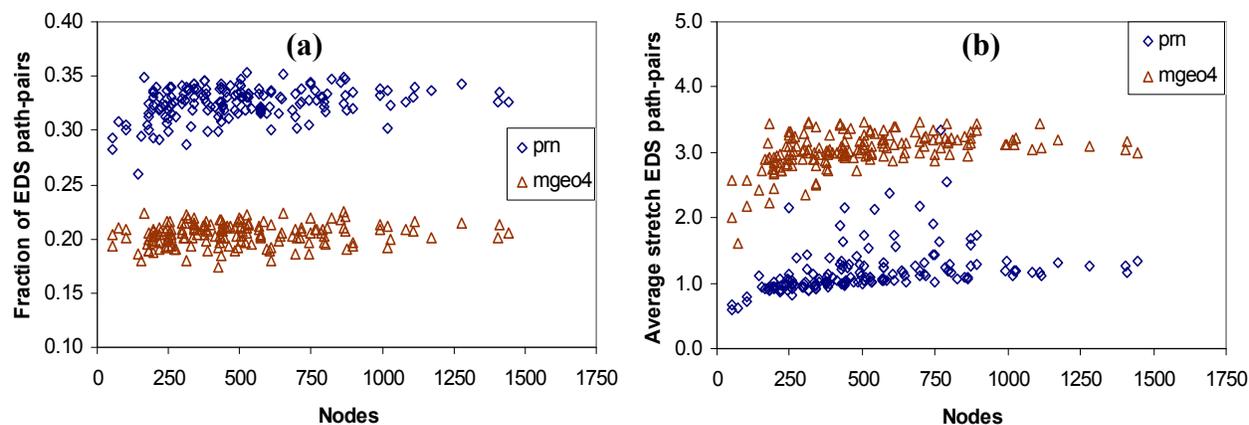

**Fig. 34 MGEO4 path diversity and stretch.** (a) A significantly smaller fraction of MGEO4 EDS path-pairs are edge independent. (b) MGEO4 EDS path-pairs have significantly larger stretch on average than PRN EDS path-pairs.

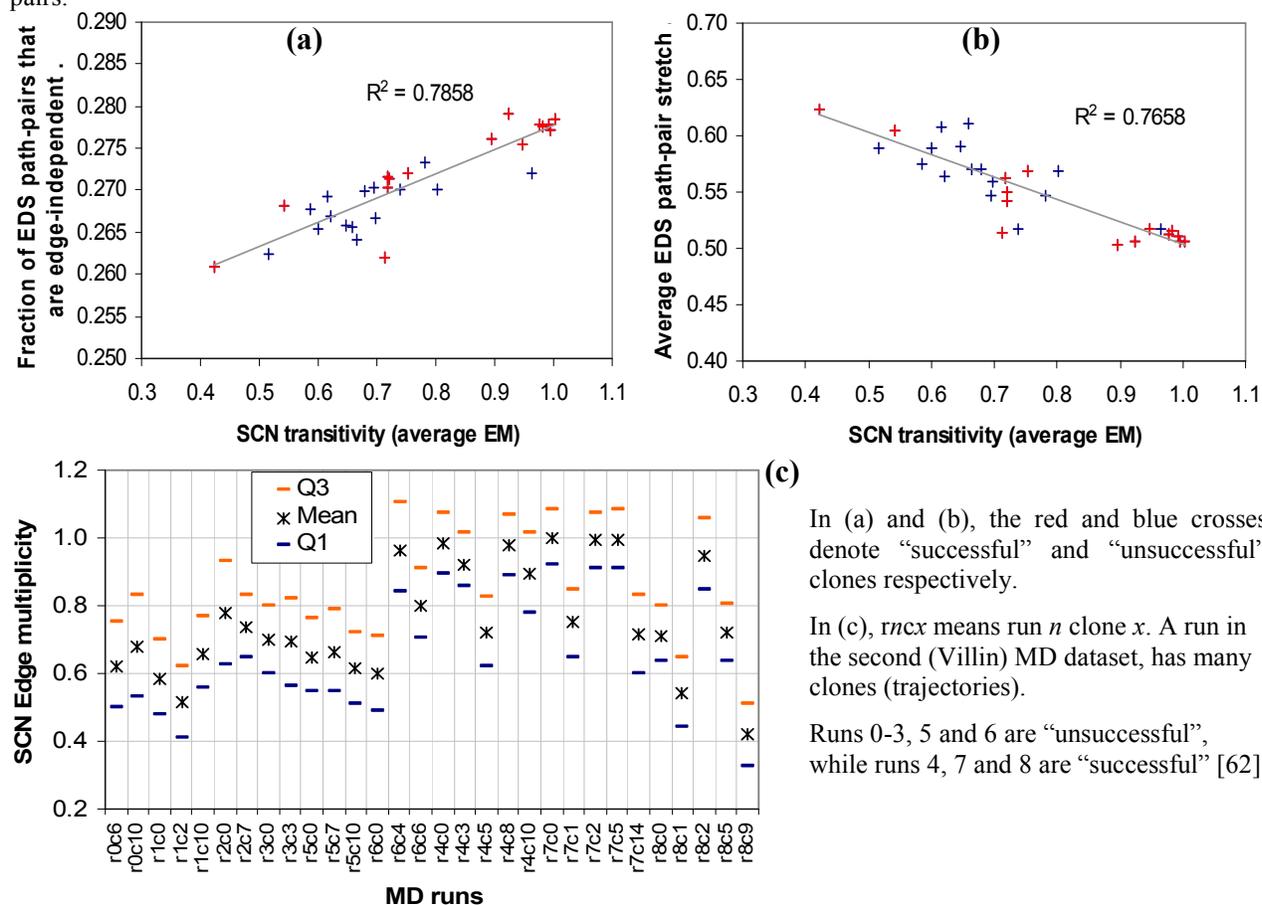

In (a) and (b), the red and blue crosses denote "successful" and "unsuccessful" clones respectively.

In (c), r*n*c*x* means run *n* clone *x*. A run in the second (Villin) MD dataset, has many clones (trajectories).

Runs 0-3, 5 and 6 are "unsuccessful", while runs 4, 7 and 8 are "successful" [62].

**Fig. 35 SCN structure and function for configurations (PRNs) from the HP-35 NLE NLE MD dataset.** (a) SCN transitivity correlates positively with path diversity. (b) SCN transitivity correlates negatively with EDS path-pair stretch. (c) SCNs from "successful" runs have significantly stronger transitivity than SCNs from "unsuccessful runs.

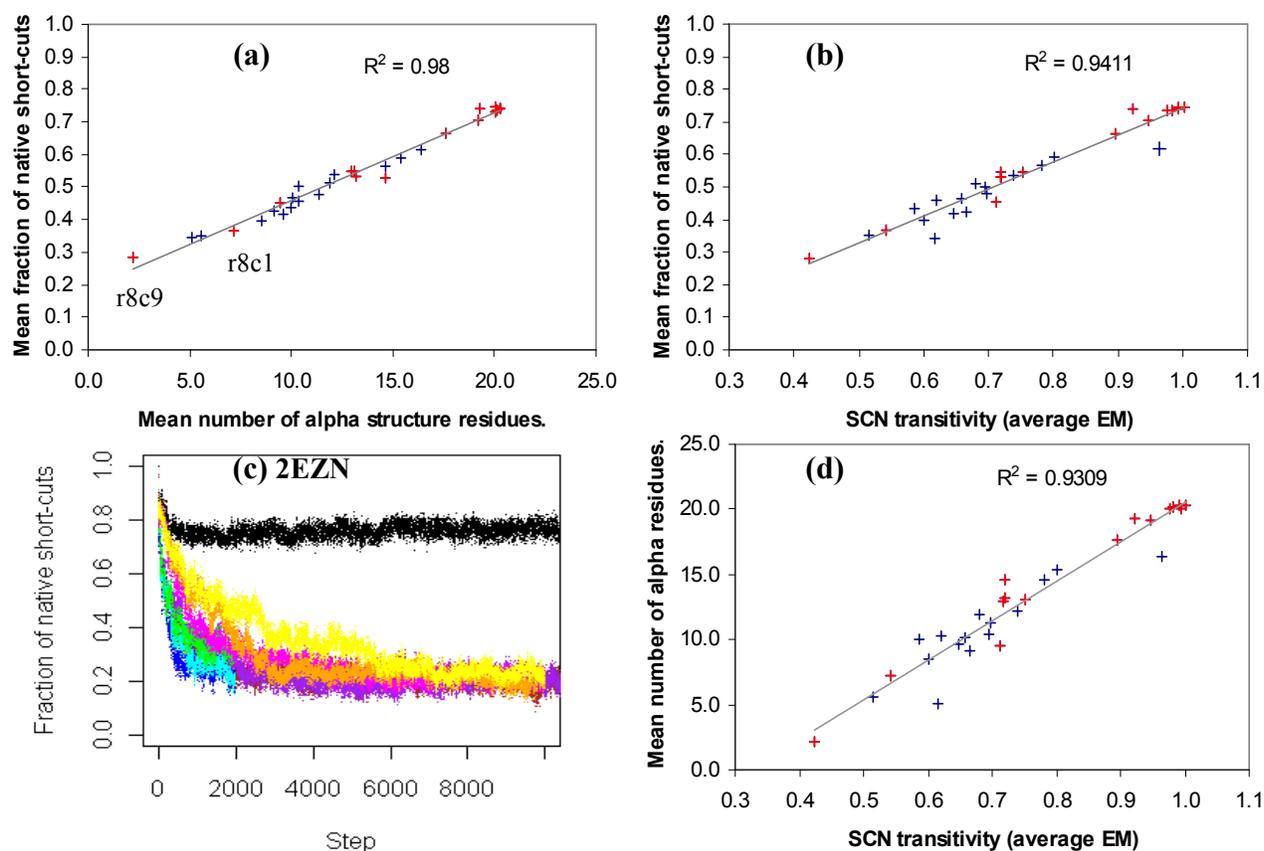

**Fig. 36 (a, b, d) Relationship between fraction of native short-cuts, alpha structure residues and SCN transitivity observed with the Villin MD dataset.** Native short-cuts are the short-cut edges in SCN0, the SCN of the 2F4K PDB structure. Mean fraction of native short-cuts for a clone (trajectory) is the fraction of edges in an SCN that are native short-cuts averaged over all SCNs (steps) in the trajectory. 2F4K comprises three helical structures. This is why we count the number of residues identified with alpha structure in the PRNs. **(a)** The mean fraction of native short-cuts is strongly and positively correlated with the mean number of alpha structure residues. Except for r8c9 and r8c1, the unsuccessful trajectories tend to cluster in the lower left quadrant of the plot. **(b)** The mean fraction of native short-cuts is strongly and positively correlated with SCN transitivity. **(c)** The fraction of native short-cuts in each SCN increases steadily as the 2EZN protein folds, and holds steady for native dynamics. **(d)** SCN transitivity is strongly and positively correlated with mean number of alpha residues.

**Table 5 Significant differences between "successful" and "unsuccessful" Villin MD runs.** SCNs of "successful" runs are better formed in the sense that they have significantly more edges, larger gSCNs and stronger transitivity. SCNs of "successful" runs have significantly more native short-cuts and their PRNs have significantly more alpha structure residues. EDS path-pairs in PRNs of "successful" runs are significantly more edge-independent and have significantly smaller stretch. *one-sided, unpaired.

| Metric | "successful" runs avg ± std. dev. | "unsuccessful" runs avg ± std. dev. | t-test* p-value |
|---|---|---|---|
| Number of short-cuts | $53.36 \pm 2.15$ | $51.41 \pm 1.89$ | 0.0067 |
| Number of nodes in the largest SCN component | $33.80 \pm 0.50$ | $33.15 \pm 0.67$ | 0.0029 |
| SCN transitivity (average edge multiplicity) | $0.82 \pm 0.18$ | $0.68 \pm 0.11$ | 0.0101 |
| Fraction of EDS path-pairs that are edge-independent | $0.2727 \pm 0.0059$ | $0.2680 \pm 0.0031$ | 0.0070 |
| EDS path-pair stretch | $0.5377 \pm 0.0394$ | $0.5682 \pm 0.0279$ | 0.0073 |
| Fraction of native short-cuts | $0.60 \pm 0.15$ | $0.47 \pm 0.08$ | 0.0037 |
| Number of alpha structure residues (DSSP) | $15.30 \pm 5.62$ | $10.70 \pm 3.17$ | 0.0056 |
| Unused deleted short-cuts (*opt_st* in Fig. 33a) | $0.1885 \pm 0.0147$ | $0.1998 \pm 0.0189$ | 0.0398 |
| Unmatched added short-cuts (*opt_st* in Fig. 33b) | $0.1799 \pm 0.0363$ | $0.1584 \pm 0.0241$ | 0.0339 |

# 5. Summary

Proteins have been abstracted as a network of interacting amino acids and much attention has been paid to the small-world property of such protein residue networks (PRNs). Hitherto, a global search strategy such as Breath-First Search (BFS) is commonly used to measure the average path length of PRNs. We propose that a local search strategy is more appropriate because the inverse relationship between clustering and average path length in a local search better fits the notion that amino acids get closer to each other as a protein becomes more compact. This inverse relationship is observed as the 2EZN protein (un)folds in a molecular dynamics (MD) simulation.

To study local search on PRNs, we devised a greedy local search algorithm called EDS, which is Euclidean distance based and allows backtracking, and compared the characteristics of BFS paths with EDS paths. While the paths are different in terms of variation in path length, search cost, link usage, path diversity, path stability and communication propensity; they exhibit similarities in terms of hierarchy and centrality. We argue that the differences are preferable as they suggest EDS paths make a better model of intra-protein communication. The similarities are also preferable as they imply the transferability of existing methods based on BFS centrality.

EDS identifies a set of edges for each PRN that helps the local search algorithm to avoid backtracking and hence keep EDS paths short. These short-cut edges are readily available in PRNs thanks to their high levels of clustering and strong transitivity. The number of short-cuts in a PRN scales linearly with protein size (number of amino acids). Short-cut edges are dominated by short-range contacts, see higher usage (are more central) and have stronger local clustering but weaker local community structure.

The short-cut edges of a PRN form a short-cut network (SCN) that, in native state PRNs, spans most of a PRN's nodes, forms a giant component, and is strongly transitive. SCNs grow in size and become more strongly transitive as the 2EZN protein folds. SCNs influence both edge-independence and stretch of EDS path-pairs significantly. As a consequence, short-cut edge sets are more volatile, i.e. they undergo significantly more additions and deletions from step to step in a MD simulation, than non-short-cut sets. Despite their volatility, deleted short-cuts are not random and appear in a larger fraction on EDS path-pairs with positive stretch, it is possible to identify deleted short-cuts from a short-cut set, and a relationship exists between the pair of deleted and added short-cut sets of a MD step. The majority of added short-cuts are found in the edge cut-set of at least one deleted short-cut, and the majority of edge cut-sets of deleted short-cuts contain at least one added short-cut. This high edge replacement rate helps to maintain connectivity, and coupled with strong SCN transitivity, fosters the growth of the largest connected component of a SCN.

We propose that well-formed protein configurations have well-formed SCNs. A structure is well-formed if it is well-suited to its function. A comparison between "successful" and "unsuccessful" MD trajectories of the mutated Villin headpiece protein supports this hypothesis. Compared with SCNs from "unsuccessful" runs, SCNs from "successful" runs are significantly larger and more strongly transitive on average, and are therefore better formed. Accordingly, EDS path-pairs from "successful" runs are more diverse and have smaller stretch on average. SCN transitivity also correlates strongly and positively with increase in native SCN contacts and with the formation of secondary structures.

We have thus far examined *atemporal* network statistics (measurements that do not involve time as a component) of sequences of static PRNs. A logical extension of this work is to utilize the formalisms and methods emerging from the currently active field of time varying graphs and dynamic networks e.g. [67], and examine temporal aspects of PRNs as a protein folds.

## Acknowledgements

This work was made possible by the facilities of the Shared Hierarchical Academic Research Computing Network (SHARCNET:www.sharcnet.ca) and Compute/Calcul Canada. Thanks to members of MUN for various discussions, and to the Dynameomics group for providing data access.

## References

1  Watts DJ and Strogatz SH (1998) Collective dynamics of 'small-world' networks. Nature 393, 440-442.
2  Vendruscolo M, Dokholyan NV, Paci E and Karplus M (2002) Small-world view of the amino acids that play a key role in protein folding. Physical Review E 65 061910-1.
3  Bagler G and Sinha S (2005). Network properties of protein structures. Physica A: Statistics Mechanics and its Apps. 346(1-2) pp. 27—33.
4  Emerson IA and Gothandam KM (2012). Network analysis of transmembrane protein structures. Physica A 391:905-916.
5  Kleinberg J (2000) Navigation in a small world. *Nature* 406, 845
6  Kleinberg J (2000) The small-world phenomenon: an algorithmic perspective. Proc. Of the 32nd Annual ACM Symposium on Theory of Computing, pp. 163-170.
7  Kasturirangan R (1999) Multiple scales in small-world graphs. cond-mat/9904055.
8  Newman MEJ (2003) The structure and function of complex networks. SIAM Review 45:167-256.
9  Newman MEJ (2010) Networks: An introduction. Oxford University Press.
10  Cohen R and Havlin S (2003) Scale-free networks are ultrasmall. Phys. Rev. Lett 90, 058701.
11  Whitley MJ and Lee AL (2009) Frameworks for understanding long-range intra-protein communication. Curr Protein Pept Sci. April, 10(2):116-127.
12  Jackson MB (2006). Molecular and cellular biophysics. Cambridge University Press.
13  Leitner DM (2008). Energy flow in proteins. Annu. Rev. Phys. Chem. 59:233-259.
14  Dokholyan NV, Li L, Ding F and Shakhnovich EI (2002) Topological determinants of protein folding. PNAS (13): 8637-8641.
15  Atilgan AR, Akan P and Baysal C (2004) Small-world communication of residues and significance for protein dynamics. Biophysical Journal 86:85-91.
16  Del Sol A, Fujihashi H, Amoros D and Nussinov R (2006). Residues crucial for maintaining short paths in network communication mediate signaling in proteins. Mol. Sys. Biol. Doi:10.1038/msb4100063

17 Atilgan AR, Turgut D and Atilgan C (2007) Screened non-bonded interactions in native proteins manipulate optimal paths for robust residue communication. Biophys. J. 92(9):3052-3062.

18 Chatterjee S, Ghosh S and Vishveshwara S (2013). Network properties of decoys and CASP predicted models: a comparison with native protein structures. Mol. BioSyst. 9:1774-1788.

19 Bhattacharyya M, Bhat, CR and Vishveshwara S (2013). An automated approach to network features of protein structure ensembles. Protein Science 22:1399-1416.

20 Maiya AS and Berger-Wolf TY (2010) Expansion and search in networks. ACM Int. Conf. on Info. and Knowledge Mgmt. (CIKM). Toronto, Canada.

21 Park K and Kim D (2011) Modeling allosteric signal propagation using protein structure networks. BMC Bioinformatics 12. From The 9th Asia Pacific Bioinformatics Conference (APBC 2011) Inchon, Korea.

22 Li G, Magana D and Dyer RB (2014) Anisotropic energy flow and allosteric ligand binding in albumin. Nature Communications 5:3100.

23 Suel GM, Lockless SW, Wall MA and Ranganathan R (2003). Evolutionarily conserved networks of residues mediate allosteric communication in proteins. Nature Structural Biology 10(1):59-69.

24 Berman HM, Westbrook J, Feng Z, Gilliland G, Bhat TN, Weissig H, Shindyalov IN and Bourne PE (2000). The Protein Data Bank. Nucleic Acids Research 28: 235-242. http://www.rcsb.org/pdb

25 Chatterjee S, Ghosh S and Vishveshwara S (2013). Network properties of decoys and CASP predicted models: a comparison with native protein structures. Mol. BioSyst. 9:1774-1788.

26 Kannan N and Vishveshwara S (1999). Identification of side-chain clusters in protein structures by a graph spectral method. J. Mol. Biol. 292:441-464.

27 Mosca R, Ceol A, Stein A, Olivella R and Aloy P (2013). 3did: a catalog of domain-based interactions of known three-dimensional structure. Nucleic Acids Research 1-6.

28 Greene LH and Higman VA (2003). Uncovering network systems within protein structures. Journal of Molecular Biology 334:781-791.

29 Milenkovic T, Filippis I, Lappe M and Przulj N (2009) Optimized null model for protein structure networks. PLoS ONE 4(6): e5967.

30 Gaci O and Balev S (2009) Node degree distribution in amino acid interaction networks. Computational Structural Bioinformatics Workshop, Washington DC, USA.

31 Serrano MA and Boguna M (2006) Clustering in complex networks. I. General formalism. Phys. Rev. E 74, 056114

32 Colomer-de-Simon P, Serrano MA, Beiro MG, Alvarez-Hamelin I and Boguna M. (2013) Deciphering the global organization of clustering in real complex networks. Scientific Reports 3:2517.

33 Serrano MA and Boguna M (2006) Clustering in complex networks. II. Percolation properties. Phys. Rev. E 74, 056115.

34 Atilgan AR and Atilgan C (2012) Local motifs in proteins combine to generate global functional moves. Briefings in Functional Genomics 2(6):479-488.

35 Bagler G and Sinha S (2007). Assortative mixing in Protein Contact Networks and protein folding kinetics. Structural Bioinformatics 23(14) pp. 1760—1767.

36 Turgut D, Atilgan AR and Atilgan C (2010) Assortative mixing in close-packed spatial networks. PLoS ONE 5(12):e15551

37 Bartoli L, Fariselli P and Casadio, R (2007) The effect of backbone on the small-world properties of Protein Contact Maps. Physical Biology 4:L1-L5.

38 Estrada E. (2010) Universality in protein residue networks. Biophysical Journal 98:890-900.

39 Beck DAC, Jonsson AL, Schaeffer RD, Scott KA, Day R, Toofanny RD, Alonso DOV and Daggett V (2008) Dynameomics: Mass Annotation of Protein Dynamics by All-Atom Molecular Dynamics Simulations. Protein Engineering Design & Selection 21: 353-368.

40 Jonsson AL, Scott KA and Daggett V (2009) Dynameomics: A consensus view of the protein unfolding/folding transition state ensemble across a diverse set of protein folds. Biophysical Journal 97:2958-2966.

41   Van der Kamp MW, Schaeffer RD, Jonsson AL, Scouras AD, Simms AM, Toofanny RD, Benson NC, Anderson PC, Merkley ED, Rysavy S, Bromley D, Beck DAC and Daggett V (2010) Dynameomics: A comprehensive database of protein dynamics. Structure, 18: 423-435.

42   Gao L (2001). On inferring autonomous system relationships in the Internet. IEEE/ACM Trans. On Networking 9: 733-745.

43   Boguna M, Krioukov D and Claffy KC (2008). Navigability of complex networks. Nature Physics 5:74-80.

44   Amitai G, Shemesh A, Sitbon E, Shklar M, Netanely D, Venger I and Pietrokovski S. (2004) Network analysis of protein structures identifies functional residues. *Journal of Molecular Biology* 344 1135-1146.

45   Li J, Wang J and Wang W. (2008) Identifying folding nucleus based on residue contact networks of proteins. Proteins 71: 1899 – 1907.

46   Del Sol A, Fujihashi H, Amoros D and Nussinov R (2006) Residue centrality, functionally important residues and active site shape: Analysis of enzyme and non-enzyme families. Protein Science 15:2120-2128. Cold Spring Harbor Laboratory Press.

47   Plaxco KW, Simons KT and Baker D (1998) Contact order, transition state placement and the refolding rates of single domain proteins. J. Mol. Biol. 277:985-994.

48   Gromiha MM and Selvaraj S (2001) Comparison between long-range interactions and contact order in determining the folding rate of two-state proteins: Application of long-range order to folding rate prediction. J. Mol. Biol. 310:27-32.

49   Ivankov DN, Garbuzynskiy SO, Alm E, Plaxco KW, Baker D and Finkelstein AV (2003) Contact order revisited: Influence of protein size on the folding rate

50   Baker D (2000) A surprising simplicity to protein folding. Nature 405, 39-42.

51   Go N. (1983) Theoretical Studies of Protein Folding. Ann. Rev. Biophys. Bioeng. 12:183-210.

52   Barthelemy M (2010) Spatial networks. arXiv:1010.0302.

53   Botan V, et. al. (2007) Energy transport in peptide helices. PNAS 104(31):12749-12754.

54   Young HT, Edwards SA and Grater F (2013) How fast does a signal propagate through proteins? PLoS ONE 8(6):e64746.

55   Belykh I, Hasler M, Lauret M and Nijmeijer H. (2005) Synchronization and graph topology. Int. Journal of Bifurcation and Chaos 15(100):3423-3433.

56   Cannistraci CV, Alanis-Lobato G and Ravasi T (2013) From link-prediction in brain connectomes and protein interactomes to the local-community-paradigm in complex networks. Scientific Reports 3:1613. doi: 10.1038/srep01613.

57   Goyal N, Rademacher L and Vempala S (2009) Expanders via random spanning trees. Proceedings of the 20[th] Annual ACM-SIAM Symposium on Discrete Algorithms (SODA), pp. 576-585.

58   Sreenivasan S, Cohen R and Lopez E. (2006) Communication bottlenecks in scale-free networks. arXiv:cs/0604023.

59   Danila B, Yu Y, Marsh JA and Bassler KE. (2006) Optimal transport on complex networks. *Physical Review E* 74 046106.

60   Kapron BM, King V and Mountjoy B. (2013) Dynamic graph connectivity in polylogarithmic worst case time. Proceedings of the 24[th] Annual ACM-SIAM Symposium on Discrete Algorithms (SODA), pp. 1131-1142.

61   Cormen TH, Leiserson CE and Rivest RL. (1998) Introduction to algorithms. The MIT Press.

62   Ensign DL, Kasson PM and Pande VS. (2007) Heterogeneity even at the speed limit of folding: large-scale molecular dynamics study of a fast-folding variant of the Villin Headpiece. J. Mol. Biol. 374:806-816.

63   Best RB, Hummer G and Eaton WA (2013) Native contacts determine protein folding mechanisms in atomistic simulations. PNAS 110(44):17874-17879.

64   Kabsch, W., and Sander, C. (1983) Dictionary of Protein Secondary Structure:  Pattern Recognition of Hydrogen-bonded and Geometrical Features. *Biopolymers* 22(12), pp. 2577-2637.

65    Laine E, Auclair C, Tchertanov L. (2012) Allosteric communication across the native and mutated KIT receptor Tyrosine Kinase. PLoS Comput. Biol 8(8):e1002661. oi:10.1371/journal.pcbi.1002661

66    Morra G, Verkhivher G and Colombo G. (2009) Modeling signal propagation mechanisms and ligand-based conformational dynamics of the Hsp90 molecular chaperone full-length dimmer. PLoS Comput. Biol. 5(3): e1000323. doi:10.1371/journal.pcbi.1000323

67    Casteigts A, Flocchini P, Quattrociocchi W and Santoro N (2012) Time-Varying Graphs and dynamic networks. International Journal of Parallel, Emergent and Distributed Systems 27(5): 387-408.

## Appendix A

Normalization values for the 20 residue types from [26], expanded with alternatives (marked with *) for the MD simulation datasets. Alternative A is used where applicable.

| Residue Type | Norm | | Residue Type | Norm | |
|---|---|---|---|---|---|
| ALA | 55.7551 | | LEU | 72.2517 | |
| ARG | 93.7891 | | NLEU | 72.2517 | * |
| ASN | 73.4097 | | NLE | 72.2517 | * |
| ASP | 75.1507 | | LYS | 69.6096 | |
| CYS | 54.9528 | | LYP | 69.6096 | * |
| CYH | 54.9528 | * | MET | 69.2569 | |
| GLN | 78.1301 | | PHE | 93.3082 | |
| GLU | 78.8288 | | CPHE | 93.3082 | * |
| GLY | 47.3129 | | CPH | 93.3082 | * |
| HIS | 83.7357 | | PRO | 51.3310 | |
| HIE | 83.7357 | * | SER | 61.3946 | |
| ILE | 67.9452 | | THR | 63.7075 | |
| | | | TRP | 106.7030 | |
| | | | TYR | 100.719 | |
| | | | VAL | 62.3673 | |

## Appendix B

*Random geometric graphs* (RGG) have many network statistics that are quantitatively similar to RINs [29]. Nodes of a RGG are situated at random 3D coordinates within a boundary and links are established between nodes that are within a certain Euclidean distance apart from each other. The *geo* networks discussed in this Appendix are a specialized version of RGGs in that they are constructed in the manner of MGEO networks but with their nodes situated at random 3D coordinates. Like the MGEO networks, the geo networks have a "backbone" and a number of potential links are skipped to tune for clustering as described in section 2.2. The clustering coefficient and average edge multiplicity of geo4 (skip=4) networks are much smaller than those of geo0 (skip=0) networks (Figs. B2a & B2b). The geo0 and geo4 contact maps for protein 1B19 are shown in Fig. B1 (compare with the PRN and MGEO4 contact maps in Fig. 4).

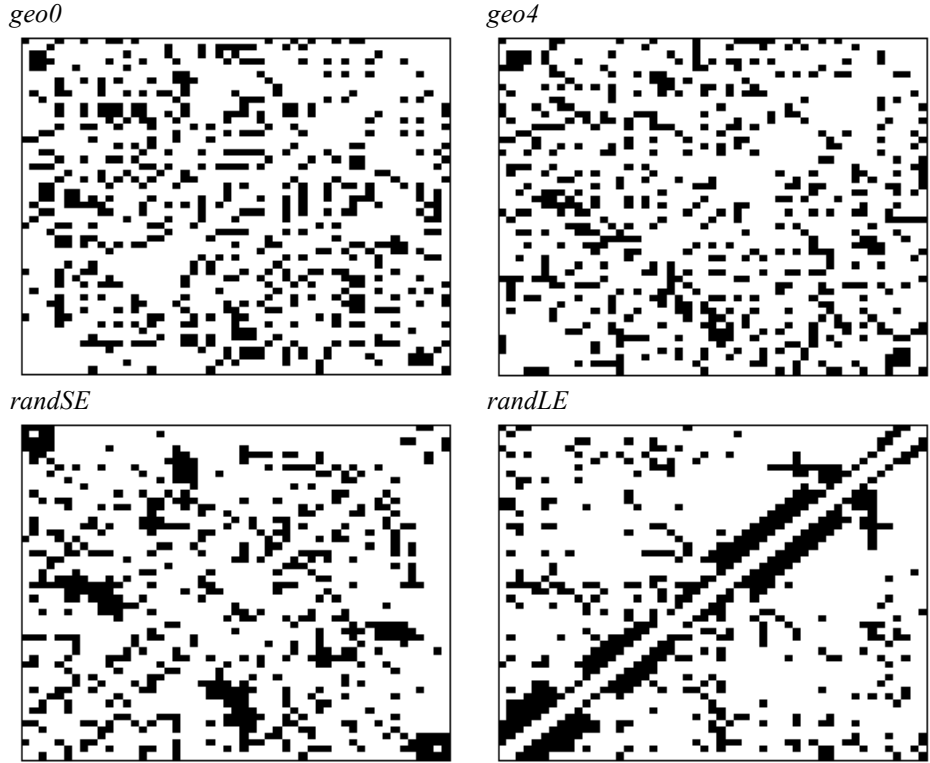

**Fig. B1**. Contact maps of geo, geo4, randSE and randLE networks for 1B19. A dark cell denotes an edge. A white cell denotes a non-edge.

Several distinct differences arise between the geo networks and PRNs. First, the $|LE|/|SE|$ ratios of geo networks are larger than that for PRNs, and increase linearly with $N$ (Fig. B2e). This has consequences for the distribution of link lengths as we discussed later. Second, the use of random 3D coordinates destroys any relationship between link Euclidean distance and link sequence distance. The average Spearman correlation between link Euclidean distance and link sequence distance for PRNs is 0.4075 (std. dev. 0.1163), and it is 0.3627 (std. dev. 0.0823) for MGEO4 networks. All correlations are significant. Navigability in a small-world network rests on the assumption that there are local (short-range) and global (long-range) links, and that local links connect more similar nodes while global links connect less similar ones. In social networks, similarity may be assessed in terms of geography, language, occupation, ethnicity, education level, interests and so forth. In many theoretical small-world models, locality is measured as distance, e.g. Manhattan distance in Kleinberg's grid model [5]. And the same distance norm is used to direct local searches in the network. In PRNs, we use sequence distance as a measure of locality, but we use Euclidean distance to direct EDS. Hence some positive correlation between sequence distance and Euclidean distance of links is necessary to study navigability of PRNs with EDS. By reason of this fundamental difference, we consider geo networks to be a different class of networks than PRNs for EDS.

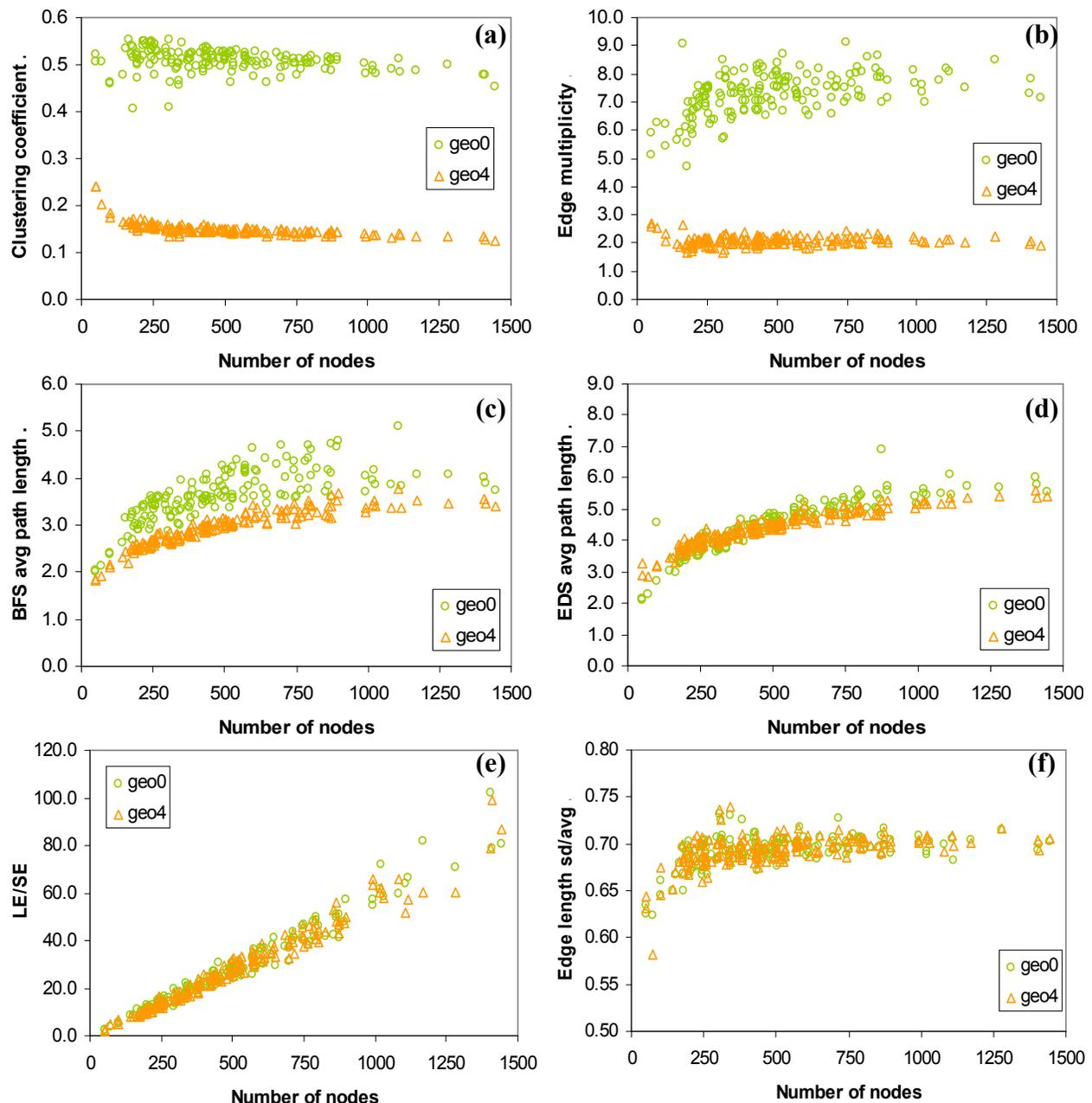

**Fig. B2** The geo0 (skip=0) and geo4 (skip=4) networks are constructed the same way as MGEO networks except with random x, y, z coordinates. (d) Average EDS path length is not significantly affected by change in clustering. (e) The geo networks have larger |LE|/|SE| ratios than PRNs, and the ratio increases with *N*. (f) Edge length distribution is not significantly affected by change in clustering.

Unlike the results reported for PRNs in section 3.1, decreased clustering did not increase average EDS path length significantly for geo networks (Fig. B2d), but the average BFS path length still decreased significantly (Fig. B2c). Navigability of geo networks was not significantly adversely affected because the reduction in clustering did not significantly change the distribution of link lengths (Fig. B2f). Navigability of a small-world network is sensitive to the distribution of link lengths, and a right skewed distribution of link lengths (the probability of a link is inversely related to its length) has been found conducive to navigability of small-world networks [4]. By contrast, the reduction in clustering for PRNs significantly changed the distributions of link lengths by making them less right skewed (Fig. B3f). We use the coefficient of variation (std. dev. / mean) of edge lengths as a proxy for skew-ness of link length distribution. Since median edge length is less than the mean edge length for the networks, a larger edge length coefficient of variation implies more skew to the right, or a longer right-tail. While the edge length coefficient of variation for PRNs range between 1.0 and 2.0 (Fig. B3f), the edge length coefficient of variation for the geo networks are less than 1.0, indicating not much skew at all (Fig. B2f). From this last observation, it appears that the geo networks deviate from the aforementioned condition for navigability in small-world networks. Nonetheless, this condition may not apply to geo networks since they have almost zero correlation between link Euclidean distance and link sequence distance.

The role link length distribution plays in navigability of PRNs is illustrated more clearly with the MGEO4, randSE and randLE networks. Even though the MGEO4 networks are less strongly clustered (Figs. B3a & B3b), they are still navigable (Fig. B3d). The edge length coefficient of variation of MGEO4 networks are smaller than PRNs, but most remain larger than 1.0 (Fig. B3f). The effect of reduced clustering on navigability of PRNs is more dramatic in randSE networks. A randSE network is a PRN with its short-range links (SE) shuffled. Link shuffling is achieved by repeatedly choosing two edges $(p, q)$ and $(r, s)$ uniformly at random with replacement from the pool of links to be shuffled, and rewiring their endpoints where permitted, i.e. without introducing links between consecutive node-pairs or creating multiple links between a node-pair. The shuffling of SE increases the |LE|/|SE| ratio of a PRN (Fig. B3e) to the extent that its edge length distribution is disturbed enough (coefficient of variation falls significantly) to adversely affect PRN navigability. The average EDS path length of randSE networks no longer increase logarithmically with $N$ (Fig. B3d), i.e. randSE networks are not navigable.

However, strong clustering is not the *sine qua non* of PRN navigability by EDS. The randLE networks are navigable in spite of them having clustering at randSE levels because shuffling of LE did not disturb PRN link length distribution enough to negatively affect local search (Fig. B3f). A randLE network is a PRN with its long-range links (LE) shuffled. The randSE and randLE contact maps for protein 1B19 are shown in Fig. B1. The average Spearman correlation between link Euclidean distance and link sequence distance is 0.5129 (std. dev. 0.1143) for randSE networks, and 0.8011 (std. dev. 0.0570) for randLE networks. All correlations are significant. The randSE and randLE results demonstrate the importance of link length distribution to navigability and highlight the important role secondary structures play in PRN navigation.

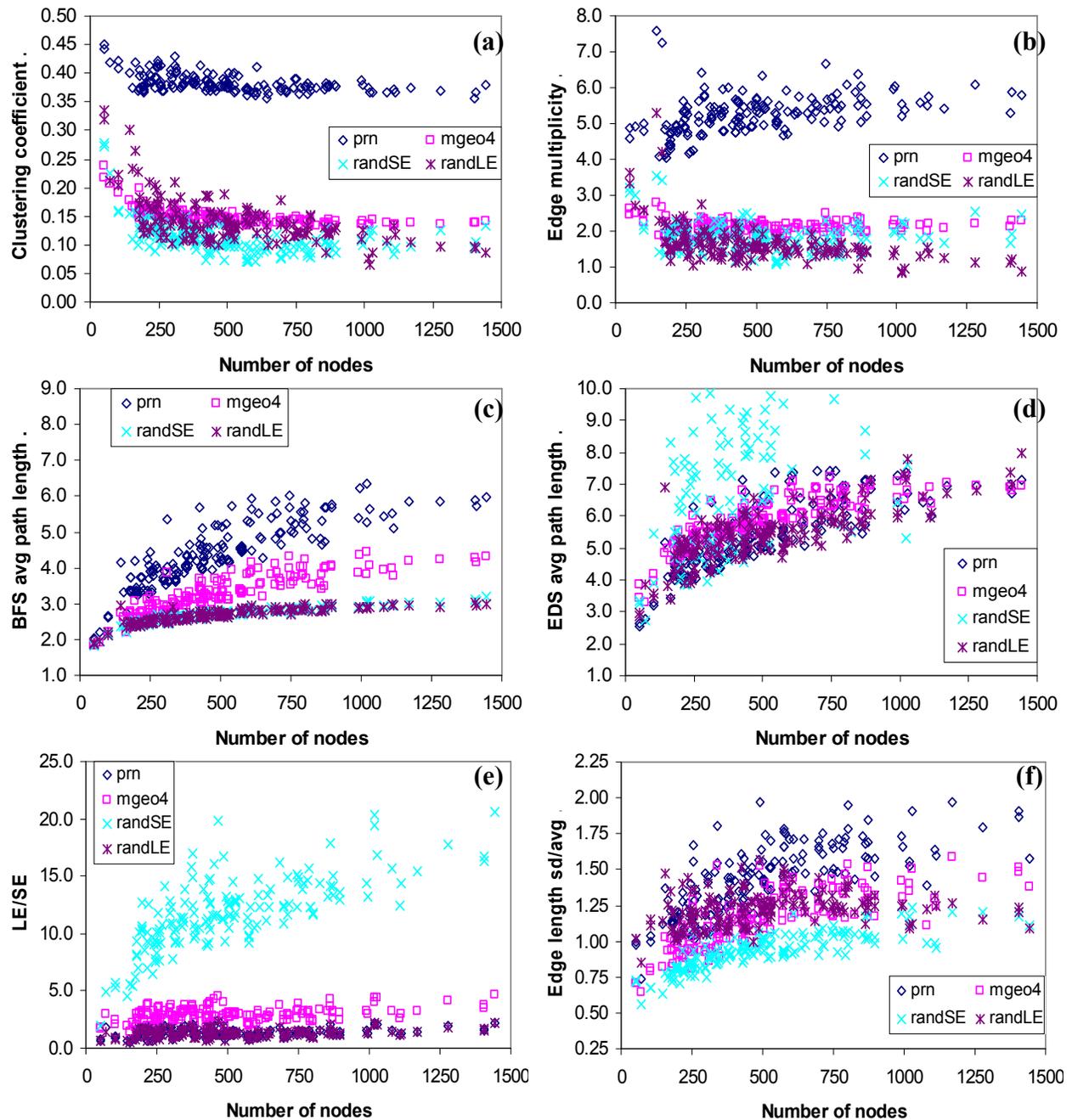

**Fig. B3** The randSE networks are PRNs with their short-range links (SE) shuffled. The randLE networks are PRNs with their long-range links (LE) shuffled. Link shuffling reduces (a) clustering, (b) edge multiplicity and (c) average BFS path length of PRNs. (d) While randLE networks are navigable, randSE networks are not. The top part of plot (d) is truncated. The randSE networks have a maximum average EDS path length of 55. (e) SE shuffling increases the |LE|/|SE| ratio much more than LE shuffling. (f) SE shuffling reduces the coefficient of variation (std. dev. / mean) of edge lengths to the extent that navigability of PRNs is adversely affected.

**Appendix C** Overview of protein selection for PRN & MGEO network construction.

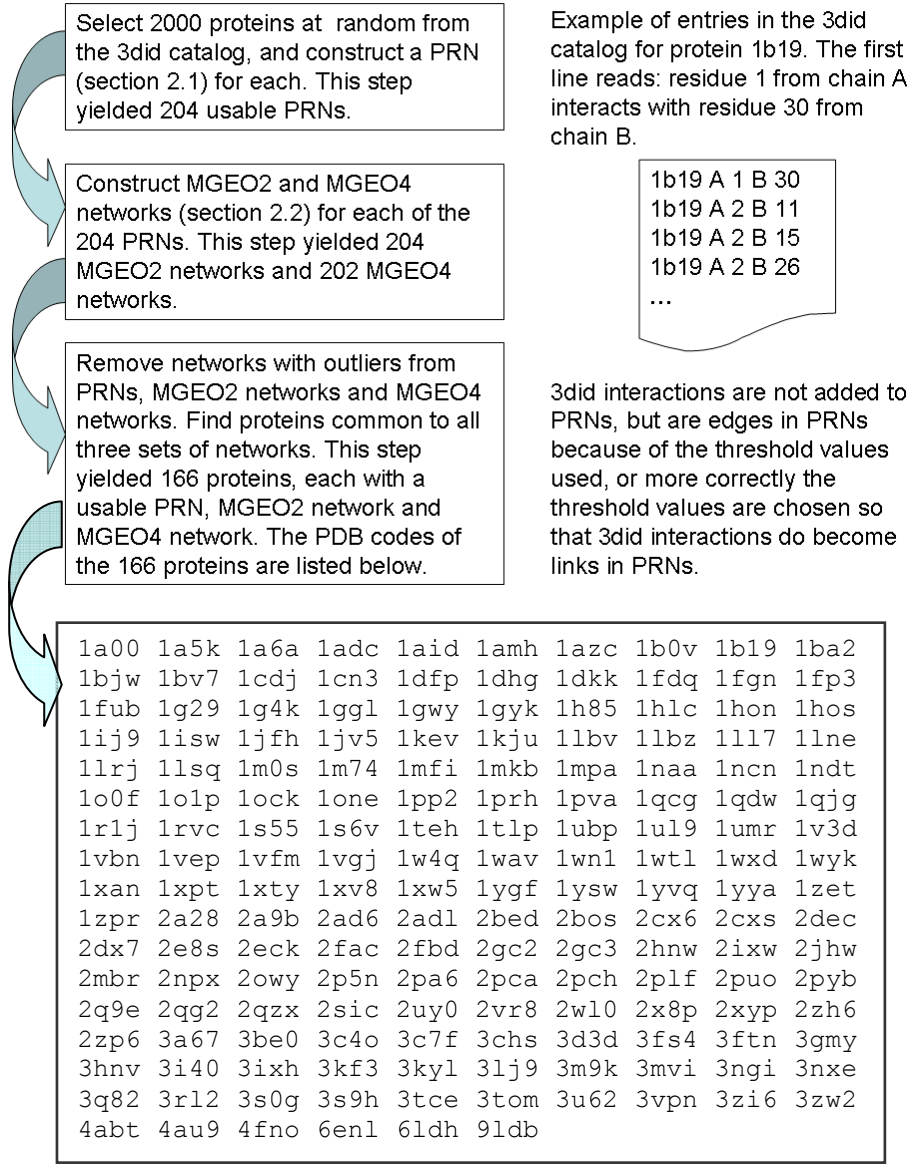

Select 2000 proteins at random from the 3did catalog, and construct a PRN (section 2.1) for each. This step yielded 204 usable PRNs.

Construct MGEO2 and MGEO4 networks (section 2.2) for each of the 204 PRNs. This step yielded 204 MGEO2 networks and 202 MGEO4 networks.

Remove networks with outliers from PRNs, MGEO2 networks and MGEO4 networks. Find proteins common to all three sets of networks. This step yielded 166 proteins, each with a usable PRN, MGEO2 network and MGEO4 network. The PDB codes of the 166 proteins are listed below.

Example of entries in the 3did catalog for protein 1b19. The first line reads: residue 1 from chain A interacts with residue 30 from chain B.

1b19 A 1 B 30
1b19 A 2 B 11
1b19 A 2 B 15
1b19 A 2 B 26
...

3did interactions are not added to PRNs, but are edges in PRNs because of the threshold values used, or more correctly the threshold values are chosen so that 3did interactions do become links in PRNs.

```
1a00 1a5k 1a6a 1adc 1aid 1amh 1azc 1b0v 1b19 1ba2
1bjw 1bv7 1cdj 1cn3 1dfp 1dhg 1dkk 1fdq 1fgn 1fp3
1fub 1g29 1g4k 1ggl 1gwy 1gyk 1h85 1hlc 1hon 1hos
1ij9 1isw 1jfh 1jv5 1kev 1kju 1lbv 1lbz 1ll7 1lne
1lrj 1lsq 1m0s 1m74 1mfi 1mkb 1mpa 1naa 1ncn 1ndt
1o0f 1o1p 1ock 1one 1pp2 1prh 1pva 1qcg 1qdw 1qjg
1r1j 1rvc 1s55 1s6v 1teh 1tlp 1ubp 1ul9 1umr 1v3d
1vbn 1vep 1vfm 1vgj 1w4q 1wav 1wn1 1wtl 1wxd 1wyk
1xan 1xpt 1xty 1xv8 1xw5 1ygf 1ysw 1yvq 1yya 1zet
1zpr 2a28 2a9b 2ad6 2adl 2bed 2bos 2cx6 2cxs 2dec
2dx7 2e8s 2eck 2fac 2fbd 2gc2 2gc3 2hnw 2ixw 2jhw
2mbr 2npx 2owy 2p5n 2pa6 2pca 2pch 2plf 2puo 2pyb
2q9e 2qg2 2qzx 2sic 2uy0 2vr8 2wl0 2x8p 2xyp 2zh6
2zp6 3a67 3be0 3c4o 3c7f 3chs 3d3d 3fs4 3ftn 3gmy
3hnv 3i40 3ixh 3kf3 3kyl 3lj9 3m9k 3mvi 3ngi 3nxe
3q82 3rl2 3s0g 3s9h 3tce 3tom 3u62 3vpn 3zi6 3zw2
4abt 4au9 4fno 6enl 6ldh 9ldb
```

The 3did links are the intra- or inter-chain residue-residue interactions between contacting PFAM domains of a protein as listed in the 3did catalog [27]. A pair of PFAM domains in a protein is deemed able to interact with each other if they have at least five estimated contacts (hydrogen bonds, electrostatic or van der Waals interaction) between them.

**Appendix D** Odd subgraph-centrality and network classes

The expansion factor $\gamma$ of a graph $G$ with $|V|$ nodes is the smallest $\frac{|\mathcal{N}(S)|}{|S|}$ ratio found over all node subsets $S$ where $0 < |S| < |V|/2$, and the boundary set $\mathcal{N}(S)$ comprises all nodes in $V \setminus S$ that are first neighbors of nodes in $S$. Calculating $\gamma$ directly from its definition is a hard problem, but techniques from spectral graph theory can be used to approximate. The expansion factor of a graph $G$ denoted $\gamma(G)$, is related to the spectral gap $\Delta\lambda$, i.e. the difference between the largest ($\lambda_1$) and the second largest ($\lambda_2$) eigenvalues, of $G$'s adjacency matrix by the Tanner-Alon-Milman inequality as $\frac{\Delta\lambda}{2} \leq \gamma(G) \leq \sqrt{2\lambda_1\Delta\lambda}$. A network with a large enough spectral gap is a good expander. Intuitively, good expanders are sparse (bounded node degree), well-connected (small diameter) and robust to small cuts. The spectral scaling method in [38] provides a way to evaluate the expansion property of graphs without defining "large enough". The method builds on the notion of odd-subgraph centrality which measures the weighted participation of a node in closed walks of odd lengths in a network.

Let $EE_{odd}(i)$ be the odd-subgraph centrality for node $i$, $EC(i)$ be the $i^{th}$ component of the principal eigenvector (the eigenvector associated with $\lambda_1$), and $xj(i)$ be the $i^{th}$ component of the eigenvector associated with the $j^{th}$ eigenvalue $\lambda_j$. $EE_{odd}(i) = [EC(i)]^2 \sinh(\lambda_1) + \sum_{j=2}^{N}[xj(i)]^2 \sinh(\lambda_j)$ [Eq. 2 in 34]. When $\lambda_1 >> \lambda_2$, the first term dominates, yielding a power-law relationship between odd-subgraph centrality and the principal eigenvector. Let $EE_{homo}(i)$ represent $EE_{odd}(i)$ when $\lambda_1 >> \lambda_2$. Then a log-log plot of $EE_{homo}(i)$ vs. $EC(i)$ is a straight line with a slop of 0.5 and a y-intercept at $\log(\sinh^{-0.5}(\lambda_1))$. In graphs whose spectral gap is not "large enough", the odd-subgraph centrality values $EE_{odd}(i)$ will deviate from this straight line. The characterization of these deviations, i.e. whether $\log[EE_{homo}(i)/(EE_{odd}(i)]$ is positive or negative, forms the basis upon which the other three network classes are defined. These deviations arise due to the sign of the eigenvalues. Networks with zero deviations, i.e. $EE_{homo}(i) = EE_{odd}(i)$ for all $i$, are Class I networks

Class II networks are those with deviations but only of the negative kind, i.e. $EE_{homo}(i) < EE_{odd}(i)$. Class III networks are those with deviations but only of the positive kind, i.e. $EE_{homo}(i) > EE_{odd}(i)$. Class IV networks are those with deviations of both the positive and the negative kinds. Class II networks correspond to networks with modular structure, i.e. their nodes can be partitioned into two or more clusters such that intra-cluster connectivity is stronger than inter-cluster connectivity. Class III networks, in contrast, have a highly connected core of nodes that is only loosely connected to nodes outside the core, i.e. the periphery nodes. Class IV networks is a hybrid of Class II and Class III organizational patterns.

**Appendix E** EDS search for node 76 (target) starting at node 72 (source) (Fig. 8 right).

At node 72, EDS inspects the 11 direct neighbors of node 72 in random order and computes their respective Euclidean distances to node 76. Since node 75 is the closest to node 76, EDS moves from node 72 to node 75. At this stage, EDS has visited nodes 72 and 75, and has memory of node 72 and its 11 direct neighbors. At node 75, EDS inspects its 11 direct neighbors and adds them to memory. From all the unvisited nodes in memory, EDS finds node 74 to be the closest to the target node. But EDS cannot move directly to node 74 since it is not a direct neighbor of node 75. To visit node 74, EDS must first move or backtrack to node 72. At node 74, EDS inspects its 11 direct neighbors in random order, and finds node 76. The maximum cost of this EDS search is 23 (at most 23 unique nodes are visited or inspected in the course of this search). The EDS path is $\langle 72, 75, 72, 74, 76 \rangle$ which has length four. This path is an example of an EDS path with backtracking that is also hierarchical (section 3.3) since its degree path which is $\langle 11, 11, 11, 11, 4 \rangle$ decreases monotonically. The BFS path is $\langle 72, 74, 76 \rangle$. In the reverse direction, the EDS path is $\langle 76, 74, 72 \rangle$ which is also hierarchical since its degree path increases monotonically.

| Visited node n | Distance to target node | Inspected nodes (direct neighbors of n) | Distance to target node |
|---|---|---|---|
| 72 Degree = 11 | 8.93095 | 75 | 3.81547 |
| | | 74 | 6.33784 |
| | | 49 | 8.08382 |
| | | 70 | 11.71590 |
| | | 2 | 12.84950 |
| | | 69 | 13.61700 |
| | | 68 | 13.97110 |
| | | 67 | 15.32500 |
| | | 6 | 15.50050 |
| | | 3 | 16.14170 |
| | | 65 | 18.82700 |
| 75 Degree = 11 | 3.81547 | 51 | 7.76938 |
| | | 50 | 7.80372 |
| | | 49 | 8.08382 |
| | | 73 | 8.90625 |
| | | 72 (x) | 8.93095 |
| | | 71 | 9.10019 |
| | | 77 | 11.26160 |
| | | 48 | 11.65110 |
| | | 70 | 11.71590 |
| | | 2 | 12.84950 |
| | | 5 | 15.67090 |
| 74 Degree = 11 | 6.33784 | 76 | 0.00000 |
| | | 51 | 7.76938 |
| | | 50 | 7.80372 |
| | | 49 | 8.08382 |
| | | 72 (x) | 8.93095 |
| | | 71 | 9.10019 |
| | | 52 | 11.11780 |
| | | 70 | 11.71590 |
| | | 53 | 13.05500 |
| | | 69 | 13.61700 |
| | | 57 | 18.29190 |
| 76 Degree = 4 | 0.00000 | | |

## Appendix F

The analysis in section 3.7 is extended to 11 other protein structures (PDB codes in Table F1) whose native dynamics (298K) is available in the Dynameomics database [39, 40]. PRN0s were constructed for the chain within the residue range simulated in Dynameomics. Unlike 2EZN, where the entire 6250 simulation was used, stability and commute times of edges in the 11 PRN0s were computed using the first $x$ of the $y$ MD snapshots available (this is due to data download constraints and may be remedied later). We experimented with fewer snapshots for the 2EZN PRN0 and could arrive at the same general conclusion (nonetheless using the whole native ensemble is preferable). A long-range edge ($LE$) is a PRN edge that connects nodes more than 10 residues apart on the protein sequence (section 2.1). A long-range path ($LP$) is a (BFS or EDS) path whose source and destination nodes are more than 10 sequence distance apart from each other (section 3.7).

**Table F1 The 12 PRN0s, including 2EZN.**

| PDB structure (residues) | MD snapshots used/total | PRN0 nodes | PRN0 links | | Paths with > 1 edge in PRN0 | |
|---|---|---|---|---|---|---|
| | | | Short-range ($SE$) | Long-range ($LE$) | Short-range ($SP$) | Long-range ($LP$) |
| 1CUK-A (156-203) | 20,000/51,163 | 48 | 178 | 65 | 494 | 1,276 |
| 1EZG-A (2-83) | 20,000/52,490 | 82 | 326 | 324 | 878 | 4,464 |
| 1ELP-A (1-83) | 20,000/52,318 | 83 | 215 | 328 | 1,120 | 4,600 |
| **2EZN-A (1-101, MODEL 1, 1511 atoms)** | **51,000/51,000** | **101** | **346** | **491** | **1,218** | **7,208** |
| 3GRS-A (366-478) | 20,000/53,650 | 113 | 330 | 329 | 1,490 | 9,848 |
| 1EBD-A (155-271) | 20,000/53,224 | 117 | 335 | 354 | 1,560 | 10,634 |
| 1D0N-A (27-159) | 20,000/51,311 | 133 | 386 | 431 | 1,778 | 14,144 |
| 1IHB-A (5-160) | 20,000/51,867 | 156 | 587 | 489 | 1,836 | 20,192 |
| 1BFD-A (2-181) | 20,000/52,997 | 180 | 561 | 712 | 2,368 | 27,306 |
| 1ESJ-A (1-272) | 20,000/52,274 | 272 | 939 | 1,065 | 3,452 | 66,252 |
| 1BS2-A (136-482) | 12,000/51,989 | 347 | 1,208 | 1,164 | 4,414 | 110,904 |
| 1EHE-A (5-404) | 12,000/51,560 | 399 | 1,430 | 1,391 | 5,010 | 148,150 |

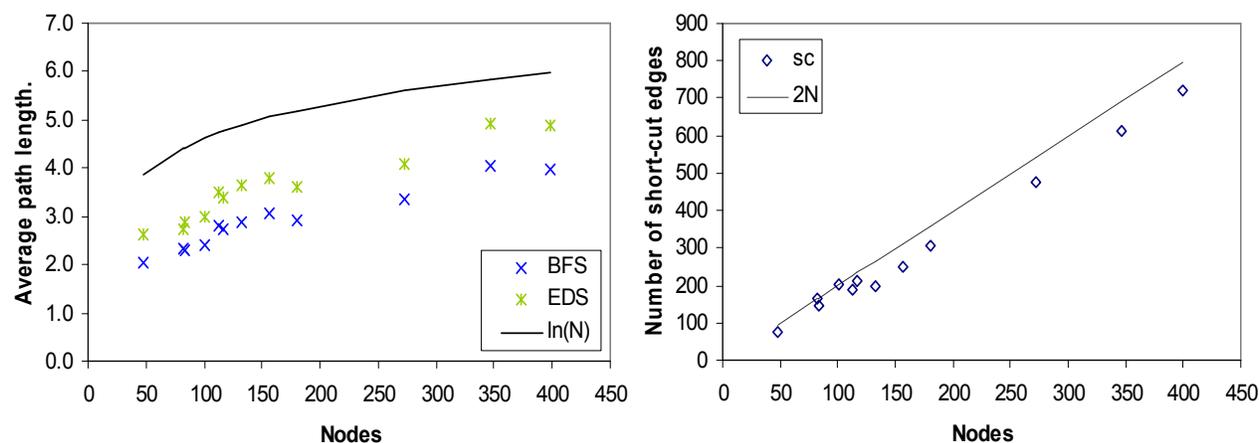

**Fig. F1 Average length of BFS and EDS paths (a) and number of short-cuts (b) for the 12 PRN0s.**

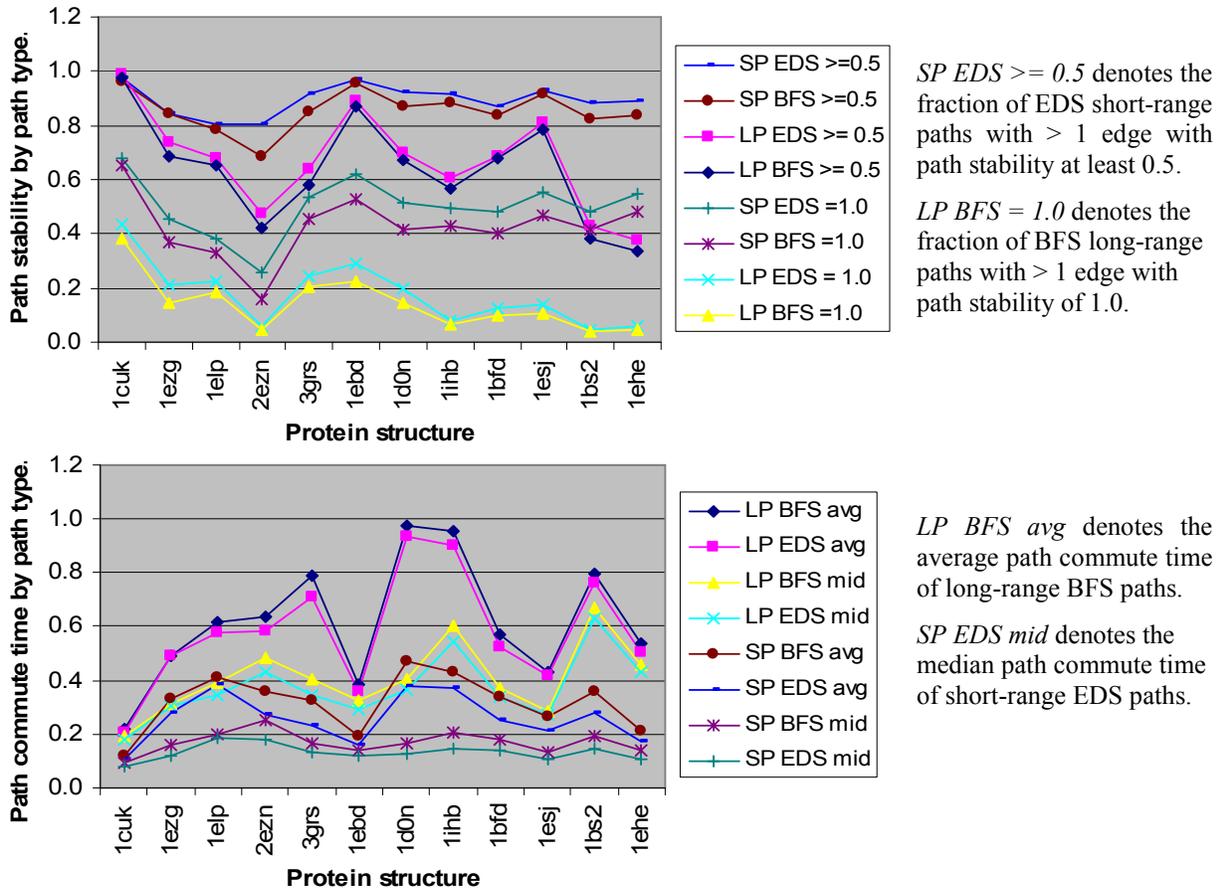

*SP EDS >= 0.5* denotes the fraction of EDS short-range paths with > 1 edge with path stability at least 0.5.

*LP BFS = 1.0* denotes the fraction of BFS long-range paths with > 1 edge with path stability of 1.0.

*LP BFS avg* denotes the average path commute time of long-range BFS paths.

*SP EDS mid* denotes the median path commute time of short-range EDS paths.

**Fig. F2 Top:** Regardless of search type (BFS or EDS), short-range paths are more stable than long-range paths. Regardless of path range (short or long), EDS paths are more stable than BFS paths. **Bottom:** Regardless of search type (EDS or BFS), long-range paths have longer commute time than short-range paths. Regardless of path range (short or long), BFS paths have longer commute time than EDS paths.

**Table F2 p-values generated with R's Wilcoxon one-sided test, paired when possible (path comparisons).** For all the 12 PRN0s, *LP* EDS paths are significantly (p-value < 0.05) more stable than *LP* BFS paths, and *LP* EDS paths have significantly smaller path commute time than *LP* BFS paths. Except for 1EZG, short-range links (*SE*) in PRN0s are significantly more stable and have significantly smaller commute time than long-range links (*LE*) in PRN0s. Except for 1EZG, short-cut links (*SC*) are significantly more stable and have significantly smaller commute time than non short-cut links (*NSC*).

| Protein structure | *LP* path stability | *LP* path commute time | Edge stability | | Edge commute time | |
|---|---|---|---|---|---|---|
| | BFS < EDS | BFS > EDS | *SE > LE* | *SC > NSC* | *SE < LE* | *SC < NSC* |
| 1CUK-A | 2.07E-02 | 5.20E-19 | 2.68E-09 | 2.05E-03 | 1.32E-25 | 2.66E-07 |
| 1EZG-A | 9.02E-29 | 1.45E-02 | 3.68E-01 | 1.82E-09 | 2.57E-07 | 1.92E-01 |
| 1ELP-A | 2.13E-09 | 4.21E-25 | 1.83E-03 | 2.77E-10 | 1.39E-09 | 1.98E-05 |
| **2EZN-A** | **9.45E-23** | **2.53E-82** | **8.40E-31** | **1.54E-42** | **1.80E-52** | **1.19E-27** |
| 3GRS-A | 4.42E-51 | 1.90E-106 | 4.94E-14 | 6.15E-14 | 1.87E-24 | 1.10E-14 |
| 1EBD-A | 5.35E-60 | 3.57E-115 | 3.31E-10 | 7.49E-11 | 9.06E-31 | 1.74E-15 |
| 1D0N-A | 5.25E-46 | 7.34E-86 | 3.05E-19 | 1.99E-19 | 9.84E-29 | 7.12E-23 |
| 1HB-A | 1.18E-59 | 2.11E-140 | 1.46E-73 | 6.89E-21 | 2.40E-103 | 2.28E-24 |
| 1BFD-A | 1.30E-25 | 4.61E-236 | 2.68E-43 | 3.89E-24 | 5.09E-63 | 3.24E-26 |
| 1ESJ-A | 1.57E-165 | 2.22E-267 | 8.06E-67 | 2.43E-30 | 9.46E-147 | 7.79E-43 |
| 1BS2-A | 2.33E-281 | 0.00E+00 | 1.01E-113 | 2.07E-54 | 5.28E-187 | 1.54E-67 |
| 1EHE-A | 0.00E+00 | 0.00E+00 | 4.92E-143 | 7.89E-65 | 8.34E-247 | 2.57E-76 |

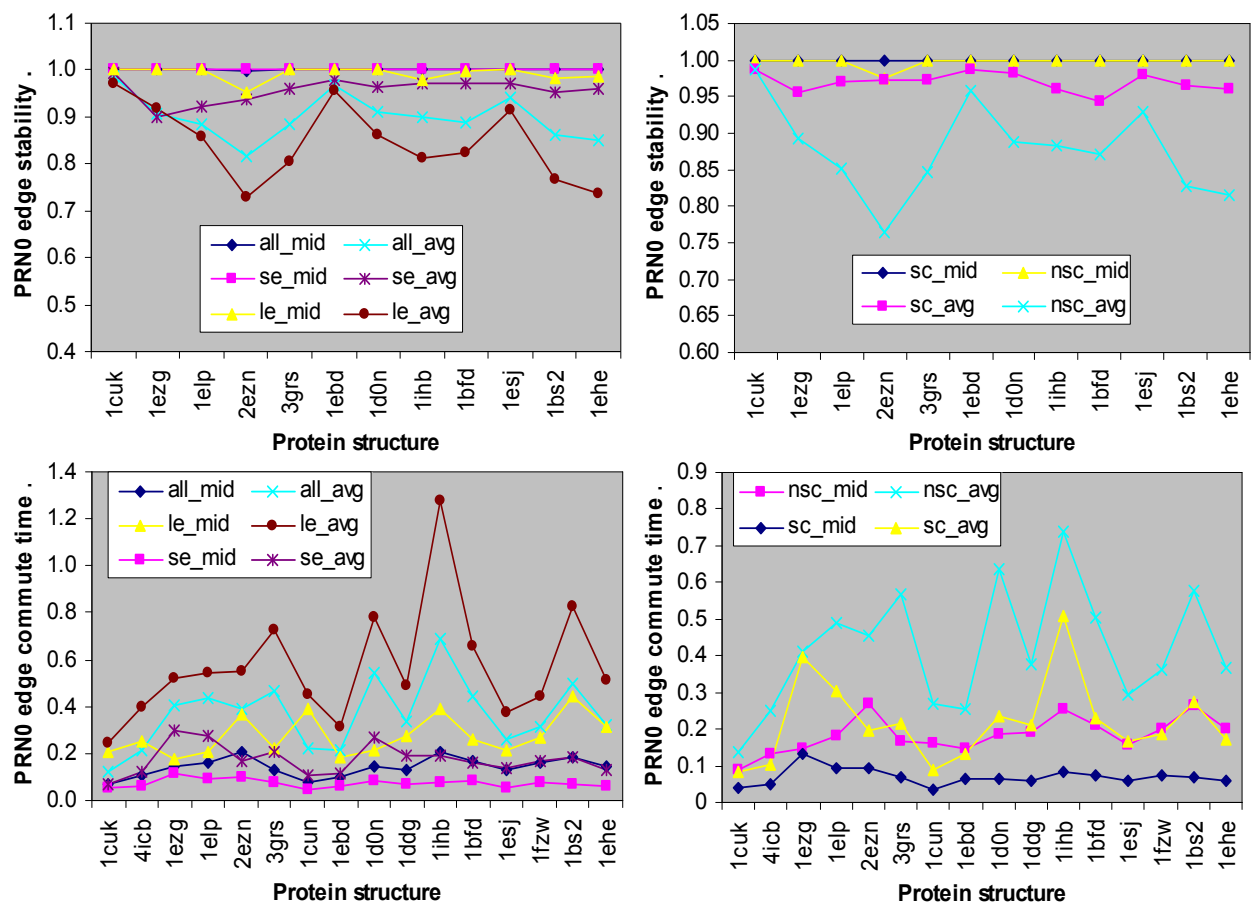

**Fig. F3 (a)** Median (mid) and mean (avg) stability values of edges in PRN0. Short-range edges (*SE*) are more stable than long-range (*LE*) edges. **(b)** Short-cut edges (*SC*) are more stable than non short-cuts (*NSC*). **(c)** Median (mid) and mean (avg) commute times of edges in PRN0. Long-range edges (*LE*) have larger commute time than short-range (SE) edges. A pair of residues with small commute time means the Euclidean distance between the pair of carbon-alpha atoms has not varied much over time (the MD native dynamics snapshots). It stands to reason that if a link exists between such a pair, the link is expected to be highly stable. **(d)** Short-cut edges (*SC*) have higher communication propensity (smaller commute time) than non short-cuts (*NSC*).

## Appendix G

The algorithm to match deleted with added short-cuts in section 4.3 assumes that most of the short-cut changes take place within the largest SCN component (gSCN), which they do. For the 2EZN MD dataset, the average number of short-cuts removed in a step (Fig. G1a) is almost balanced by the average number of short-cuts added in a step (Fig. G1b); almost all of the deleted short-cuts (edges in $delSC_{t+1}$) are in $gSCN_t$ (Fig. G1c); and about 85% of the added short-cuts (edges in $addSC_{t+1}$) can be found in $gSCN'_t$ (Fig. G1d).

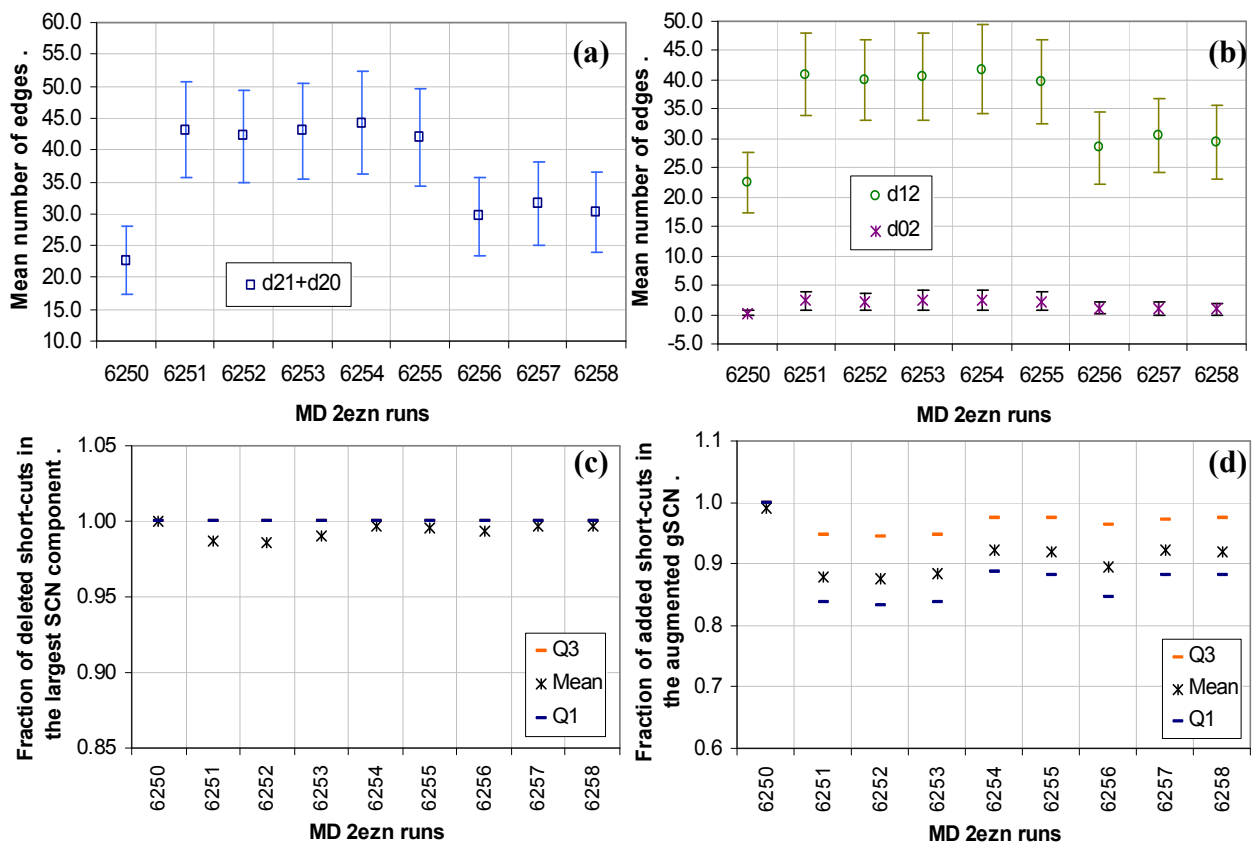

**Fig. G1** **(a)** The average number of short-cuts removed (edges making either a d21 or a d20 transition) in a step. **(b)** The average number of short-cuts added in a step. d12 denotes non-short-cut edges that become short-cuts in the next step/snapshot. d02 denotes non-edges that become short-cut edges in the next step. **(c)** Almost all of the deleted short-cuts (d21 and d20 edges) are found in the largest SCN component. **(d)** An average of at least 85% of the added short-cuts (d12 edges) can be found in the augmented largest SCN component. In (a) & (b), error bars denote standard deviation about the mean. In (c) & (d), Q1 and Q3 denote the first and third quartiles respectively.

*Notes on the algorithm to match deleted with added short-cuts in section 4.3.* We adopted Kruskal's approach [61 p. 504] of constructing spanning trees in step 2 of section 4.3. We found that starting with a forest of trees made up of deleted short-cuts worked better (yielded lower unused and unmatched rates) for both MD datasets than growing the spanning tree as a single tree starting from a single edge (Figs. G2 & G3). Some deleted short-cuts in gSCN will still be left out of *ST* but this is unavoidable (SCNs are strongly transitive). To remedy this, such cycles of deleted short-cuts are broken up by removing an edge chosen at random from the cycle, and the analysis is repeated on five *ST*s constructed with a different random number seed each time. Results from the different *ST*s did not vary significantly from each other. For an added short-cut *f* to be a replacement edge of a deleted short-cut *e* = (*u*, *v*) with respect to a *ST*, *e* must be part of *ST*, and *f* needs to "straddle" *e*, i.e. *f* must be part of some path from *u* to *v* in gSCN' that does not involve *e*. To increase the chance of this happening, we try to place nodes incident to many added short-cuts on the periphery of *ST*. Observe that leaf nodes in a spanning tree have more possible edge partners than non-leaf nodes, and edges between nodes in the same sub-tree cannot contribute to a cut-set.

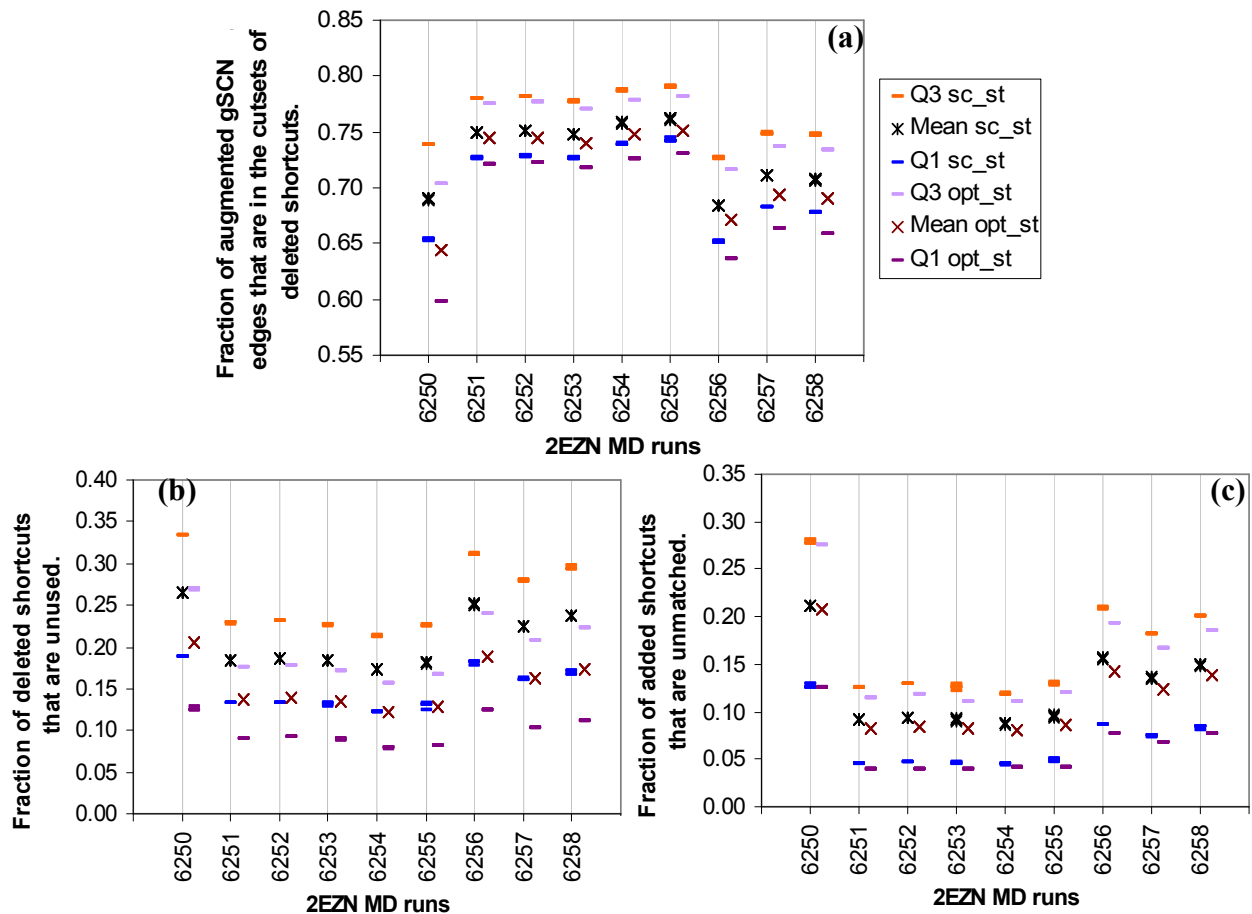

**Fig. G2** Results from our first attempt (growing *ST* as a tree from a single starting edge) are suffixed *sc_st*. Results from our second attempt (growing *ST* from a forest of starter trees with the heuristic described in the notes) are suffixed *opt_st*. Despite producing significantly smaller *CUT*s (union of the cutsets of deleted edges in gSCN) (a), *opt_st* outperformed *sc_st* by yielding lower unused rates (b) and lower unmatched rates (c).

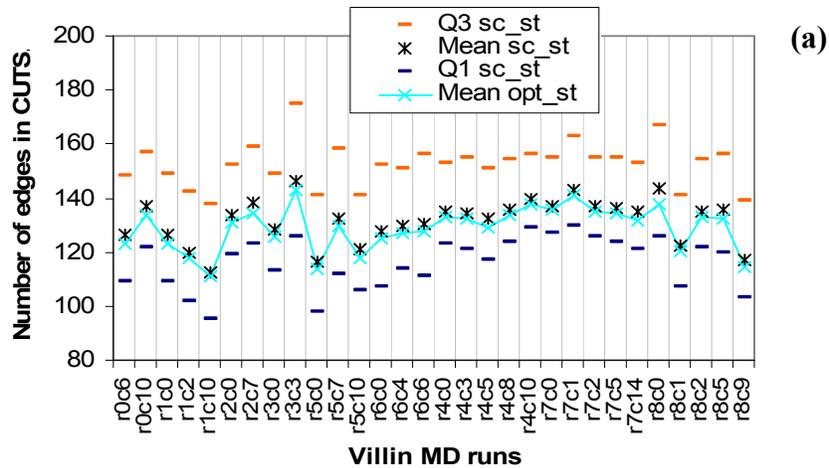

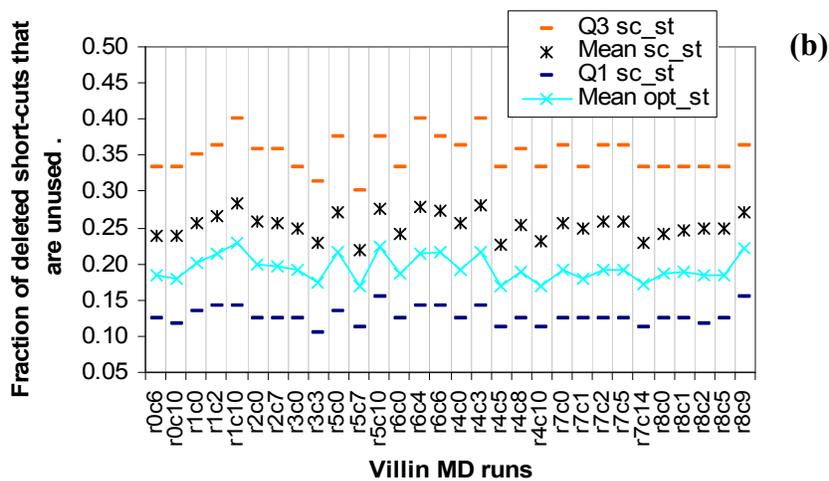

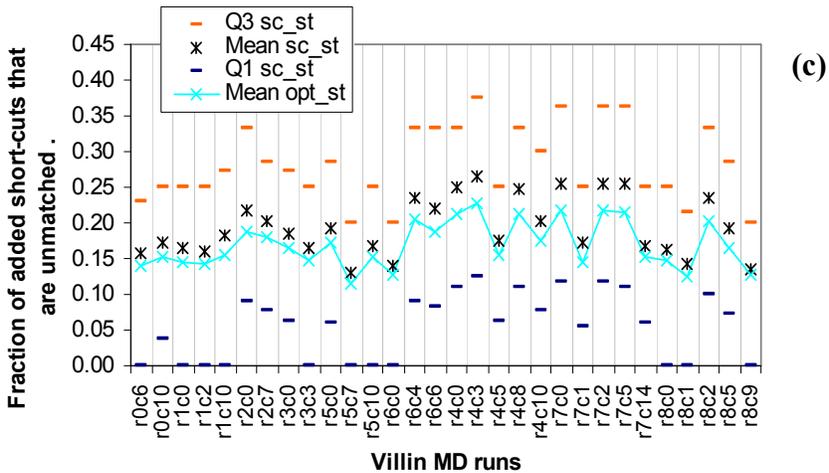

**Fig. G3** The light blue line (*opt_st*) plots the results from our second attempt (growing *ST* from a forest of starter trees with the heuristic described in the notes). It is compared with results from our first attempt (*sc_st*) (growing *ST* as a tree from a single starting edge). Despite producing significantly smaller *CUTs* (union of the cutsets of deleted edges) (a), the second approach outperformed the first by yielding lower unused rates (b) and lower unmatched rates (c).

**Appendix H** Including peptide bonds, (*x*, *x*+1) links, in PRNs.

The plots in Fig. H highlight the major differences (beyond magnitude) when peptide bonds are included into the 2EZN PRNs. There is no clear gap in clustering between native and non-native runs (a) but average edge multiplicity still increases as 2EZN folds, and a clear gap between native and non-native runs is still discernable (b). The number of short-cut edges is much fewer and does not increase as 2EZN folds (c).

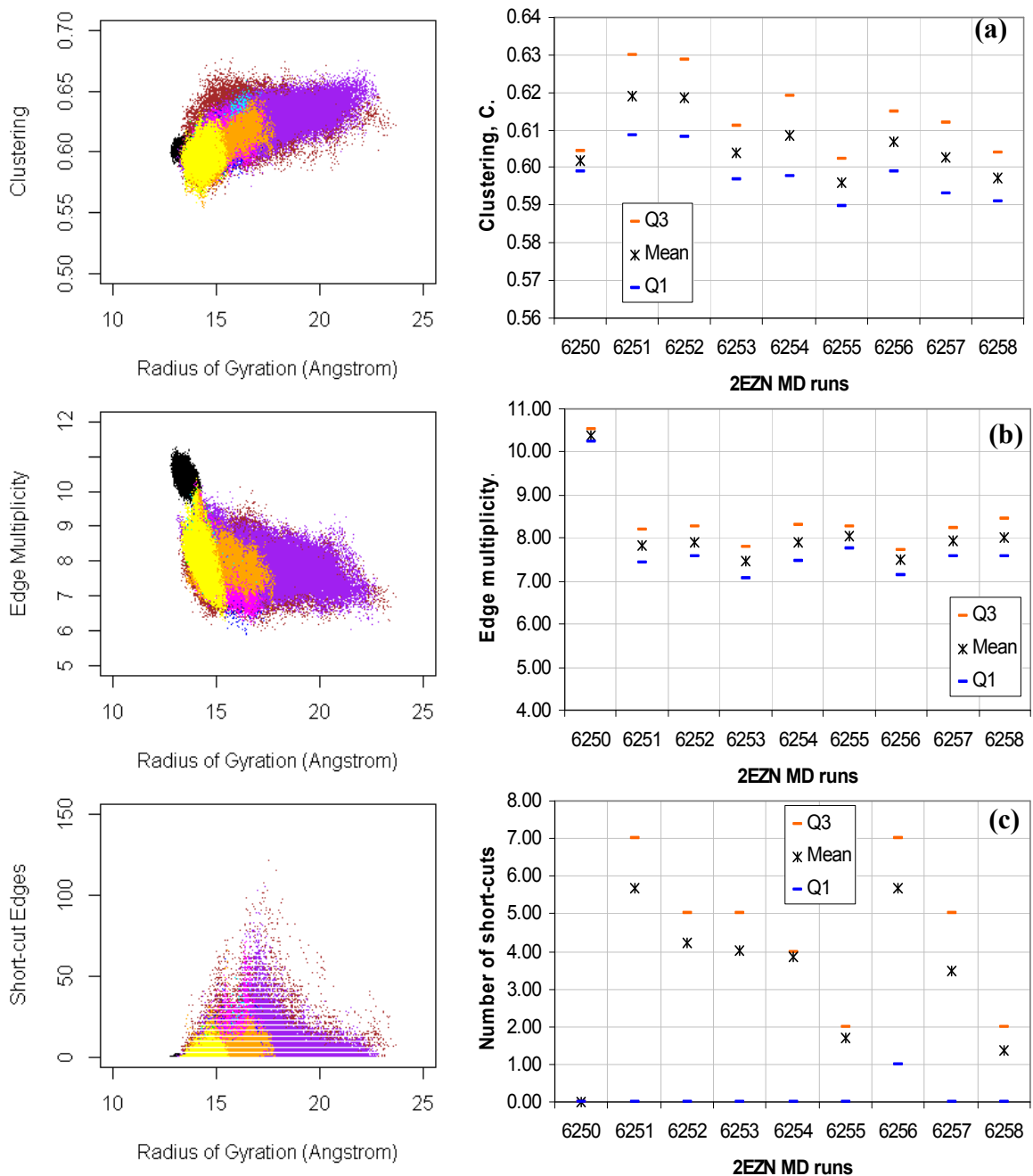

**Fig. H1** Main effects of including peptide bonds |*x* − *y*| > 0 into the 2EZN PRNs. Compare (a) & (b) with Figs. 1a & 1b respectively, and (c) with Fig. 24a.